\documentclass[12pt]{article}
\usepackage{amssymb}
\usepackage{amsmath}
\usepackage{youngtab}
\usepackage{cite}
\usepackage[]{hyperref}
\usepackage{multirow}
\input xy
\xyoption{all}

\numberwithin{equation}{section}

\begin{document}

\begin{center}

{\large\bf More nonabelian mirrors and some two-dimensional dualities}

\vspace{0.2in}

Wei Gu, Hadi Parsian, Eric Sharpe

Dep't of Physics\\
Virginia Tech\\
850 West Campus Dr.\\
Blacksburg, VA  24061\\

{\tt weig8@vt.edu}, {\tt varzi61@vt.edu},
{\tt ersharpe@vt.edu}

$\,$

\end{center}

In this paper we extend the nonabelian mirror proposal of two of the
authors from two-dimensional gauge theories with
connected gauge groups to the case of $O(k)$ gauge groups with
discrete theta angles.
We check our proposed extension by counting and comparing vacua in 
mirrors to known dual
two-dimensional $(S)O(k)$ gauge theories.  The mirrors in question are
Landau-Ginzburg orbifolds, and for mirrors to $O(k)$ gauge theories,
the critical loci of the mirror superpotential often intersect
fixed-point loci, so that to count vacua, one must take into account twisted
sector contributions.  This is a technical novelty relative to mirrors of
gauge theories with connected gauge groups, for which critical loci
do not intersect fixed-point loci and so no orbifold twisted sector
contributions are pertinent.  The vacuum computations turn out to be a rather
intricate test of the proposed mirrors, in particular as untwisted
sector states in the mirror to one theory are often exchanged 
with twisted sector states in the
mirror to the dual.
In cases with nontrivial IR limits, we also check that central charges
computed from the Landau-Ginzburg mirrors match those expected for the
IR SCFTs.

\begin{flushleft}
July 2019
\end{flushleft}

\newpage

\tableofcontents

\newpage

\section{Introduction}

In the paper \cite{Gu:2018fpm}, two of the authors proposed a
description of mirrors to nonabelian two-dimensional gauge theories,
extending work on mirrors to abelian two-dimensional gauge theories
in \cite{Hori:2000kt}.  The paper \cite{Gu:2018fpm} only considered
mirrors to gauge theories with connected gauge groups:
$SO(k)$ but not $O(k)$, for example.  The later work
\cite{Chen:2018wep} also only considered mirrors of theories with connected
gauge groups.  The purpose of this paper is to remedy this omission
by extending
the nonabelian mirror proposal of \cite{Gu:2018fpm} to
mirrors to gauge theories with non-connected gauge group $O(k)$, 
which we will check
by comparing mirrors to dual gauge theories described in
\cite{Hori:2011pd}.  The dualities described there
relate, for example, theories with gauge
group $SO(k)$ to theories with gauge group $O(k)$ with various mod two theta
angles and ${\mathbb Z}_2$ orbifold actions, denoted $O_{\pm}(k)$ 
in \cite{Hori:2011pd},
and so by studying the mirrors to those duals
we can get some nontrivial tests of
our extension of the nonabelian mirror proposal of \cite{Gu:2018fpm}.

The dual gauge theories discussed in \cite{Hori:2011pd}
are Seiberg dual:  equivalent in the IR, but not necessarily in the UV.
We will compute mirrors to gauge theories on each side of the
dualities in \cite{Hori:2011pd}, and compare the number of vacua
in the mirrors to either side.  
This is a rather intricate computation, and we will
see that in every case, our (extended) nonabelian mirror construction
correctly duplicates both the number of vacua of the original
gauge theory as well as the number of vacua of the dual.

On a more technical level, this paper also extends the nonabelian
mirror proposal of \cite{Gu:2018fpm} in another way.
The nonabelian mirrors constructed are Landau-Ginzburg orbifolds,
but for mirrors to theories with
connected gauge groups, the orbifold group acts freely on the
non-excluded critical loci, so that there are no twisted sector 
contributions to low-energy analyses.  By contrast, mirrors to
$O_{\pm}(k)$ theories will often involve orbifolds whose fixed point
locus does intersect the critical locus, and so it will be necessary
to consider twisted sector contributions.  Furthermore, the mirrors to
the dualities of \cite{Hori:2011pd} will often relate
untwisted sector vacua on one side of the duality, to a combination of
untwisted and twisted sector vacua on the other side of the duality.
Global symmetries in one theory become quantum symmetries \cite{Vafa:1989ih}
in the other.
This is part of the reason why we described the vacuum counting and
comparisons as a rather intricate test of the nonabelian mirror proposal:
to get the number of vacua right is a very sensitive test of whether
we have utilized the correct orbifold in the mirror theory.

We begin in section~\ref{sect:proposal} by reviewing the nonabelian
mirrors proposal, and describing its extension to gauge theories
with gauge groups $O_{\pm}(k)$.  We also very briefly review
the notion of regularity, defined in \cite{Hori:2011pd}, as the
dualities we will test later relate regular gauge theories in that
sense, and we include a table (taken from \cite{Hori:2011pd}) listing
the number of vacua expected in each (regular) gauge theory pertinent to
the dualities explored here.  Later in this
paper we will see mirror computations reproducing every entry of
that table. 

In section~\ref{sect:so:o:duality}, we then test the
dualities described in \cite{Hori:2011pd} relating $SO$ and $O_+$ gauge
theories, by computing the number of vacua in the mirrors to the gauge
theories on either side, and verifying that the number of vacua in the
mirrors match on either side of the duality, and also match the number of
vacua in the original gauge theory.  (As the original gauge theories
are related in the IR, so too do we expect the mirrors to match in the IR.)
In the mirrors to the $O_{\pm}$
gauge theories, the fixed-point locus will often intersect the critical locus
of the superpotential, so that there can be twisted sector contributions to
the vacua, resulting in a rather intricate computation.  To assist
the reader, rather than just giving general arguments, we also examine in
detail three prototypical special cases:  the mirrors to the dualities
$O_+(2) \leftrightarrow SO(2)$, $O_+(1) \leftrightarrow SO(N)$,
and $O_+(3) \leftrightarrow SO(2)$ (for various matter contents).  It is our
hope that the details in these special cases are easier to follow explicitly,
and so we include them to improve readability of the paper.

In addition to counting vacua, for cases in which the theory is
expected to flow to a nontrivial SCFT, we also check that the central
charge one computes from the Landau-Ginzburg mirror matches that one
expects for the IR SCFT.  That central charge comparison applies to a variety
of cases, so we only describe it once.

In section~\ref{sect:o-o-}, we perform the analogous computations and
verifications for the gauge dualities in \cite{Hori:2011pd} relating
$O_-$ gauge theories to other $O_-$ gauge theories.  In each case,
we compute mirrors to the gauge theories on each side of the duality,
and verify that the number of vacua in the mirror matches the number of
vacua of the original gauge theory as well as in the mirror to the
dual gauge theory.
Here also, the fixed-point locus of the Landau-Ginzburg orbifold of the
mirror can intersect the critical locus, and so again this results
in an intricate intermixing between untwisted and twisted sector
contributions.

In sections~\ref{sect:other}, \ref{sect:sp-sp} we check some additional
claims about gauge theories in \cite{Hori:2011pd} using the nonabelian
mirror construction, such as supersymmetry breaking in cases with too
few vectors, and dualities between gauge theories with symplectic
gauge groups.  These computations do not specifically involve
$O_{\pm}$ mirrors, and for the most part can be performed using just
the original nonabelian mirror construction for connected gauge
groups \cite{Gu:2018fpm}.  We include them here as this seems the
natural place to discuss mirrors of these other results from \cite{Hori:2011pd}.

Finally, in appendix~\ref{app:chiral} we briefly make some observations
about mirrors to chirals rings.

\section{Proposal}
\label{sect:proposal}

\subsection{Basics}

In \cite{Gu:2018fpm}, it was proposed that the mirror to a
(2,2) supersymmetric GLSM with connected gauge group $G$, with
matter in some representation ${\cal R}$ and with twisted
masses $\tilde{m}_i$, is a Weyl-group-orbifold of
a Landau-Ginzburg model with fields
\begin{itemize}
\item $(\dim {\cal R})$ chiral superfields $Y_i$, each of periodicity
$2 \pi i$, mirror to the matter fields,
\item $(\dim G - {\rm rank}\, G)$ chiral superfields $X_{\tilde{\mu}}$,
\item $({\rm rank}\, G)$ chiral superfields $\sigma_a$,
\end{itemize}
and a superpotential
\begin{eqnarray}
W & = & \sum_{a=1}^r \sigma_a \left( 
\sum_{i=1}^N \rho_i^a Y_i \: - \: \sum_{\tilde{\mu}=1}^{n-r} \alpha_{\tilde{\mu}}^a 
\ln X_{\tilde{\mu}}
\: - \: t_a \right) 
\nonumber \\
& & 
\: - \: \sum_{i=1}^N \tilde{m}_i Y_i
\: + \: \sum_{i=1}^N \exp\left(-Y_i\right) \: + \:
\sum_{\tilde{\mu}=1}^{n-r} X_{\tilde{\mu}}, 
\label{eq:proposal-w}
\end{eqnarray}
generalizing the construction of mirrors to abelian GLSMs in
\cite{Hori:2000kt},
where the $\rho_i^a$ are the weight vectors for the representation
${\cal R}$, and the $\alpha_{\tilde{\mu}}^a$ are root vectors for the
Lie algebra of $G$.  

The purpose of this paper is to extend this proposal to cases in
which the group $G$ is not connected.  Specifically, in this section
we will propose the form of mirrors to theories with gauge
group $O_{\pm}(n)$, which we will check by comparing mirrors of the
dual $SO$, $O_{\pm}$ theories discussed in \cite{Hori:2011pd}.
(In addition, we will also discuss a handful of other results
in \cite{Hori:2011pd} which do not involve $O_{\pm}$ mirrors, but
which will enable us to test the nonabelian mirror proposal of
\cite{Gu:2018fpm}.)

The mirrors to theories with gauge group $O_{\pm}(n)$ for $n$ even
will have a somewhat different form from mirrors to 
$O_{\pm}(n)$ gauge theories with $n$ odd.  We discuss each case in
turn below.

\subsection{$O_{\pm}$(even)}
\label{sect:orb-even}

First, let us recall the details of the Weyl group orbifold action
on the mirror that already exists for the mirror of $SO(2k)$.
The Weyl group $W$ of $SO(2k)$ is an extension
\begin{equation}
1 \: \longrightarrow \: K \: \longrightarrow \: W \: \longrightarrow \:
S_k \: \longrightarrow \: 1,
\end{equation}
where $K$ is the subset of $({\mathbb Z}_2)^k$ with an even number of
generators.  The action on fields is as follows
(in the notation of \cite{Gu:2018fpm}[sections 9, 10]):
\begin{itemize}
\item Elements of $S_k$ permute the $\sigma_a$ and also permute 
corresponding blocks $\{ Y_{i, 2a-1}, Y_{i, 2a} \}$ (in mirrors to
vector-valued fields),
\item Elements of $K$ act by sign flips on $\sigma$'s:
\begin{equation}
\sigma_a \: \mapsto \: \epsilon_a \sigma_a,
\end{equation}
where $\epsilon_a \in \{\pm 1\}$, and $\epsilon_1 \epsilon_2 \cdots \epsilon_k
= +1$.  At the same time, for each sign flip on a $\sigma_a$,
one also exchanges 
\begin{eqnarray}
Y_{i,2a-1} & \leftrightarrow & Y_{i,2a},
\\
X_{\mu,2a-1} & \leftrightarrow & X_{\mu, 2a} \: \: \:
\mbox{ for } \mu < 2a-1,
\\
X_{2a-1,\nu} & \leftrightarrow & X_{2a,\nu} \: \: \:
\mbox{ for } \nu > 2a.
\end{eqnarray}
\end{itemize}

We propose that the mirror to $O(2k)$ is defined by an additional
${\mathbb Z}_2$ orbifold.  We can think of this as one that
acts as $\sigma_k \mapsto - \sigma_k$, with a corresponding
action $Y_{i,2k-1} \leftrightarrow Y_{i,2k}$ (for mirrors to vector-valued
fields), as well as $X_{\mu,2k-1} \leftrightarrow X_{\mu,2k}$.
More invariantly,
we can think of combining this with the Weyl group of $SO(2k)$ to
form a $W'$ orbifold where
\begin{equation}
1 \: \longrightarrow \: ({\mathbb Z}_2)^k \: \longrightarrow \:
W' \: \longrightarrow \: S_k \: \longrightarrow \: 1.
\end{equation}
In essence, we drop the constraint on $K$, and allow all elements
of $({\mathbb Z}_2)^k$.

Now, strictly speaking, there are two different $O(2k)$ gauge
theories, distinguished by whether the extra ${\mathbb Z}_2$ orbifold
is $\tau$ or $\tau (-)^F$, and this choice distinguishes
$O_+(2k)$ from $O_-(2k)$, in the notation of \cite{Hori:2011pd}.
We will carefully define and distinguish $\tau$ from $\tau (-)^F$ in
section~\ref{sect:proto:o+2}, when we discuss details of twisted sector
contributions.  Briefly, in our conventions, for a single free chiral
superfield, $\tau$ flips the sign of the RR Fock vacuum, whereas
$\tau (-)^F$ leaves it invariant.

Now, we need to specify which of these two extra
${\mathbb Z}_2$ orbifolds, $\tau$ or $\tau (-)^F$, appears in the
mirror.
To further confuse matters, when one
speaks of Hori-Vafa-type mirror constructions, sometimes one takes
the mirror superpotential to be a twisted superpotential depending upon
twisted chirals, and sometimes one takes it to be an ordinary superpotential
depending upon ordinary chirals, and these two matters are linked,
essentially because this distinguishes complex from twisted masses,
which can have different numbers of twisted sector ground states.
Our proposal for the extra ${\mathbb Z}_2$ orbifold above is as follows:
\begin{itemize}
\item In conventions in which the mirror superpotential is an ordinary
superpotential, depending upon ordinary chiral superfields,
we take the extra orbifold appearing in the mirror to be 
$\tau (-)^F$.
\item In conventions in which the mirror superpotential is a twisted
superpotential, depending upon twisted chiral superfields,
we take the mirror of the orbifold to be just $\tau$.
\end{itemize}
(It may be helpful for the reader to recall that T-duality performs
an analogous swap of ${\mathbb Z}_2$s, see for example
\cite{Seiberg:1986by}.)

In this paper, we will take the mirror superpotential to
an $O_+(2k)$ gauge theory to be an
ordinary superpotential, a function of ordinary chiral superfields.
In this convention, the extra ${\mathbb Z}_2$ will be
$\tau (-)^F$, in the notation of \cite{Hori:2011pd}.
Similarly, we take the mirror to the $O_-(2k)$ gauge theory
to be an ordinary superpotential, depending upon ordinary chiral superfields
with the extra ${\mathbb Z}_2$ given by $\tau$.

\subsection{$O_{\pm}$(odd)}
\label{sect:orb-odd}

We have just argued that in the mirror of an $O(2k)$ gauge theory,
we essentially orbifold by the Weyl group of $SO(2k+1)$.
This then begs the question, how do we describe the mirror of
$O(2k+1)$?

Another puzzle in describing this case arises from the fact that it
includes $O(1) = {\mathbb Z}_2$ as a special case.  As discussed
in \cite{Pantev:2005zs,Pantev:2005rh,Pantev:2005wj,Hellerman:2006zs,
Sharpe:2014tca}, orbifolds with trivially-acting ${\mathbb Z}_2$'s
`decompose' into two copies of one underlying theory (a notion
that generalizes to other two-dimensional gauge theories whose
gauge groups have nontrivial finite centers).  Our mirror
proposal for $O$(odd) gauge theories
needs to account for the possibility of such decompositions.

Finally, the difference between $O_{\pm}$(odd) gauge theories
depends upon both a choice of orbifold ($\tau$ or $\tau (-)^F$),
as well as the number $N$ of vectors in the gauge theory.

With the questions above in mind, 
we propose that the mirror of an $O_{\pm}(k)$ gauge theory for
$k$ odd, for generic twisted masses, is
either one or two copies of the mirror to the corresponding
$SO(k)$ gauge theory (with the same matter),
depending upon the $\pm$ as well as upon whether
$N$ is even or odd.  The precise dictionary we give in the table below:
\begin{center}
\begin{tabular}{c|cc}
& $N$ even & $N$ odd \\ \hline
$O_+$(odd) & 2 copies & 1 copy \\
$O_-$(odd) & 1 copy & 2 copies
\end{tabular}
\end{center}
As a consistency check, this correctly reproduces relations between
vacua in $O_{\pm}$(odd) and $SO$(odd) gauge theories listed in
\cite{Hori:2011pd}[table~(4.20)], computed for generic nonzero
twisted masses, and we will check extensively that it is also consistent with 
(mirrors to) dualities of two-dimensional theories later in this paper.

In particular, for $O_+$(odd) gauge theories with $N$ even,
and $O_-$(odd) gauge theories with $N$ odd, our proposal is that
the mirror decomposes into two disjoint theories, which suggests
that, at least for generic nonzero masses,
the original gauge theories also decompose into a disjoint union
of simpler theories, much as in \cite{Sharpe:2014tca}.
We will not pursue
this predicted decomposition of certain $O$(odd) gauge theories here,
but instead leave it for future work.

\subsection{Regularity}
\label{sect:regular-review}

The dual gauge theories described
in \cite{Hori:2011pd} are all assumed to be `regular.'
This term is defined in that reference, and for completeness,
we repeat that definition here.  Briefly,
an $(S)O(k)$ gauge theory with $N$ chirals in the vector representation
is regular when
\begin{itemize}
\item $N-k$ is odd and the mod 2 theta angle is turned off, or
\item $N-k$ is even and the mod 2 theta angle is turned on.
\end{itemize}

\subsection{Number of vacua}
\label{sect:vacua-review}

In much of this paper we will duplicate results for the number
of vacua in $SO(k)$ and $O_{\pm}(k)$ gauge theories via
mirror computations.  Results for the original gauge theories
are listed in \cite{Hori:2011pd}[table~(4.20)].  Because we will
refer to this table extensively, to make this paper
self-contained, and also to visually illustrate the complexity of
the vacuum computations we are duplicating in our mirror construction, 
we reproduce here
the vacuum counts of \cite{Hori:2011pd}[table~(4.20)].

Specifically, consider a regular $SO(k)$ or $O_{\pm}(k)$ gauge theory
with $N$ vectors.  For $N \geq k-1$, the number of vacua
is as listed in table~\ref{table:vac}.
We will reproduce the listed number of vacua in each case in mirror
computations later in this paper.  (For non-regular theories, vacuum
counts are not listed.)

\begin{table}[h]
\begin{center}
\begin{tabular}{c|cc|c||cc||c}
& $k$ & $N$ & Number of vacua & $k$ & $N$ & Number of vacua \\ \hline
$O_+(k)$ & even & even & 
$\left( \begin{array}{c} N/2 \\ k/2 \end{array} \right)$
& odd & even & 
$2 \left( \begin{array}{c} N/2 \\ (k-1)/2 \end{array} \right)$ \\
$O_+(k)$ & even & odd &
$\left( \begin{array}{c} (N-1)/2 \\ k/2 \end{array} \right) \: + \:
2 \left( \begin{array}{c} (N-1)/2 \\ (k/2)-1 \end{array} \right)$
& odd & odd &
$\left( \begin{array}{c} (N-1)/2 \\ (k-1)/2 \end{array} \right)$ \\
\hline
$O_-(k)$ & even & even & 
$\left( \begin{array}{c} N/2 \\ k/2 \end{array} \right)$ &
odd & even &
$\left( \begin{array}{c} N/2 \\ (k-1)/2 \end{array} \right)$ \\
$O_-(k)$ & even & odd &
$\left( \begin{array}{c} (N+1)/2 \\ k/2 \end{array} \right)$ &
odd & odd &
$2 \left( \begin{array}{c} (N-1)/2 \\ (k-1)/2 \end{array} \right)$ \\
\hline
$SO(k)$ & even & even & 
$2 \left( \begin{array}{c} N/2 \\ k/2 \end{array} \right)$ &
odd & even & 
$\left( \begin{array}{c} N/2 \\ (k-1)/2 \end{array} \right)$ \\
$SO(k)$ & even & odd &
$2 \left( \begin{array}{c} (N-1)/2 \\ k/2 \end{array} \right) \: + \:
\left( \begin{array}{c} (N-1)/2 \\ (k/2)-1 \end{array} \right)$ &
odd & odd &
$\left( \begin{array}{c} (N-1)/2 \\ (k-1)/2 \end{array} \right)$
\end{tabular}
\caption{Results for number of vacua of regular
$SO(k)$ and $O_{\pm}(k)$ gauge theories,
taken from \cite{Hori:2011pd}[table~(4.20)].}
\label{table:vac}
\end{center}
\end{table}

\section{$SO$-$O$ duality}
\label{sect:so:o:duality}

In \cite{Hori:2011pd}[section 4.6], it was proposed that for $N \geq k$,
there exist IR dualities 
\begin{eqnarray}
O_+(k) & \leftrightarrow & SO(N-k+1), \\
SO(k) & \leftrightarrow & O_+(N-k+1), \\
O_-(k) & \leftrightarrow &  O_-(N-k+1),
\end{eqnarray}
where in each case
\begin{itemize}
\item the theory on the left has
$N$ massless vectors $x_1, \cdots, x_N$,
with twisted masses $\tilde{m}_i$,
and 
\item
the theory on the right has $N$ vectors
$\tilde{x}^1, \cdots, \tilde{x}^N$, of twisted masses $\tilde{m}_i$,
along with
$(1/2)N(N+1)$ singlets $s_{ij} = + s_{ji}$,
$1 \leq i, j \leq N$, of twisted mass $-\tilde{m}_i - \tilde{m}_j$,
and a superpotential
\begin{equation}
W \: = \: \sum_{i,j} s_{ij} \tilde{x}^i \cdot \tilde{x}^j.
\end{equation}
\end{itemize}
The mesons in the two theories are related by
\begin{equation}
s_{ij} \: = \: x_i \cdot x_j.
\end{equation}
The dualities are only claimed to exist when all the theories
in question are regular, in the sense of \cite{Hori:2011pd}
(and as reviewed in section~\ref{sect:regular-review}), which constrains the
discrete theta angle.

In this section, we will examine the mirrors to each side
of the first two dualities above, using an extension of the
nonabelian mirrors proposal of \cite{Gu:2018fpm},
and argue that the mirrors of the duals are equivalent.

The general case of the $O_+(k) \leftrightarrow SO(N-k+1)$ duality
will be discussed in subsection~\ref{sect:o+k:son-k+1}.  However,
this section is rather technical, so to improve readability,
we have separate discussions of three special cases.
We begin with
detailed discussions of the simplest case, the
mirror to $O_+(2)$ with $N=3$ and its dual,
in subsection~\ref{sect:o+2:so2}, and include a detailed 
review of pertinent orbifold
twisted sector contributions that will be utilized later in this paper.
We then later discuss
the mirror to $O_+(1)$ in subsection~\ref{sect:proto:o+1},
and the mirror to $O_+(3)$ and its dual in subsection~\ref{sect:proto:o+3},
to better illustrate some of the subtleties of mirrors to $O_+$(odd)
gauge theories, that the reader may find obscure from the general analysis
alone.

Finally, in subsection~\ref{sect:sok:o+n-k+1}, 
we consider the general case of the
(mirror to the) $SO(k) \leftrightarrow O_+(N-k+1)$ duality.
Because singlets appear asymmetrically in the duality, this is not
quite the same as the $O_+(k) \leftrightarrow SO(N-k+1)$ duality,
though it is clearly very closely related. In any event, since it is
not quite the same, to be thorough we include its analysis separately.

We perform the analogous analysis for the $O_--O_-$ duality
in section~\ref{sect:o-o-}.

\subsection{Prototype:  $O_+(2) \leftrightarrow SO(2)$}
\label{sect:o+2:so2}

In this section we will study the simplest duality proposed
in \cite{Hori:2011pd}[section 4.6], namely a duality between
\begin{itemize}
\item an $O_+(2)$ gauge theory with three chiral multiplets $x_i^a$
in the doublet representation ($i$ a flavor index, $i \in \{1, 2, 3\}$,
$a$ a color index, $a \in \{1, 2\}$), with twisted masses $\tilde{m}_i$,
\item an $SO(2)$ gauge theory with three chiral multiplets
$\tilde{x}^{i a}$ in the doublet representation,
of twisted mass $- \tilde{m}_i$ and R-charge $1$,
six singlets $s_{ij} = +s_{ji}$, 
of twisted mass $\tilde{m}_i + \tilde{m}_j$ and R-charge $0$, 
and a superpotential
\begin{equation}
W \: = \: \sum_{ij} s_{ij} \tilde{x}^{i a} \tilde{x}^{j b} \delta_{ab}.
\end{equation}
\end{itemize}
Our analysis of the mirror to this duality
will serve as a prototype for our
analysis of mirrors to more general dualities.
(Note also that since the groups are abelian, this is
an exercise in abelian mirrors \cite{Hori:2000kt} 
rather than nonabelian mirrors \cite{Gu:2018fpm},
though it will serve as a prototype for work with the latter.)

\subsubsection{Mirror to $O_+(2)$ gauge theory, and invariant states}
\label{sect:proto:o+2}

Let us first consider the mirror to the $O_+(2)$ gauge theory
with $N=3$ doublets.
In principle, this is an orbifold of a $U(1)$ gauge theory, but
we will follow the same analysis here as for $O(k)$ gauge theories for
larger $k$, so we will interpret the matter as being in a representation of
the complexification of the gauge group.  As a result, we will speak
of doublets under $SO(2)$, pairs of chiral fields in which one has
charge $+1$ and the other has charge $-1$.

The mirror Landau-Ginzburg orbifold has six fields $Y^i_a$ as well as
one $\sigma$, with a superpotential
\begin{equation}  \label{eq:o+2:mirrorsup}
W \: = \: \sigma \left( - \sum_{i=1}^3 Y_1^i + \sum_{i=1}^3 Y_2^i
\: + \: t \right)
\: - \: \sum_{i=1}^3 \tilde{m}_i \left(  Y_1^i + Y_2^i \right)
\: + \: \sum_{i=1}^3 \exp\left( - Y_1^i \right) 
\: + \: \sum_{i=1}^3 \exp\left( - Y_2^i \right).
\end{equation}
As discussed in section~\ref{sect:orb-even},
the disconnected component of $O_+(2)$ is realized in the mirror by a 
${\mathbb Z}_2$ orbifold that acts as
\begin{equation}  \label{eq:o+2:orb1}
\sigma \: \mapsto \: - \sigma, \: \: \:
Y_1^i \: \leftrightarrow Y_2^i.
\end{equation}
(Note that although the original gauge theory is an abelian gauge theory,
the ${\mathbb Z}_2$ orbifold above restricts allowed values of the
Fayet-Iliopoulos parameter to $\{0, \pi i\}$.
For this example, regularity in the sense of \cite{Hori:2011pd}
implies that $t$ should vanish, but we include the possibility of
a nonzero $t$ to better illustrate certain features of other 
$O_+(k)$ cases.)

Integrating out the $\sigma$ field, we get the constraint
\begin{equation}  
\sum_i Y_1^i \: = \: \sum_i Y_2^i \: + \: t.
\end{equation}
At this point, it is conceptually useful to change variables to
\begin{eqnarray}
Y_+^i & = & \frac{1}{2} \left(Y_1^i \: + \: Y_2^i \right), \\
Y_-^i & = & \frac{1}{2} \left( - Y_1^i \: + \: Y_2^i \right),
\end{eqnarray}
so that the constraint arising from the $\sigma$ field is
\begin{equation}
\sum_i Y_-^i \: = \: -t/2.
\end{equation}
Note that the periodicity of $Y_{\pm}^i$ is a bit different from the
original $Y$ fields:  under $Y_1^i \mapsto Y_1^i + 2 \pi i$, for example,
$Y_{\pm}^i \mapsto Y_{\pm}^i + \pi i$, which effectively eliminates the
square-root branch cut ambiguity implied implicitly in ``$t/2$.''

Since we are changing variables, we should also check Jacobians, at least
formally, to determine if the true fundamental fields have
changed, as in \cite{Gu:2018fpm}.  Here the pertinent Jacobian is of the form
\begin{equation}
\det \left[ \begin{array}{cc}
1/2 & 1/2 \\
-1/2 & 1/2 \end{array} \right] \: = \: 1/2,
\end{equation}
a constant.  Since it is constant, we can ignore it -- there is no meaningful
change in the fundamental fields -- but in the next
section, it will play a more important role.

Using this constraint to eliminate\footnote{
We are eliminating a $Y_-$ field, which is antiinvariant under ${\mathbb Z}_2$,
instead of an invariant field because we are also integrating out $\sigma$,
which also picks up a sign under ${\mathbb Z}_2$.  The combination of
$\sigma$ and a $Y_-$ can form a ${\mathbb Z}_2$-invariant combination, which
is surely what we want to integrate out of an orbifold.
} $Y^3_-$ as
\begin{equation}
Y^3_- \: = \: - Y_-^1 \: - \: Y_-^2 \: - \: t/2,
\end{equation}
the superpotential~(\ref{eq:o+2:mirrorsup}) can be rewritten as
\begin{eqnarray}
W & = & - 2 \sum_{i=1}^3 \tilde{m}_i Y_+^i \: + \: 
\sum_{i=1}^3 \exp\left( - Y_+^i + Y_-^i \right) \: + \:
\sum_{i=1}^3 \exp\left( - Y_+^i - Y_-^i \right),
\\
& = & - 2 \sum_{i=1}^3 \tilde{m}_i Y_+^i \: + \:
\sum_{i=1}^2 \exp\left( - Y_+^i + Y_-^i \right) \: + \:
\sum_{i=1}^2 \exp\left( - Y_+^i - Y_-^i \right) 
\nonumber \\
& & 
\: + \:  \exp\left( - Y_+^3 - Y_-^1 - Y_-^2 - t/2\right) \: + \:
 \exp\left( - Y_+^3 + Y_-^1 + Y_-^2 + t/2\right).
\end{eqnarray}

The (untwisted) critical locus of this superpotential is
defined by the following equations:
\begin{eqnarray*}
\lefteqn{
\exp\left( - Y_+^i + Y_-^i \right) \: + \:
 \exp\left( - Y_+^3 + Y_-^1 + Y_-^2 + t/2 \right)
} \nonumber \\
& = &
\exp\left( - Y_+^i - Y_-^i \right) \: + \:
 \exp\left( - Y_+^3 - Y_-^1 - Y_-^2 - t/2 \right)
\: \: \: \mbox{ for } i=1, 2,
\end{eqnarray*}
\begin{eqnarray*}
\exp\left( - Y_+^i + Y_-^i \right) \: + \:
\exp\left( - Y_+^i - Y_-^i \right) & = &
-2 \tilde{m}_i
\: \: \: \mbox{ for } i=1, 2,
\\
 \exp\left( - Y_+^3 - Y_-^1 - Y_-^2 - t/2 \right) \: + \:
 \exp\left( - Y_+^3 + Y_-^1 + Y_-^2 + t/2 \right)
& = & -2 \tilde{m}_3.
\end{eqnarray*}
Along the critical locus, define
\begin{eqnarray}
X & = & \frac{1}{2} \left( \exp\left( - Y_+^i + Y_-^i \right) \: - \:
\exp\left( - Y_+^i - Y_-^i \right) \right)
\: \: \: \mbox{ for }i=1, 2, \\
& = & \frac{1}{2} \left(
\exp\left( - Y_+^3 - Y_-^1 - Y_-^2 - t/2 \right) \: - \:
\exp\left( - Y_+^3 + Y_-^1 + Y_-^2 + t/2 \right) \right),
\end{eqnarray}
where we have used the first constraint above.
It is then straightforward to verify that in terms of $X$, along
the critical locus,
\begin{eqnarray}
\prod_{i=1}^3 \left( X - \tilde{m}_i \right) & = &
 q^{-1} \prod_{i=1}^3 \left( -X - \tilde{m}_i \right),
\label{eq:o+2:qc1}\\
& = &  q^{-1/2} \prod_{i=1}^3 \exp\left( - Y_+^i \right),
\end{eqnarray}
where $q = \exp(-t)$.  (Note that for the allowed values of $t$,
$q = q^{-1}$.)  For the remainder of this section, we will
restrict to the regular case, which in this case means $q = +1$.

Now, equation~(\ref{eq:o+2:qc1}) is a cubic polynomial in $X$,
that is symmetric under $X \mapsto -X$.  It has roots at
$0$ and $\pm X_0$ 
where
\begin{equation}
X_0^2 \: = \:  - \sum_{i < j} \tilde{m}_i \tilde{m}_j .
\end{equation}
For simplicity, we will assume that $X_0 \neq 0$, so that these
three roots are distinct.

It is straightforward to check that the ${\mathbb Z}_2$ 
orbifold~(\ref{eq:o+2:orb1}) acts as $X \mapsto -X$,
exchanging the two nonzero roots above, but leaving the
zero root fixed.  As a result, in this model,
the critical locus of the mirror superpotential does intersect
an orbifold fixed point, namely at $X=0$.

Expanding about that point, we find that all of the $Y$ fields 
are massive.  Specificaly, we find that after evaluating 
along the critical locus,
at $X=0$, the only nonzero second derivatives of the superpotential are
\begin{eqnarray}
\frac{\partial^2 W}{\partial Y_+^j \partial Y_+^i} & = & -2 \tilde{m}_i
\delta_{ij},
\\
\frac{\partial^2 W}{\partial Y_+^3 \partial Y_+^3} & = & - 2 \tilde{m}_3,
\\
\frac{\partial^2 W}{\partial Y_-^j \partial Y_-^i} & = &
-2 \tilde{m}_i \delta_{ij} - 2 \tilde{m}_3,
\end{eqnarray}
where $i, j \in \{1, 2\}$.
As a result, for generic twisted masses, the five $Y$ fields are all
massive.

Now, we need to compute the number of twisted sector states.
To make this more interesting, the ${\mathbb Z}_2$ orbifold is only
acting on the two $Y_-^i$ fields, which are massive.  The reader
might well ask, how should an orbifold that acts only on massive
states modify the number of ground states?

This question is closely related to a question of the behavior
of orbifolds in which the orbifold group acts trivially on massless
matter, which was discussed in 
\cite{Pantev:2005zs,Pantev:2005rh,Pantev:2005wj,Hellerman:2006zs,Sharpe:2014tca}.
Briefly, despite the fact that the orbifold group only acts nontrivially
on massive matter, its presence can nonetheless be observed within
the theory.  The precise version we need here is in
\cite{Hori:2011pd}[section 2.2].  Applying it, one should bear in mind
that the twisted masses $\tilde{m}_i$ of the original gauge
theory, are complex masses in the mirror, as they arise as
superpotential terms.

Let us quickly review the results of \cite{Hori:2011pd}[section 2.2].
Consider a single chiral superfield $(\phi, \psi_{\pm})$
of some nonzero mass, either complex (appearing in a superpotential)
or twisted.  Following \cite{Hori:2011pd},
we distinguish two ${\mathbb Z}_2$ orbifolds:
\begin{itemize}
\item $\tau$,
\item $\tau (-1)^F$.
\end{itemize} 
When acting on a single chiral superfield with components
$(\phi, \psi_{\pm})$, they both map
\begin{displaymath}
(\phi, \psi_{\pm}) \mapsto (- \phi, - \psi_{\pm}),
\end{displaymath}
and are distinguished by their action on the vacuum.

In the RR sector, let $| \Omega \rangle_{RR}$ denote the untwisted
ground state
of the theory of a single chiral superfield with complex mass $m$, 
and $| \tilde{\Omega} \rangle_{RR}$ denote the
untwisted ground state of the theory of a single chiral
superfield with twisted mass $\tilde{m}$,
so that \cite{Hori:2011pd}[section 2.2]:
\begin{eqnarray}
| \Omega \rangle_{RR} & = & | 0 \rangle_0 \: + \:
\frac{ \overline{m} }{ | m | } \overline{\psi}_{+ 0} \overline{\psi}_{- 0}
| 0 \rangle_0,
\\
| \tilde{\Omega} \rangle_{RR} & = & \overline{\psi}_{+ 0} | 0 \rangle_0
\: + \: \frac{\tilde{m} }{| \tilde{m} | } \overline{\psi}_{- 0} 
| 0 \rangle_0.
\end{eqnarray}
Under the action of the orbifolds above, we define and distinguish
$\tau$ and $\tau (-)^F$ as follows: 
\begin{itemize}
\item $\tau$ flips the sign of $| \Omega \rangle_{RR}$, but leaves
$| \tilde{\Omega} \rangle_{RR}$ invariant,
\item $\tau (-)^F$ leaves $| \Omega \rangle_{RR}$ invariant,
but flips the sign of $| \tilde{\Omega} \rangle_{RR}$.
\end{itemize}
The twisted sector ground state is invariant under both orbifolds
(since all fields are half-integrally moded, so that there are no
zero modes).  As a result, in the RR sector, ground states exist as
follows:
\begin{center}
\begin{tabular}{c|cc}
& Complex mass & Twisted mass \\ \hline
$\tau$ & Twisted sector only & Untwisted and twisted sector \\
$\tau (-)^F$ & Untwisted and twisted sector & Twisted sector only
\end{tabular}
\end{center}

In a NS-NS sector, the analysis is precisely reversed:  the untwisted
sector ground state is invariant for both complex and twisted masses,
since the fields are half-integrally moded.  In the twisted sector,
on the other hand, the fields have zero modes, and then the analysis
is identical to RR sector untwisted sectors.  As a result, ground
states in the NS-NS sector exist as follows:
\begin{center}
\begin{tabular}{c|cc}
& Complex mass & Twisted mass \\ \hline
$\tau$ & Untwisted sector only & Untwisted and twisted sector \\
$\tau (-)^F$ & Untwisted and twisted sector  & Untwisted sector only
\end{tabular}
\end{center}

Following \cite{Hori:2011pd}[section 2.2],
we can derive similar results for multiple massive fields by tensoring
together the various ground states.  In a system with $n$ massive
fermions, if the ground state of a one-fermion system is not invariant
under a given ${\mathbb Z}_2$ orbifold but $n$ is even, the ground
state of the $n$-fermion system will be invariant.
For example, in a $\tau (-)^F$ orbifold of a system of
$n$ chiral multiplets with nonzero complex mass and $m$
chiral multiplets with nonzero twisted mass, 
\begin{itemize}
\item if $m$ is even, there will be an invariant
ground state in both the twisted
and untwisted sectors for both RR and NS-NS,
\item if $m$ is odd, only the RR twisted sector and NS-NS untwisted
sector will have an invariant ground state.
\end{itemize}
By contrast, in a $\tau$ orbifold, 
\begin{itemize}
\item if $n$ is even, there will be an invariant ground state in
both the twisted and untwisted sectors for both RR and NS-NS,
\item if $n$ is odd, only the RR twisted sector and NS-NS untwisted
sector will have an invariant ground state.
\end{itemize}

Applying this to the present case, the mirror of the $O_+(2)$ gauge
theory, described by ordinary chiral superfields and a
$\tau (-)^F$ orbifold, we find that for our
orbifold acting on chiral superfields of
nonzero complex mass (mirror to chirals of nonzero twisted mass), 
we see that there is an invariant ground state
in both the twisted and untwisted sectors for both RR and NS-NS,
independent of the number of massive chirals.
As a result, there is a total of $2+1=3$ vacua, in agreement with
the number of vacua in the original gauge theory, as listed in
\cite{Hori:2011pd}[table (4.20)] and table~\ref{table:vac}.

\subsubsection{Mirror to $SO(2)$ gauge theory}

Next, we consider the mirror to the $SO(2)$ gauge theory
with three doublets, six singlets, and a superpotential.
This mirror has six chiral multiplets 
$W^{i a} =  \exp\left( - (1/2) \tilde{Y}^{i a} \right)$,
six chiral multiplets $T_{ij}$, and a superpotential
\begin{eqnarray}   \label{eq:o2:init-sup}
W & = & - \sum_{i=1}^3 \sigma \left( - \ln (W^{i 1})^2 + \ln (W^{i 2})^2 \right)
\: - \: \sum_i \tilde{m}_i \left( \ln (W^{i 1})^2 + \ln (W^{i 2})^2 \right)
\nonumber \\
& &
\: + \: \sum_i (W^{i 1})^2 
\: + \: \sum_i (W^{i 2})^2
\nonumber \\
& & 
\: - \: \sum_{i \leq j} \left( \tilde{m}_i + \tilde{m}_j \right) T_{ij}
\: + \: \sum_{i \leq j} \exp\left( - T_{ij} \right). 
\end{eqnarray}
Because the original six fields $\tilde{x}^{ia}$ each have R-charge $1$,
the mirror has a $({\mathbb Z}_2)^6$ orbifold, in which each 
${\mathbb Z}_2$ maps $W^{i a} \mapsto - W^{i a}$.

In principle, since the critical locus is defined by vanishing of a
first derivative, and the superpotential (when written in terms of
$\tilde{Y}^{ia}$) has the same form as for the $O_+(2)$ model, so the
critical locus in the untwisted sector should have the same form as there.
Nevertheless, let us briefly walk through the highlights in the 
new variables $W^{ia}$.

First, integrating out $\sigma$ gives the constraint
\begin{equation}
\sum_{i=1}^3 \ln \left( W^{i 1} \right)^2 \: = \:
\sum_{i=1}^3 \ln \left( W^{i 2} \right)^2,
\end{equation}
or more simply,
\begin{equation}
\prod_{i=1}^3 \left( \frac{ W^{i 1} }{ W^{i 2} } \right)^2 \: = \: 1.
\end{equation}
We define
\begin{eqnarray}
W_+^i & \equiv & W^{i 1} W^{i 2}, \label{eq:so2:defn1}\\
W_-^i & \equiv & \frac{ W^{i 1} }{ W^{i 2} },  \label{eq:so2:defn2}
\end{eqnarray}
in terms of which, the constraint from $\sigma$ becomes
\begin{equation}
\prod_{i=1}^3 \left( W_-^i \right)^2 \: = \: 1.
\end{equation}
We eliminate $W_-^3$ as
\begin{equation}
\left( W_-^3 \right)^2 \: = \: \frac{1}{\left( W_-^1 W_-^2 \right)^2 },
\end{equation}
which has solutions
\begin{equation}
W_-^3 \: = \: \pm \frac{1}{W_-^1 W_-^2 }.
\end{equation}

In principle, we also need to compute the Jacobian of the
coordinate transformation from $(W^{i1}, W^{i2})$ to
$(W_+^i, W_-^i)$, to determine the true fundamental fields,
as explained in \cite{Gu:2018fpm}.
Here, the Jacobian is 
\begin{equation}
\det \left[ \begin{array}{cc} W^{i2} & W^{i1} \\
1/W^{i2} & - W^{i1}/(W^{i2})^2 \end{array} \right]
\: = \:
-2 W^i_-,
\end{equation}
hence the fundamental fields are slightly different.
We can write
\begin{equation}
(-4)^{-1} d W_+^i d \ln( W_-^i)^2 \: = \: dW^{i1} dW^{i2},
\end{equation}
and so after changing variables, we can take the fundamental fields
to be $W_+^i$ and $\ln ( W_-^i)^2$.  This will not change the 
critical locus or vacuum computation, which we will leave in terms of
$W_{\pm}^i$, but will be important for e.g. central charge computations.

In principle, we should solve for critical loci for each
of these two roots separately.
For the moment, we shall focus on the $+$ root, for which
\begin{equation}
W_-^3 \: = \: + \frac{1}{W_-^1 W_-^2 }.
\end{equation}
The superpotential~(\ref{eq:o2:init-sup}) becomes
\begin{eqnarray}
W & = & - \sum_{i=1}^2 \tilde{m}_i \ln \left( W_+^i \right)^2
\: - \: \tilde{m}_3 \ln \left( W_+^3 \right)^2
\: + \: 
\sum_{i=1}^2 W_+^i W_-^i \: + \:
\sum_{i=1}^2 \frac{ W_+^i }{ W_-^i }
\nonumber \\
& & 
\: + \:
\frac{ W_+^3 }{ W_-^1 W_-^2 } \: + \:
W_+^3 W_-^1 W_-^2
\nonumber \\
& & 
\: - \: \sum_{i \leq j} \left( \tilde{m}_i + \tilde{m}_j \right) T_{ij}
\: + \: \sum_{i \leq j} \exp\left( - T_{ij} \right). 
\end{eqnarray}
The critical locus is defined by the equations
\begin{eqnarray}
W_+^i W_-^i \: + \: \frac{W_+^i}{W_-^i} & = & 2 \tilde{m}_i
\: \: \: \mbox{ for } i=1, 2,   \label{eq:so2:crit1}
\\ 
\frac{W_+^3}{W_-^1 W_-^2} \: + \: W_+^3 W_-^1 W_-^2 & = &
2 \tilde{m}_3,   \label{eq:so2:crit2}
\\
W_+^i W_-^i \: - \: \frac{W_+^i}{W_-^i} & = &
\frac{W_+^3}{W_-^1 W_-^2} \: - \: W_+^3 W_-^1 W_-^2
\: \: \: \mbox{ for } i=1, 2,  \label{eq:so2:crit3}
\\
\exp\left( -T_{ij}\right) & = & - \left( \tilde{m}_i + \tilde{m}_j
\right).  \label{eq:so2:crit4}
\end{eqnarray}
The last equation completely determines $\exp( -T_{ij})$ on the critical locus.
For the others, on the critical locus, define
\begin{eqnarray}
X & = & \frac{1}{2} \left( W_+^i W_-^i \: - \: \frac{W_+^i}{W_-^i} \right)
\: \: \: \mbox{ for } i=1, 2, \\
& = & \frac{1}{2} \left( \frac{W_+^3}{W_-^1 W_-^2} \: - \: W_+^3 W_-^1 W_-^2
\right).
\end{eqnarray}
(On the critical locus, these various definitions all match.)
Then, it is straightforward to show that
\begin{equation}
\prod_{i=1}^3 \left( X - \tilde{m}_i \right) \: = \:
\prod_{i=1}^3 \left( - X - \tilde{m}_i \right).
\end{equation}

This has all been for the $+$ root of the solution for $W_-^3$.
If we pick the other root,
\begin{equation}
W_-^3 \: = \: - \frac{1}{W_-^1 W_-^2},
\end{equation}
then after a very similar analysis, for
\begin{eqnarray}
W & = & \frac{1}{2} \left( W_+^i W_-^i \: - \: \frac{W_+^i}{W_-^i} \right)
\: \: \: \mbox{ for } i=1, 2,
\\
& = & - \frac{1}{2} \left( \frac{ W_+^3}{W_-^1 W_-^2} \: - \:
W_+^3 W_-^1 W_-^2 \right),
\end{eqnarray}
we again find
\begin{displaymath}
\prod_{i=1}^3 \left( X - \tilde{m}_i \right) \: = \:
\prod_{i=1}^3 \left( - X - \tilde{m}_i \right)
\end{displaymath}
describes the critical loci.

Now, we have to take into account the ${\mathbb Z}_2$ orbifolds that
act by signs on each $W^i$.  Those orbifolds exchange the $+$ and $-$
roots of the solution for $W_-^3$, so we see that we do not need
to consider the other root in detail. 

Furthermore, note that for generic twisted masses,
the critical loci cannot intersect the fixed points of these
${\mathbb Z}_2$ orbifolds.  For example, from~(\ref{eq:so2:crit1}),
we see that $W_+^i \neq 0$ along the critical locus, so long as
$\tilde{m}_i \neq 0$.  Similarly, from~(\ref{eq:so2:crit2}),
we see that $W_+^3 \neq 0$ along the critical locus, so long as
$\tilde{m}_3 \neq 0$.  Similarly, from~(\ref{eq:so2:crit2}),
neither $W_-^i$ can vanish so long as $\tilde{m}_3$ is finite.
As a result, since the critical loci do not intersect the fixed points
of the ${\mathbb Z}_2$ orbifolds, there should not be any twisted sector
ground states to consider.

To summarize, in this theory, the mirror to the $SO(2)$ theory,
the ground states are defined by the solutions to the equation
\begin{displaymath}
\prod_{i=1}^3 \left( X - \tilde{m}_i \right) \: = \:
\prod_{i=1}^3 \left( - X - \tilde{m}_i \right),
\end{displaymath}
much as in the mirror to the $O_+(2)$ theory.  This has three solutions,
at $X=0$ and $X = \pm X_0$, where
\begin{displaymath}
X_0^2 \: = \: - \sum_{i < j} \tilde{m}_i \tilde{m}_j.
\end{displaymath}
Although there are six ${\mathbb Z}_2$ orbifolds, they leave
$X$ invariant.  For example, from the definitions~(\ref{eq:so2:defn1}),
(\ref{eq:so2:defn2}),
$W_+^i$ will be anti-invariant if and only if $W_-^i$ is antiinvariant,
hence combinations such as $W_+^i W_-^i$ and $W_+^i/W_-^i$ are
invariant.

\subsubsection{Comparison of vacua}

Now, let us compare the vacua in the mirror to the $O_+(2)$
theory and the mirror to the dual $SO(2)$ theory.  They both have
the same quantum cohomology relation,
\begin{equation}
\prod_{i=1}^3 \left( X - \tilde{m}_i \right) \: = \:
\prod_{i=1}^3 \left( - X - \tilde{m}_i \right),
\end{equation}
but in one case there is an orbifold whose fixed points intersect critical loci
and generates a twisted sector state, whereas in the other, orbifolds
leave the vacua invariant, and their fixed points do not intersect critical
loci.
\begin{itemize}
\item In the mirror to the $O_+(2)$ theory, we have two untwisted
sector states, corresponding to $X=0$ and $X = \pm X_0$.
The two nonzero solutions are exchanged by the ${\mathbb Z}_2$ orbifold.
Sitting over $X=0$ is one twisted sector state. 
\item In the mirror to the dual $SO(2)$ theory, we have 
three untwisted sector states, corresponding to $X=0$ and
$X = \pm X_0$.  There is a global symmetry sending $X \mapsto -X$,
but this is not gauged, it does not correspond to an orbifold group
action.
\end{itemize}

In addition, there is a global ${\mathbb Z}_2$ symmetry in both
theories:
\begin{itemize}
\item In the mirror to the $O_+(2)$ theory, there is a ${\mathbb Z}_2$
quantum symmetry \cite{Vafa:1989ih}, 
which acts by a sign flip on the twisted sector state.
Under the quantum symmetry, the states
\begin{displaymath}
| X=0, {\rm untwisted} \rangle \: \pm \:
| X=0, {\rm twisted} \rangle
\end{displaymath}
are exchanged.
\item In the mirror to the dual $SO(2)$ theory, the critical locus
is invariant under a global ${\mathbb Z}_2$ that maps $X \mapsto -X$.
This symmetry exchanges the states $X = \pm X_0$.
\end{itemize}

Based on the global symmetries above, we believe that the
ground states
\begin{displaymath}
| X=0, {\rm untwisted} \rangle \: \pm \:
| X=0, {\rm twisted} \rangle
\end{displaymath}
of the $O_+(2)$ theory are dual to the $X = \pm X_0$ states of the
$SO(2)$ theory, and that the single $\pm X_0$ state of the $O_+(2)$
theory is dual to the $X=0$ state of the $SO(2)$ theory.
Thus, in this example, the proposed mirror is compatible with the
gauge theory duality.

\subsection{$O_+(k) \leftrightarrow SO(N-k+1)$ duality}
\label{sect:o+k:son-k+1}

In this section, we will discuss mirrors to each side of the duality
relating
\begin{itemize}
\item an $O_+(k)$ gauge theory with $N$ vectors
$x_1, \cdots, x_N$ of twisted masses $\tilde{m}_i$, and
\item an $SO(N-k+1)$ gauge theory with $N$ vectors
$\tilde{x}^1, \cdots, \tilde{x}^N$ of twisted masses $\tilde{m}_i$,
along with $(1/2) N (N+1)$ singlets $s_{ij} = + s_{ji}$ of twisted
masses $- \tilde{m}_i - \tilde{m}_j$, and a superpotential
\begin{equation}
W \: = \: \sum_{ij} s_{ij} \tilde{x}^i \cdot \tilde{x}^j.
\end{equation}
\end{itemize}
We will compute the mirrors to either side of the duality, and check
that the mirrors have the same number of vacua as the original theories,
and each other.  We will also compare central charges of the corresponding
IR SCFTs in the special case that the twisted masses all vanish.

\subsubsection{Mirror to $O_+(k)$ gauge theory}
\label{sect:o+k:mirror}

In this section we will compute the mirror to an
$O_+(k)$ gauge theory with $N \geq k$ massless vectors $x_1, \cdots, x_N$
with twisted masses $+ \tilde{m}_i$.  This mirror will be closely related
to the mirror of $SO(k)$ -- for $k$ even, it will have a slightly more
complicated orbifold group, and for $k$ odd, it will be either one or two
copies of the $SO(k)$ mirror, as discussed in sections~\ref{sect:orb-even},
\ref{sect:orb-odd}.  We will therefore begin by describing
the mirror to the $SO(k)$ gauge theory with $N \geq k$ massless vectors
and twisted masses as above.

Following \cite{Gu:2018fpm}[sections 9, 10], the mirror is an 
orbifold of a Landau-Ginzburg model with fields
\begin{itemize}
\item $Y_{i \alpha}$, $i \in \{1, \cdots, N\}$, $\alpha \in \{1, \cdots, k \}$,
\item $X_{\mu \nu} = \exp\left(-Z_{\mu \nu} \right)$, 
$X_{\mu \nu} = X_{\nu \mu}^{-1}$, $\mu \nu \in \{1, \cdots, k\}$
(excluding $X_{2a-1,2a}$),
\item $\sigma_a$, $a \in \{1, \cdots, M\}$ where for $k$ even,
$k=2M$, and for $k$ odd, $k = 2M+1$,
\end{itemize}
with superpotential
\begin{eqnarray}
W & = &
\sum_{a=1}^M \sigma_a \left(
\sum_{i \alpha \beta} \rho^a_{i \alpha \beta} Y_{i \beta} \: + \:
\sum_{\mu < \nu; \mu' , \nu'} \alpha^a_{\mu \nu, \mu' \nu'} Z_{\mu' \nu'}
\: - \: t \right)
\nonumber \\
& & \: + \:
\sum_{i \alpha} \exp\left( - Y_{i \alpha} \right) 
\: + \: 
\sum_{\mu < \nu} X_{\mu \nu}
\nonumber \\
& & \: - \:
\sum_{i, \alpha} \tilde{m}_i  Y_{i \alpha}.
\end{eqnarray}
There is no continuous FI parameter, but we retain $t$ to allow for
the possibility of a discrete theta angle.  In the superpotential above,
we take
\begin{eqnarray}
\rho^a_{i \alpha \beta} & = &  \delta_{\alpha, 2a-1}
\delta_{\beta, 2a} - \delta_{\beta, 2a-1} \delta_{\alpha, 2a} ,
\label{eq:so2k1:rho-defn}
\\
\alpha^a_{\mu \nu, \mu' \nu'} & = &
 \delta_{\nu \nu'} \left( \delta_{\mu, 2a-1} \delta_{\mu', 2a} -
\delta_{\mu, 2a} \delta_{\mu', 2a-1} \right)
\nonumber \\
& & \hspace*{0.5in}
\: + \:
 \delta_{\mu \mu'} \left( \delta_{\nu, 2a-1} \delta_{\nu', 2a} - 
\delta_{\nu, 2a} \delta_{\nu', 2a-1} \right).
\label{eq:so2k1:alpha-defn}
\end{eqnarray}

We can analyze the untwisted sector of this theory in the
same fashion as \cite{Gu:2018fpm}[sections 9, 10].  
For completeness, we outline the highlights here.
Integrating out the $\sigma_a$, we have
the constraints
\begin{equation}
\sum_{i=1}^N \left( Y_{i,2a} - Y_{i,2a-1} \right) \: + \:
\sum_{\nu > 2a} \ln \left( \frac{ X_{2a-1,\nu} }{ X_{2a,\nu} } \right)
\: + \:
\sum_{\mu < 2a-1} \ln \left( \frac{ X_{\mu, 2a-1} }{ X_{\mu,2a} } \right)
\: = \: t.
\end{equation}
To eliminate fields in a fashion somewhat consistent with the orbifold,
we define
\begin{eqnarray}
Y_{i, a}^+ & \equiv & \frac{1}{2} \left( Y_{i,2a} + Y_{i, 2a-1} \right),
\\
Y_{i,a}^- & \equiv & \frac{1}{2} \left( Y_{i,2a} - Y_{i, 2a-1} \right),
\end{eqnarray}
and then use the constraints above to eliminate $Y_{N,a}^-$:
\begin{equation}
Y_{N,a}^- \: = \: - \sum_{i=1}^{N-1} Y_{i,a}^- 
 \: - \: \frac{1}{2} \sum_{\nu > 2a} \ln \left( \frac{ X_{2a-1,\nu} }{ X_{2a,\nu} } \right)
\: - \: \frac{1}{2} \sum_{\mu < 2a-1} \ln \left( \frac{ X_{\mu, 2a-1} }{ X_{\mu,2a} } \right)
\: + \: \frac{t}{2}.
\end{equation}

We define
\begin{eqnarray}
\Upsilon_a & \equiv & \exp\left( - Y_{N,a}^- \right), 
\\
& = & q^{1/2} \left( \prod_{i=1}^{N-1} \exp\left( + Y_{i,a}^- \right) \right)
\cdot
\nonumber \\
& & \hspace*{0.5in} \cdot
\left( \prod_{\nu > 2a}  \frac{ X_{2a-1,\nu} }{ X_{2a,\nu} } \right)^{1/2}
\left( \prod_{\mu < 2a-1}  \frac{ X_{\mu, 2a-1} }{ X_{\mu,2a} }
\right)^{1/2},     \label{eq:o+k:pi-defn}
\end{eqnarray}
where $q = \exp(-t)$.
As in the previous section, the square root branch cut ambiguities are 
absorbed by the periodicities of $Y^{\pm}$.

The superpotential then reduces to
\begin{eqnarray}
W & = &
\sum_{i=1}^{N-1} \sum_{a=1}^M \left( \exp\left( - Y_{i, a}^+ - Y_{i,a}^- \right)
\: + \: \exp\left( - Y_{i,a}^+ + Y_{i,a}^- \right) \right)
\nonumber \\
& &
\: + \: \sum_{a=1}^M \exp\left( - Y_{N,a}^+ \right) \left(
\Upsilon_a + \Upsilon_a^{-1} \right)
\nonumber \\
& & \: + \:
\sum_{\mu < \nu} X_{\mu \nu} 
\: - \:
\sum_{i=1}^{N} \sum_{a=1}^M 2 \tilde{m}_i Y_{i, a}^+
\nonumber \\
& & \: + \:
\left\{ \begin{array}{cl}
\sum_{i=1}^N \exp\left( - Y_{i, 2M+1} \right)
\: - \: \sum_{i=1}^N \tilde{m}_i Y_{i, 2M+1} & k \mbox{ odd}, \\
0 & {\rm else}.
\end{array}
\right.    \label{eq:o+k:mirror:sup-final}
\end{eqnarray}

The critical locus is then given by
\begin{eqnarray}
Y_{i,a}^+: & &
\exp\left( - Y_{i,a}^+ - Y_{i,a}^- \right) \: + \:
\exp\left( - Y_{i,a}^+ + Y_{i,a}^- \right) \: = \: -2 \tilde{m}_i
\: \: \: \mbox{ for }i < N,
\\
Y_{N,a}^+: & &
\exp\left( - Y_{N,a}^+\right) \left( \Upsilon_a + \Upsilon_a^{-1}
\right) \: = \: -2 \tilde{m}_N,
\\
Y_{i,a}^-: & &
\exp\left( - Y_{i,a}^+ - Y_{i,a}^- \right) \: - \:
\exp\left( - Y_{i,a}^+ + Y_{i,a}^- \right) 
\nonumber \\
& & \hspace*{0.5in} \: = \:
\exp\left( - Y_{N,a}^+ \right) \left( \Upsilon_a - \Upsilon_a^{-1} \right)
\: \: \: \mbox{ for }i < N,
\label{eq:o+k:mirror:critical:3}
\\
Y_{i,2M+1}: & &
\exp\left( - Y_{i,2M+1} \right) \: = \: - \tilde{m}_i
\: \: \: \mbox{ for $k$ odd},
\\
X_{\mu \nu}: & &
2 X_{2a,2b} \: = \: \exp\left( - Y_{N,a}^+ \right) \left(  \Upsilon_a -
\Upsilon_a^{-1} \right) \: + \:
\exp\left( - Y_{N,b}^+ \right) \left( \Upsilon_b - \Upsilon_b^{-1} \right),
\\
 & &
2 X_{2a,2b-1} \: = \: \exp\left(  - Y_{N,a}^+ \right) \left( \Upsilon_a -
\Upsilon_a^{-1} \right) \: + \:
\exp\left( - Y_{N,b}^+ \right) \left( - \Upsilon_b + \Upsilon_b^{-1} \right),
\\
 & &
2 X_{2a-1,2b} \: = \:
\exp\left(  - Y_{N,a}^+ \right) \left( - \Upsilon_a + \Upsilon_a^{-1} \right)
 \: + \:
\exp\left( - Y_{N,b}^+ \right) \left( \Upsilon_b - \Upsilon_b^{-1} \right),
\\
 & &
2 X_{2a-1,2b-1} \: = \: 
\exp\left(  - Y_{N,a}^+ \right) \left(- \Upsilon_a + \Upsilon_a^{-1} \right)
\nonumber \\
& & \hspace*{1in}
 \: + \:
\exp\left( - Y_{N,b}^+ \right) \left( -\Upsilon_b + \Upsilon_b^{-1} \right),
\\
& & 
2 X_{2a,2M+1} \: = \: \exp\left( - Y_{N,a}^+ \right) \left( \Upsilon_a
- \Upsilon_a^{-1} \right)
\: \: \: \mbox{ for $k$ odd},
\\
& & 
2 X_{2a-1,2M+1} \: = \: \exp\left( - Y_{N,a}^+ \right) \left( - \Upsilon_a
+ \Upsilon_a^{-1} \right)
\: \: \: \mbox{ for $k$ odd}.
\end{eqnarray}

On the critical locus, define
\begin{eqnarray}
\sigma_a & \equiv & \frac{1}{2} \exp\left( - Y_{i, a}^+ - Y_{i,a}^- \right)
\: - \: \frac{1}{2} \exp\left( - Y_{i,a}^+ + Y_{i,a}^- \right)
\: \: \: \mbox{ for any }i < N,
\\
& = & \frac{1}{2} \exp\left( - Y_{N,a}^+ \right) \left( \Upsilon_a - 
\Upsilon_a^{-1}
\right).
\end{eqnarray}
(The fact that these expressions are all equal is a consequence of
critical locus equation~(\ref{eq:o+k:mirror:critical:3}).)
Since $\exp(-Y) \neq 0$, we have from the first two critical locus
equations that
\begin{equation}
\sigma_a \: \neq \: \tilde{m}_i,
\end{equation}
for all $i$.
Since $X_{\mu \nu} \neq 0$, we find for $a \neq b$ that
\begin{equation}   \label{eq:o+k:even:excluded1}
\sigma_a \: \neq \: \pm \sigma_b,
\end{equation}
and in the special case that $k$ is odd,
\begin{equation}
\sigma_a \: \neq \: 0.
\end{equation}

It is also straightforward to demonstrate that on the critical locus,
\begin{equation}  \label{eq:o+k:mirror:qc}
\prod_{i=1}^N \left( \sigma_a - \tilde{m}_i \right) \: = \: (-)^k q
\prod_{i=1}^N \left( - \sigma_a - \tilde{m}_i \right).
\end{equation}

Before moving on, we should also observe that the operator
mirror map described in \cite{Gu:2018fpm} relates the quantity we
have labelled ``$\sigma_a$'' in the algebra above, to the
$\sigma_a$ of the original theory; our notation was deliberately
chosen to encode the operator mirror map.

Now that we have reviewed the basics of the mirror to an $SO(k)$
gauge theory, we will discuss the mirror of $O_+(k)$.

Let us first consider the case that $k$ is even, $k=2M$.
In this case, the orbifold group for the mirror of $O_+(k)$
is effectively the Weyl group of
$SO(2M+1)$.  This has a subgroup corresopnding to the Weyl group
of $SO(2M)$, which as discussed in \cite{Gu:2018fpm}[section 9],
has fixed points which do not intersect the critical loci, and so does
not generate any twisted sector contributions to ground states.
The remainder is generated by a ${\mathbb Z}_2$ which can be taken to
act as
\begin{equation}
\sigma_M \: \mapsto \: - \sigma_M,
\: \: \:
Y_{i, 2M-1} \: \leftrightarrow \: Y_{i,2M},
\: \: \:
X_{\mu,2M-1} \: \leftrightarrow \: X_{\mu,2M}.
\end{equation}
(This also encodes the action on the critical locus of the
map $Y_{i,2M} \leftrightarrow Y_{i,2M-1}$.)
This orbifold has a fixed point locus at
\begin{equation}
\sigma_a \: = \: 0,
\end{equation}
and since $k$ is even, this is not on the excluded locus.
Furthermore, for $k$ even, depending upon the number $N$ of
vectors and the value of $q$,
this fixed point locus can also be on the critical locus,
hence we have twisted sector contributions to the ground states.
In fact, after combining this ${\mathbb Z}_2$ with other elements of
the Weyl group of $SO(2M)$, for $q=+1$ and $k$ even,
we have intersections of orbifold fixed points
and critical loci where
\begin{equation}
\sigma_a \: = \: 0
\end{equation}
for all $a$.
If any one of these combinations vanishes, for any one value
of $a$, then other combinations for other $a$ cannot also vanish,
from the excluded locus condition~(\ref{eq:o+k:even:excluded1}). 
Furthermore, since the $S_M$ in the Weyl group of $SO(2M)$
exchanges these quantities, there is at most one vacuum defined by
such an intersection of the orbifold fixed point with the critical locus.

Let us back up a step and count the number of possible vacua
more systematically.

First, consider the case that $q=+1$.  For $N$ even (and even $k$),
the Coulomb branch relation~(\ref{eq:o+k:mirror:qc}) reduces, for even
$k$, to the polynomial
\begin{equation}  \label{eq:o+k:qc:n-1}
\left( \sum_i \tilde{m}_i \right) \sigma^{N-1} \: + \:
\left( \sum_{i_1 < i_2 < i_3} \tilde{m}_{i_1} \tilde{m}_{i_2}
\tilde{m}_{i_3} \right) \sigma^{N-3} \: + \: \cdots 
\: = \: 0.
\end{equation}
This equation is symmetric in $\sigma \mapsto - \sigma$, so all nonzero roots
will come in positive/negative pairs.
This equation has $N-1$ roots, of the form
\begin{equation}
\sigma \: = \: 0, \pm \tilde{\sigma}_1 , \: \cdots, \:
\pm \tilde{\sigma}_{(N-2)/2},
\end{equation}
with (for generic $\tilde{m}_i$) none vanishing.  In terms of vacua,
any two signs $\pm \tilde{\sigma}$ are related by the action of one
of the ${\mathbb Z}_2$ subgroups of the orbifold group, and so define
the same vacuum.

Since there are $M = k/2$ $\sigma_a$, and the corresponding
$\sigma$ are all
distinct (including signs) from condition~(\ref{eq:o+k:even:excluded1}),
exchanged by the action of an $S_k$ orbifold,
we see that in this case, there are 
\begin{equation}
\left( \begin{array}{c} (N-2)/2 + 1\\ M \end{array} \right)
\: = \:
\left( \begin{array}{c} N/2 \\ k/2 \end{array} \right)
\end{equation}
untwisted vacua in the system.
(Since $N \geq k = 2M$, we see that $N/2 \geq M$, and so the expression
above is well-defined.)  The root at $\sigma=0$ intersects the fixed-point
locus of the orbifold, and so there can be additional ground states
arising in the twisted sector.

As mentioned in section~\ref{sect:orb-even}, our convention is to
treat the mirror superpotential as an ordinary superpotential,
depending upon ordinary chiral superfields, and so the extra
${\mathbb Z}_2$ is realized in the mirror by $\tau (-)^F$.
In this case, since the fields have complex masses (instead of
twisted masses), there will be invariant ground states in both the
untwisted and twisted sectors for both RR and NS-NS, regardless of
the number of massive fields, following the analysis of
section~\ref{sect:proto:o+2} and \cite{Hori:2011pd}.

We can combine a twisted sector ground state with untwisted sector
vacua corresponding to nonzero roots, to get a total of
\begin{equation}
\left( \begin{array}{c} (N-2)/2 \\ k/2 - 1 \end{array} \right)
\end{equation}
twisted sector vacua, and a grand total of
\begin{equation}
\left( \begin{array}{c} N/2 \\ k/2 \end{array} \right)
\: + \:
\left( \begin{array}{c} (N-2)/2 \\ k/2 - 1 \end{array} \right)
\end{equation}
untwisted and twisted sector vacua.

For completeness, using the identity
\begin{equation}
\left( \begin{array}{c} a \\ b \end{array} \right) 
\: + \:
\left( \begin{array}{c} a \\ b-1 \end{array} \right)
\: = \:
\left( \begin{array}{c} a+1 \\ b \end{array} \right),
\end{equation}
it is easy to show that the number of vacua above equals
\begin{equation}
\left( \begin{array}{c} (N-2)/2 \\ k/2 \end{array} \right)
\: + \:
2 \left( \begin{array}{c} (N-2)/2 \\ k/2 - 1 \end{array} \right).
\end{equation}

Finally, this particular theory (the mirror of $O_+(k)$ with
$k$ even, $N$ even, and $q=+1$, is not regular in the sense of
\cite{Hori:2011pd}, and so is not listed in 
\cite{Hori:2011pd}[table~(4.20)] or table~\ref{table:vac}.

For $N$ odd (and $q=+1$), 
the Coulomb branch relation~(\ref{eq:o+k:mirror:qc}) reduces,
for even $k$, to the polynomial
\begin{equation}   \label{eq:o+k:qc:n}
\sigma^N \: + \: \left( \sum_{i<j} \tilde{m}_i \tilde{m}_j \right) \sigma^{N-2}
\: + \: \left( \sum_{i_1 < i_2 < i_3 < i_4} \tilde{m}_{i_1}
\tilde{m}_{i_2} \tilde{m}_{i_3} \tilde{m}_{i_4} \right) \sigma^{N-4}
\: + \: \cdots \: = \: 0.
\end{equation}
As before, this equation is symmetric in $\sigma \mapsto - \sigma$,
so all nonzero roots will come in positive/negative pairs.
The Coulomb branch relation has $N$ roots
of the form
\begin{equation}
\sigma \: = \: 0, \:
\pm \tilde{\sigma}_1, \: \cdots, \: 
\pm \tilde{\sigma}_{(N-1)/2}
\end{equation}
with (for generic $\tilde{m}_i$) none of the $\tilde{\sigma}$
vanishing (except the first root in the list above).
For each nonzero root $\tilde{\sigma}_p$, the two signs
$\pm \tilde{\sigma}_p$ are exchanged by a ${\mathbb Z}_2$
subgroup of the orbifold group, and so define the same vacuum.
The root at $\sigma=0$ is more interesting.  This root intersects
the fixed-point locus of the orbifold, and so there can be
an additional ground state arising in the twisted sector.

Including the zero, the number of untwisted sector ground states
is
\begin{equation}
\left( \begin{array}{c} (N+1)/2 \\ M \end{array} \right)
\: = \:
\left( \begin{array}{c} (N+1)/2 \\ k/2 \end{array} \right).
\end{equation}
Next, we consider twisted sector states.  
As mentioned earlier, in our conventions, the mirror of
an $O(2M)$ gauge theory involves a Landau-Ginzburg orbifold by
$\tau (-)^F$ (combined with the Weyl group orbifold).  In the
same conventions, we are treating the mirror superpotential as a function
of ordinary chiral superfields, with complex masses, so there are
invariant ground states in both the untwisted and twisted sectors
for both RR and NS-NS, regardless of the number of massive fields,
following section~\ref{sect:proto:o+2} and \cite{Hori:2011pd}.
We can combine a twisted sector ground state with untwisted sector
vacua corresponding to nonzero roots, to get
\begin{equation}
\left( \begin{array}{c} (N-1)/2 \\ k/2-1 \end{array} \right)
\end{equation}
twisted sector vacua.

Putting this together, for $N$ odd and $q=+1$,
there are
\begin{equation}
\left( \begin{array}{c}  (N+1)/2 \\ k/2 \end{array} \right) \: + \:
\left( \begin{array}{c} (N-1)/2 \\ k/2-1 \end{array} \right)
\: = \: 
\left( \begin{array}{c} (N-1)/2 \\ k/2 \end{array} \right)
\: + \:
2 \left( \begin{array}{c} (N-1)/2 \\ k/2-1 \end{array} \right)
\end{equation}
untwisted and twisted vacua.  
This theory is regular in the sense of \cite{Hori:2011pd},
and the result above matches an entry in 
\cite{Hori:2011pd}[table~(4.20)] and table~\ref{table:vac}.

Now let us turn to the case $q=-1$, for $k$ even.

For $N-1$ even ($N$ odd), 
the Coulomb branch relation~(\ref{eq:o+k:mirror:qc}) reduces
to the degree $N-1$ polynomial~(\ref{eq:o+k:qc:n-1}).
This equation has $N-1$ roots, of the form
\begin{equation}
\sigma \: = \: \pm \tilde{\sigma}_1, \: \cdots, \:
\pm \tilde{\sigma}_{(N-1)/2},
\end{equation}
with (for generic $\tilde{m}_i$) none of the $\tilde{\sigma}$ vanishing.
In terms of vacua, as before, any two signs $\pm \tilde{\sigma}$ 
rea related by the action of one of the ${\mathbb Z}_2$ subgroups of
the orbifold group, and so define the same vacuum.  Since there
are $M$ $\sigma_a$, and the corresponding $\sigma$ are all distinct
(including signs) from condition~(\ref{eq:o+k:even:excluded1}),
exchanged by the action of an $S_k$ orbifold, we see that in this
case, there are
\begin{equation}
\left( \begin{array}{c} 
(N-1)/2 \\ M \end{array} \right) 
\: = \:
\left( \begin{array}{c} 
(N-1)/2 \\ k/2 \end{array} \right) 
\end{equation}
vacua in the system, all in untwisted sectors.
Since $N \geq k = 2M$ and $N$ is odd, we have that $N-1 \geq 2M$,
hence $(N-1)/2 \geq M$, and so the expression above is well-defined.
This theory is not regular in the sense of
\cite{Hori:2011pd}.

Finally, we turn to the case that $N-1$ is odd ($N$ even),
again for $q=-1$ and $k$ even.  In this case, the Coulomb branch relation
reduces to the degree $N$ polynomial~(\ref{eq:o+k:qc:n}).
It has $N$ roots, of the form
\begin{equation}
\sigma \: = \:  \pm \tilde{\sigma}_1, \: \cdots, \:
\pm \tilde{\sigma}_{N/2},
\end{equation}
where (for generic $\tilde{m}_i$) the $\tilde{\sigma}$ are all nonzero.
None of these roots intersect the fixed-point locus of the orbifold.
As before, for each nonzero root $\tilde{\sigma}_p$, the
two signs $\pm \tilde{\sigma}_p$ are exchanged by a ${\mathbb Z}_2$
subgroup of the orbifold group, and so define the same vacuum.
We count
\begin{equation}
\left( \begin{array}{c} N/2 \\ M \end{array} \right)
\: = \:
\left( \begin{array}{c} N/2 \\ k/2 \end{array} \right)
\end{equation}
vacua in this case.
This theory is regular in the sense of \cite{Hori:2011pd},
and the number of vacua counted above matches the number given
in \cite{Hori:2011pd}[table~(4.20)] and table~\ref{table:vac} for this case.

Now that we have discussed the mirror to $O_+(k)$ for $k$ even,
we turn to the case that $k$ is odd, $k = 2M+1$.  As discussed in
section~\ref{sect:orb-odd}, if the number of vectors $N$ is even,
we take the mirror to be two copies of the $SO(k)$ mirror,
and if $N$ is odd, we take the mirror to be one copy of the $SO(k)$
mirror.

First, consider the case that $q=+1$.  For $N$ even (and odd $k$),
the Coulomb branch relation~(\ref{eq:o+k:mirror:qc}) reduces to
a degree $N$ polynomial, with roots
\begin{equation}
\sigma \: = \: \pm \tilde{\sigma}_1, \cdots,
\pm \tilde{\sigma}_{N/2}.
\end{equation}
The mirror to $SO(k)$ with $N$ even therefore has
\begin{equation}
\left( \begin{array}{c} N/2 \\ M \end{array} \right)
\: = \:
\left( \begin{array}{c} N/2 \\ (k-1)/2 \end{array} \right)
\end{equation}
vacua, and since the mirror of $O_+(k)$ with an even number of vectors
is two copies of the mirror of $SO(k)$, we find that in this case the
mirror to the $O_+(k)$ gauge theory has
\begin{equation}
2 \left( \begin{array}{c} N/2 \\ (k-1)/2 \end{array} \right)
\end{equation}
vacua.  This theory is regular in the sense of \cite{Hori:2011pd},
and the number of vacua computed above matches the corresponding
entry in \cite{Hori:2011pd}[table~(4.20)] and table~\ref{table:vac}.

Next, consider the case that $q=+1$ and $N$ is odd.
In this case, the Coulomb branch relation~(\ref{eq:o+k:mirror:qc})
reduces to a degree $N-1$ polynomial, with roots
\begin{equation}
\sigma \: = \: \pm \tilde{\sigma}_1, \cdots,
\pm \tilde{\sigma}_{(N-1)/2}.
\end{equation}
The mirror to $SO(k)$ with $N$ odd therefore has
\begin{equation}
\left( \begin{array}{c} (N-1)/2 \\ M \end{array} \right)
\: = \:
\left( \begin{array}{c} (N-1)/2 \\ (k-1)/2 \end{array} \right)
\end{equation}
vacua, and since the mirror of $O_+(k)$ with an odd number of vectors
is the same as the mirror of the $SO(k)$ gauge theory, we find that
in this case the mirror to the $O_+(k)$ gauge theory has
\begin{equation}
\left( \begin{array}{c} (N-1)/2 \\ (k-1)/2 \end{array} \right)
\end{equation}
vacua.  This theory is not regular in the sense of \cite{Hori:2011pd}.

Next, consider the case that $q=-1$.  For $N$ even (and odd $k$),
the Coulomb branch relation~(\ref{eq:o+k:mirror:qc}) reduces to
a degree $N-1$ polynomial, with roots
\begin{equation}
\sigma \: = \: 0, \pm \tilde{\sigma}_1, \cdots,
\pm \tilde{\sigma}_{(N-2)/2}.
\end{equation}
For the mirror to $SO(k)$, the zero root lies along the excluded locus,
so we see that the $SO(k)$ mirror for $N$ even has
\begin{equation}
\left( \begin{array}{c} (N-2)/2 \\ (k-1)/2 \end{array} \right)
\end{equation}
vacua.  The mirror to $O_+(k)$ with an even number of vectors is
two copies of the mirror to the corresponding $SO(k)$ gauge theory,
so we see that the mirror to $O_+(k)$ has
\begin{equation}
2 \left( \begin{array}{c} (N-2)/2 \\ (k-1)/2 \end{array} \right)
\end{equation}
vacua.  This theory is not regular in the sense of \cite{Hori:2011pd}.

Finally, consider the case that $q=-1$ and $N$ is odd.
In this case, the Coulomb branch relation~(\ref{eq:o+k:mirror:qc})
is a degree $N$ polynomial, with roots
\begin{equation}
\sigma \: = \: 0, \pm \tilde{\sigma}_1, \cdots,
\pm \tilde{\sigma}_{(N-1)/2}.
\end{equation}
The zero root lies along the excluded locus, so we see that the $SO(k)$
mirror for $N$ odd and $q=-1$ has
\begin{equation}
\left( \begin{array}{c} (N-1)/2 \\ (k-1)/2 \end{array} \right)
\end{equation}
vacua.  The mirror to $O_+(k)$ with $N$ odd is one copy of the
mirror to $SO(k)$, so we see that the $O_+(k)$ mirror in this case has
\begin{equation}
\left( \begin{array}{c} (N-1)/2 \\ (k-1)/2 \end{array} \right)
\end{equation}
vacua also.  This theory is regular in the sense of \cite{Hori:2011pd},
and the number of vacua matches the corresponding entry in
\cite{Hori:2011pd}[table~(4.20)] and table~\ref{table:vac}.

\subsubsection{Mirror to $SO(N-k+1)$ gauge theory}
\label{sect:sonk1:mirror}

In this section we will compute the mirror to an $SO(N-k+1)$ gauge
theory with $N \geq k$ vectors $\tilde{x}^1, \cdots, \tilde{x}^N$
of R-charge $1$ and twisted mass $+ \tilde{m}_i$, 
$(1/2)N(N+1)$ singlets $s_{ij} = + s_{ji}$ of R-charge $0$ and twisted
mass $-\tilde{m}_i - \tilde{m}_j$, and
a superpotential
\begin{equation}
W \: = \: \sum_{i \leq j} s_{ij} \tilde{x}^i \cdot \tilde{x}^j.
\end{equation}

If $N-k+1$ is even, $N-k+1 = 2M$,
then the mirror \cite{Gu:2018fpm}
is given by a Landau-Ginzburg orbifold with fields
\begin{itemize}
\item $W^{i \alpha} = \exp\left( - (1/2) \tilde{Y}^{i a} \right)$,
$i \in \{1, \cdots, N\}$, $\alpha \in \{1, \cdots, N-k+1 \}$,
\item $X_{\mu \nu} = \exp(-Z_{\mu \nu} )$,
$X_{\nu \mu} = X_{\mu \nu}^{-1}$, $\mu, \nu \in \{1, \cdots, N-k+1\}$
(excluding $X_{2j-1,2j}$),
\item $\sigma_a$, $a \in \{1, \cdots, M\}$,
\item $T_{ij} = + T_{ji}$,
\end{itemize}
and superpotential
\begin{eqnarray}
W & = & \sum_{a=1}^M \sigma_a \left( - \sum_{i, \alpha, \beta} 
\rho^a_{i \alpha \beta} \ln \left( W^{i \beta} \right)^2 \: - \:
\sum_{\mu < \nu; \mu', \nu'} \alpha^a_{\mu \nu; \mu' \nu'}
\ln X_{\mu \nu} \: - \: \tilde{t} \right) 
\nonumber \\
& & 
\: + \: \sum_{i \alpha}  \left( W^{i \alpha} \right)^2
\: + \: \sum_{\mu < \nu} X_{\mu \nu}
\: + \: \sum_{i \leq j} \exp\left( - T_{ij} \right)
\nonumber \\
& &
\: + \: \sum_{i \alpha} 2 \tilde{m}_{i} \ln W^{i \alpha} 
\: + \: \sum_{i \leq j} \left( \tilde{m}_i + \tilde{m}_j \right) T_{ij},
\end{eqnarray}
where
\begin{eqnarray}
\rho^a_{i \alpha \beta} & = &  \delta_{\alpha, 2a-1}
\delta_{\beta, 2a} - \delta_{\beta, 2a-1} \delta_{\alpha, 2a} ,
\label{eq:so2k:rho-defn}
\\
\alpha^a_{\mu \nu, \mu' \nu'} & = &
 \delta_{\nu \nu'} \left( \delta_{\mu, 2a-1} \delta_{\mu', 2a} -
\delta_{\mu, 2a} \delta_{\mu', 2a-1} \right)
\nonumber \\
& & \hspace*{0.5in}
\: + \:
 \delta_{\mu \mu'} \left( \delta_{\nu, 2a-1} \delta_{\nu', 2a} - 
\delta_{\nu, 2a} \delta_{\nu', 2a-1} \right).
\label{eq:so2k:alpha-defn}
\end{eqnarray}
(Our description is modelled on the closely related mirror described
in \cite{Gu:2018fpm}[section 9].)
Although there is no continuous FI parameter, we retain $t$ above
to allow for the possibility of a discrete theta angle.

We orbifold this model by the Weyl group $W$ of $SO(2M)$, 
where $W$ is the extension
\begin{equation}
1 \: \longrightarrow \: K \: \longrightarrow \: W \: \longrightarrow \: S_M
\: \longrightarrow \: 1,
\end{equation}
where $K$ is the subgroup of $({\mathbb Z}_2)^M$ with an even number of
nontrivial generators.  We also orbifold by additional
${\mathbb Z}_2$s, one for each $W^{i \alpha}$, mapping
$W^{i \alpha} \mapsto - W^{i \alpha}$. 

The case $N-k+1$ odd is described in a formally identical fashion
(compare e.g. \cite{Gu:2018fpm}[section 10]).
Here, we define $M$ by $N-k+1 = 2M+1$, and in the description of
the Weyl group, the kernel $K$ is taken to be all of $({\mathbb Z}_2)^M$.
Its action on the fields is identical to the case $N-k+1$ even -- meaning,
for example, that $W^{i, N-k+1} = W^{i,2M+1}$ is invariant under the
Weyl group orbifold.

Proceeding as in \cite{Gu:2018fpm}[sections 9, 10], we integrate out
the $\sigma_a$ to get constraints
\begin{equation}
2 \sum_{i=1}^N \ln\left( \frac{ W^{i,2a-1} }{ W^{i,2a} } \right)
\: + \:
\sum_{\nu > 2a} \ln \left( \frac{ X_{2a-1,\nu} }{ X_{2a,\nu} } \right)
\: + \:
\sum_{\mu < 2a-1} \ln \left( \frac{ X_{\mu, 2a-1} }{ X_{\mu,2a} } \right)
\: = \: \tilde{t}.
\end{equation}
We define 
\begin{eqnarray}
W_+^{i,a} & \equiv & W^{i, 2a-1} W^{i, 2a},
\\
W_-^{i,a} & \equiv & \frac{ W^{i,2a-1} }{ W^{i,2a} },
\end{eqnarray}
in terms of which the constraint becomes
\begin{equation}
\sum_{i=1}^N \ln \left( W_-^{i,a} \right)^2 
\: + \:
\sum_{\nu > 2a} \ln \left( \frac{ X_{2a-1,\nu} }{ X_{2a,\nu} } \right)
\: + \:
\sum_{\mu < 2a-1} \ln \left( \frac{ X_{\mu, 2a-1} }{ X_{\mu,2a} } \right)
\: = \: \tilde{t}.
\end{equation}
We use this constraint to eliminate $W_-^{N,a}$:
\begin{equation}
\ln \left( W_-^{N,a} \right)^2 \: = \: 
- \sum_{i=1}^{N-1} \ln \left( W_-^{i,a} \right)^2 
\: - \:
 \sum_{\nu > 2a} \ln \left( \frac{ X_{2a-1,\nu} }{ X_{2a,\nu} } \right)
\: - \:
\sum_{\mu < 2a-1} \ln \left( \frac{ X_{\mu, 2a-1} }{ X_{\mu,2a} } \right)
\: + \:  \tilde{t}.
\end{equation}
(In principle, there are two roots for $W_-^{N,a}$, but as we saw
previously in the discussion of the $SO(2)$ mirror, there is a 
${\mathbb Z}_2$ orbifold that relates the two signs.)
Since we will no longer treat $W_-^{N,a}$ as a propagating field, we shall
rename it to $\Upsilon_a$, following the pattern of the 
discussion of the $O_+(k)$ mirror:
\begin{eqnarray}
\Upsilon_a & \equiv & W_-^{N,a},
\\
& = &
\tilde{q}^{-1/2} \left( \prod_{i=1}^{N-1} W_-^{i,a} \right)^{-1}
\left( \prod_{\nu > 2a} \frac{ X_{2a,\nu} }{ X_{2a-1,\nu} } \right)^{1/2}
\left( \prod_{\mu < 2a-1} \frac{ X_{\mu,2a} }{ X_{\mu,2a-1} } \right)^{1/2},
\end{eqnarray}
where $\tilde{q} = \exp(- \tilde{t})$.

In passing, just as in the previous discussion of the $SO(2)$ mirror,
if we check the pertinent Jacobian, we find that the fundamental fields
are $W_+^{i,a}$ and $\ln W_-^{i,a}$.  This will not modify the 
vacuum computation, but will be relevant for the central charge
computation.

The superpotential can now be written
\begin{eqnarray}
W & = & \sum_{i=1}^{N-1} \sum_{a=1}^M W_+^{i,a} \left(
W_-^{i,a} \: + \: \frac{1}{W_-^{i,a} } \right)
\: + \:
\sum_{a=1}^M W_+^{N,a} \left( \Upsilon_a \: + \: \Upsilon_a^{-1} \right)
\nonumber \\
& & \: + \: 
\sum_{\mu < \nu} X_{\mu \nu}
\: + \: \sum_{i=1}^N \sum_{a=1}^M \tilde{m}_i \ln \left( W_+^{i,a} \right)^2
\nonumber \\
& &
\: + \: \sum_{i \leq j} \exp\left( - T_{ij} \right)
\: + \: \sum_{i \leq j} \left( \tilde{m}_i + \tilde{m}_j \right) T_{ij}
\nonumber \\
& & 
\: + \: \left\{ \begin{array}{cl}
\sum_{i=1}^N \left( W^{i,2M+1} \right)^2 \: + \:
\sum_{i=1}^N  \tilde{m}_i \ln \left( W^{i,2M+1} \right)^2 & N-k+1 \mbox{ odd},
\\
0 & \mbox{else}. \end{array} \right.
\end{eqnarray}

The critical locus is then given by
\begin{eqnarray}
W_+^{i,a} \left( W_-^{i,a} \: + \: \frac{1}{ W_-^{i,a} } \right) & = &
- 2 \tilde{m}_i
\: \: \: \mbox{ for }i < N,
\\
W_+^{N,a} \left( \Upsilon_a + \Upsilon_a^{-1} \right) & = & - 2 \tilde{m}_N,
\\
W_+^{i,a} \left( W_-^{i,a} \: - \: \frac{1}{ W_-^{i,a} } \right) & = &
W_+^{N,a} \left( \Upsilon_a - \Upsilon_a^{-1} \right)
\: \: \: \mbox{ for }i < N,
\label{eq:sonk1:mirror:critical:3}
\\
\exp\left( - T_{ij} \right) & = & \tilde{m}_i + \tilde{m}_j,
\\
2 X_{2a,2b} & = & W_+^{N,a} \left( - \Upsilon_a + \Upsilon_a^{-1} \right)
\: + \:
W_+^{N,b} \left( - \Upsilon_b + \Upsilon_b^{-1} \right),
\\
2 X_{2a,2b-1} & = & W_+^{N,a} \left( - \Upsilon_a + \Upsilon_a^{-1} \right)
\: + \:
W_+^{N,b} \left( \Upsilon_b - \Upsilon_b^{-1} \right),
\\
2 X_{2a-1, 2b} & = & W_+^{N,a} \left( \Upsilon_a  - \Upsilon_a^{-1} \right)
\: + \: W_+^{N,b} \left( - \Upsilon_b + \Upsilon_b^{-1} \right),
\\
& = & - 2 X_{2a, 2b-1}, \\
2 X_{2a-1,2b-1} & = & W_+^{N,a} \left( \Upsilon_a - \Upsilon_a^{-1} \right)
\: + \:
W_+^{N,b}\left( \Upsilon_b - \Upsilon_b^{-1} \right),
\\
& = & - 2 X_{2a,2b},
\end{eqnarray}
and for $N-k+1$ odd,
\begin{eqnarray}
\left( W^{i,2M+1} \right)^2 & = & - \tilde{m}_i,
\\
2 X_{2a,2M+1} & = & W_+^{N,a} \left( -\Upsilon_a + \Upsilon_a^{-1} \right),
\\
2 X_{2a-1,2M+1} & = & W_+^{N,a} \left( \Upsilon_a - \Upsilon_a^{-1} \right),
\\
& = & - 2 X_{2a, 2M+1}.
\end{eqnarray}

On the critical locus, define
\begin{eqnarray}
\sigma_a & \equiv & \frac{1}{2} W_+^{i,a} \left( W_-^{i,a} \: - \:
\frac{1}{ W_-^{i,a} } \right)
\: \: \: \mbox{ for any }i < N,
\\
& = & \frac{1}{2} W_+^{N,a} \left( \Upsilon_a - \Upsilon_a^{-1} \right).
\end{eqnarray}
(These expressions match due to expression~(\ref{eq:sonk1:mirror:critical:3}).)
Furthermore, from the operator mirror map, as discussed in
\cite{Gu:2018fpm}, it is straightforward to see that the
``$\sigma_a$'' defined above is mirror to the $\sigma_a$ of the
original theory.  We chose the notation to implicitly reflect the
operator mirror map.

It is straightforward to verify that on the critical locus,
one has the Coulomb branch relation
\begin{equation}   \label{eq:sonk1:mirror:qc}
\prod_{i=1}^N \left( \sigma_a \: - \: \tilde{m}_i \right)
\: = \:
(-)^{N-k+1} \tilde{q} 
\prod_{i=1}^N \left( - \sigma_a \: - \: \tilde{m}_i \right).
\end{equation}

The excluded locus computed from the critical locus relations for 
the $W$ fields above
is then
\begin{equation}
\sigma_a \: \neq \: \pm \tilde{m}_i.
\end{equation}
From the $X$ fields, for $a \neq b$,
\begin{equation}   \label{eq:sonk1:mirror:excluded2}
\sigma_a \: \neq \: \pm \sigma_b,
\end{equation}
and for $N-k+1$ odd, we also have
\begin{equation}    \label{eq:sonk1:mirror:excluded3}
\sigma_a \: \neq \: 0.
\end{equation}

Now, let us quickly review the orbifold group actions, to assist
us in counting vacua later.  Briefly,
we claim the orbifold group will exchange roots, but the critical locus
will not intersect the fixed point locus of the orbifold in this mirror,
so in this particular mirror we do not have to consider any
twisted sector ground states.

For $N-k+1$ odd, the Weyl orbifold
group will act on $\sigma$s by signs, 
exchanging positive and negative roots, and leaving
zero roots fixed.  However, from the excluded locus 
condition~(\ref{eq:sonk1:mirror:excluded3}) in this case, we see that
no zero roots are allowed.  It will also interchange the
various solutions for different $\sigma_a$.

For $N-k+1$ even, the Weyl group orbifold will exchange signs of pairs 
$(\sigma_a, \sigma_b)$, $a \neq b$, and so on the fixed-point locus of
this part of the orbifold, $\sigma_a = \sigma_b = 0$ for
$a \neq b$, which is disallowed by excluded locus
condition~(\ref{eq:sonk1:mirror:excluded2}). Any one $\sigma_a$ can
vanish along the critical locus, but not multiple $\sigma_a$ simultaneously.
The fact that signs of $\sigma$'s are flipped only in pairs means there
is a two-fold degeneracy relative to the case of $N-k+1$ odd.
For example, for $SO(2)$, given two nonzero roots $\pm \sigma_0$,
there is no second element to exchange, so both roots count separately.
For $SO(4)$, the orbifold group identifies elements of two equivalence
classes of nonzero $\sigma$:
\begin{eqnarray*}
(+ \sigma_1, + \sigma_2 ) & \sim & (- \sigma_1, - \sigma_2),
\\
(+ \sigma_1, - \sigma_2 ) & \sim & (- \sigma_1, + \sigma_2).
\end{eqnarray*}
For $SO(6)$, again we have two equivalence classes of nonzero roots
under the orbifold, which we illustrate schematically below:
\begin{eqnarray*}
(+,+,+) & \sim & (+,-,-) \: \sim \: (-,+,-) \: \sim \: (-,-,+),
\\
(+,+,-) & \sim & (+,-,+) \: \sim \: (-,+,+) \: \sim \: (-,-,-).
\end{eqnarray*}
It is straightforward to argue that the same pattern holds for any
group $SO(2M)$:  the orbifold group action (flipping pairs of signs,
not individual signs) always results in precisely two inequivalent
nonzero $\sigma$ solutions.
As before, the orbifold also interchanges the various solutions for
different $\sigma_a$.

Finally, in both cases, there are multiple additional ${\mathbb Z}_2$
orbifolds, each acting as $W^{i,\alpha} \mapsto - W^{i,\alpha}$.
However, all $W^{i,\alpha} \neq 0$ (excluded loci), hence these 
orbifolds do not have any fixed points.

Now that we have established that there are no twisted sector
ground states, we will next count vacua, as distinct roots
of the Coulomb branch relation.

First, consider the
case that
\begin{equation}
(-)^{N-k+1} \tilde{q} \: = \: +1,
\end{equation}
for $k$ even.

If $N-k+1$ is even, then for $k$ even, $N$ is odd, and the
Coulomb branch relation~(\ref{eq:sonk1:mirror:qc}) reduces to the
degree $N$ polynomial
\begin{equation}  \label{eq:sonk1:mirror:qc:n-even}
\sigma^N \: + \: \left( \sum_{i<j} \tilde{m}_i \tilde{m}_j \right) \sigma^{N-2}
\: + \: \left( \sum_{i_1 < i_2 < i_3 < i_4} \tilde{m}_{i_1}
\tilde{m}_{i_2} \tilde{m}_{i_3} \tilde{m}_{i_4} \right) \sigma^{N-4}
\: + \: \cdots \: = \: 0,
\end{equation}
where $\sigma = \Pi_a + \tilde{m}_N$.
This equation is symmetric in $\sigma \mapsto - \sigma$, so all nonzero roots
will come in positive/negative pairs.  
The $N$ roots are of the form
\begin{equation}
\sigma \: = \: 0, \:
\pm \tilde{\sigma}_1, \: \cdots, \: 
\pm \tilde{\sigma}_{(N-1)/2}
\end{equation}
with (for generic $\tilde{m}_i$) none of the $\tilde{\sigma}$
vanishing (except the first root in the list above).
For $N-k+1$ even, one zero root is allowed by the excluded locus
conditions, and so we find the
number of vacua is
\begin{equation}
\left( \begin{array}{c} (N+1)/2 \\ M \end{array} \right)
\: + \:
\left( \begin{array}{c} (N-1)/2 \\ M-1 \end{array} \right)
\end{equation}
(including possible zero roots), all in untwisted sectors.  
The second term above reflects
the two-fold degeneracy among nonzero roots arising from the fact that
the Weyl orbifold only flips signs of pairs of $\sigma_a$.
(Since $N \geq k$, and $N-k+1 = 2M$, it is straightforward to 
see that the expression above is well-defined.)
The original gauge theory in this case is regular in the
sense of \cite{Hori:2011pd}.

Next we turn to the case that $N-k+1$ is odd,
so that (since $k$ is even), $N$ is even.  In this case,
the Coulomb branch relation~(\ref{eq:sonk1:mirror:qc})
reduces to the degree $N-1$ polynomial
\begin{equation}  \label{eq:sonk1:mirror:qc:n-1}
\left( \sum_i \tilde{m}_i \right) \sigma^{N-1} \: + \:
\left( \sum_{i_1 < i_2 < i_3} \tilde{m}_{i_1} \tilde{m}_{i_2}
\tilde{m}_{i_3} \right) \sigma^{N-3} \: + \: \cdots 
\: = \: 0,
\end{equation}
where again $\sigma = \Pi_a + \tilde{m}_N$.
As before, this equation is symmetric in $\sigma \mapsto - \sigma$,
so all nonzero roots will come in positive/negative pairs.

Just as in the analysis of the last section, for $N$ even,
this equation has $N$ roots, of the form
\begin{equation}
\sigma \: = \: \pm \tilde{\sigma}_1 , \: \cdots, \:
\pm \tilde{\sigma}_{N/2},
\end{equation}
with (for generic $\tilde{m}_i$) none vanishing.
Given that the orbifold group will flip signs of 
$\sigma$s, and that absolute values of different $\sigma_a$ must
differ, and finally given the $S_{N-k+1}$ portion of the orbifold, we
count
\begin{equation}
\left( \begin{array}{c} N/2 \\ M \end{array} \right)
\: = \:
\left( \begin{array}{c} N/2 \\ (N-k)/2 \end{array} \right)
\end{equation}
vacua, all in the untwisted sector.
Since $N \geq k$ and $N-k+1 = 2M+1$, $N/2 \geq M$, and so the
expression above is well-defined.

Now, we turn to the case that
\begin{equation}
(-)^{N-k+1} \tilde{q} \: = \: -1.
\end{equation}

Consider the case that $N-k+1$ is even, so that
(since $k$ is even), $N$ is odd.  In this case,
the Coulomb branch relation~(\ref{eq:sonk1:mirror:qc})
reduces to the degree $N-1$ polynomial~(\ref{eq:sonk1:mirror:qc:n-1}).
The $N-1$ roots are of the form
\begin{equation}
\sigma \: = \: \pm \tilde{\sigma}_1, \: \cdots, \:
\pm \tilde{\sigma}_{(N-1)/2},
\end{equation}
with (for generic $\tilde{m}_i$) none of the $\tilde{\sigma}$ vanishing.
As before, we count
\begin{equation}
2 \left( \begin{array}{c}  (N-1)/2 \\ M \end{array} \right)
\: = \:
2 \left( \begin{array}{c}  (N-1)/2 \\ (N-k+1)/2 \end{array} \right)
\end{equation}
vacua.  (The factor of $2$ arises because the Weyl group orbifold
only flips signs of pairs of $\sigma_a$.)

Finally, we turn to the case that $N-k+1$ is odd, so that
(since $k$ is even), $N$ is even.  
In this case, the Coulomb branch relation~(\ref{eq:sonk1:mirror:qc})
is a degree $N$ polynomial, with roots
\begin{equation}
\sigma \: = \: \pm \tilde{\sigma}_1, \: \cdots, \:
\pm \tilde{\sigma}_{N/2},
\end{equation}
where (for generic $\tilde{m}_i$) the $\tilde{\sigma}$ are all nonzero.
The number of vacua is
\begin{equation}
\left( \begin{array}{c} N/2 \\ M \end{array} \right)
\: = \:
\left( \begin{array}{c} N/2 \\ (N-k)/2 \end{array} \right).
\end{equation}
The original gauge theory in this case is regular in the sense
of \cite{Hori:2011pd}.

So far we have discussed results for $k$ even.
Now, we turn to $k$ odd.

First, consider the case that
\begin{equation}
(-)^{N-k+1} \tilde{q} \: = \: +1
\end{equation}
for $k$ odd.

If $N-k+1$ is even, then for $k$ odd, $N$ is even, and the Coulomb
branch relation~(\ref{eq:sonk1:mirror:qc}) reduces to a degree $N-1$
polynomial, with roots
\begin{equation}
\sigma \: = \: 0, \pm \tilde{\sigma}_1, \cdots, 
\pm \tilde{\sigma}_{(N-2)/2}.
\end{equation}
The number of vacua is
\begin{equation}
\left( \begin{array}{c} N/2 \\ (N-k+1)/2 \end{array} \right)
\: + \:
\left( \begin{array}{c} (N-2)/2 \\ (N-k+1)/2 \end{array} \right),
\end{equation}
all in untwisted sectors.  The second term reflects the two-fold
degeneracy among nonzero roots, arising from the fact that the
Weyl group acts by pairs of sign flips.  This theory is not regular
in the sense of \cite{Hori:2011pd}.

If $N-k+1$ is odd, then for $k$ odd, $N$ is odd, and the
Coulomb branch relation~(\ref{eq:sonk1:mirror:qc}) reduces to 
a degree $N$ polynomial, with roots
\begin{equation}
\sigma \: = \: 0, \pm \tilde{\sigma}_1, \cdots,
\pm \tilde{\sigma}_{(N-1)/2}.
\end{equation}
Since $N-k+1$ is odd, the zero root is on the excluded locus.
The number of vacua is
\begin{equation}
\left( \begin{array}{c} (N-1)/2 \\ (N-k)/2 \end{array} \right),
\end{equation}
using the fact that the Weyl orbifold now flips individual signs,
not just by pairs.  This theory is regular in the sense of
\cite{Hori:2011pd}, and the number of vacua matches that given
in \cite{Hori:2011pd}[table~(4.20)] and table~\ref{table:vac}.

Next, consider the case that
\begin{equation}
(-)^{N-k+1} \tilde{q} \: = \: -1
\end{equation}
for $k$ odd.

If $N-k+1$ is even, then for $k$ odd, $N$ is even,
and the Coulomb branch relation~(\ref{eq:sonk1:mirror:qc}) reduces
to a degree $N$ polynomial, with roots
\begin{equation}
\sigma \: = \: \pm \tilde{\sigma}_1, \cdots,
\pm \tilde{\sigma}_{N/2}.
\end{equation}
Since $N-k+1$ is even, the Weyl group acts by even numbers of sign
flips, leading to a two-fold degeneracy, so the number of vacua is
\begin{equation}
2 \left( \begin{array}{c} N/2 \\ (N-k+1)/2 \end{array} \right).
\end{equation}
This theory is regular in the sense of \cite{Hori:2011pd},
and the number of vacua matches that given in
\cite{Hori:2011pd}[table~(4.20)] and table~\ref{table:vac}.

Finally, if $N-k+1$ is odd, then for $k$ odd, $N$ is odd,
and the Coulomb branch relation~(\ref{eq:sonk1:mirror:qc}) reduces
to a degree $N-1$ polynomial, with roots
\begin{equation}
\sigma \: = \: \pm \tilde{\sigma}_1, \cdots,
\pm \tilde{\sigma}_{(N-1)/2}.
\end{equation}
Since $N-k+1$ is odd, the Weyl group acts by flipping individual
signs, so the number of vacua is
\begin{equation}
\left( \begin{array}{c} (N-1)/2 \\ (N-k)/2 \end{array} \right).
\end{equation}
This theory is not regular in the sense of \cite{Hori:2011pd}.

\subsubsection{Comparison of vacua}

Now, we shall compare vacua in the mirrors to the $O_+(k)$ and
$SO(N-k+1)$ gauge theories.  We shall compare regular theories
(in the sense of \cite{Hori:2011pd}, and will see that mirrors to
dual regular theories have the same number of vacua, as expected.
Non-regular theories are not related by the duality in the same 
fashion, and so do not exhibit matching vacua.

First, we shall assume $k$ even.

Consider the $O_+(k)$ gauge theory with $N$ odd and $q=+1$.
We have seen that the mirror has
\begin{equation}
\left( \begin{array}{c} (N+1)/2 \\ k/2 \end{array} \right) 
\: + \:
\left( \begin{array}{c} (N-1)/2 \\ k/2 - 1 \end{array} \right)
\end{equation}
vacua,
the same number as the mirror of the corresponding
$SO(N-k+1)$ gauge theory with $\tilde{q}=+1$.
In the $O_+(k)$ mirror, the second set of vacua arise in twisted
sectors, whereas in the $SO(N-k+1)$ mirror, all of the vacua arise
in an untwisted sector.

Much as in the prototype case, there is a ${\mathbb Z}_2$
symmetry on both sides.  In the $O_+(k)$ mirror, there is a quantum
${\mathbb Z}_2$ symmetry, acting by a sign on the 
\begin{displaymath}
\left( \begin{array}{c} (N-1)/2 \\ k/2 - 1 \end{array} \right)
\end{displaymath}
states in twisted sectors.  In the $SO(N-k+1)$ mirror,
there is an ordinary global ${\mathbb Z}_2$ symmetry, flipping the
sign of a single $\sigma$, and the corresponding set of states (reflecting
a two-fold degeneracy amongst nonzero roots, as the Weyl orbifold only
acts on pairs of $\sigma$s) are
again anti-invariant.

This case specializes to the $O_+(2) \leftrightarrow SO(2)$ prototype
matching, for $k=2$, $N=3$, and $q = \tilde{q} = +1$.
In this case,
\begin{equation}
\left( \begin{array}{c} (N+1)/2 \\ k/2 \end{array} \right) 
\: + \:
\left( \begin{array}{c} (N-1)/2 \\ k/2 - 1 \end{array} \right)
\: = \:
\left( \begin{array}{c} 2 \\ 1 \end{array} \right) \: + \:
\left( \begin{array}{c} 1 \\ 0 \end{array} \right) \: = \: 3.
\end{equation}
The comparison of ${\mathbb Z}_2$ symmetries also specializes.

Next, for $k$ even, consider the $O_+(k)$ gauge theory with $N$ even
and $q=-1$.  We have seen that the mirror has 
\begin{equation}
\left( \begin{array}{c} N/2 \\ k/2 \end{array} \right)
\end{equation}
vacua,
the same number as the mirror of the corresponding
$SO(N-k+1)$ gauge theory with $\tilde{q}=+1$.  In both cases,
all of the vacua arise in an untwisted sector; in neither case are
there any twisted sector ground states.

The other theories with $k$ even are not regular, and it is easy to
check that the number of vacua in mirrors to corresponding
$O_+(k)$ and $SO(N-k+1)$ gauge theories do not match.
The lesson we take is that the duality is only meant to apply to
regular theories, where regularity determines the value of the
discrete theta angle on either side of the duality.

Next, consider the case $k$ odd.  
We have seen that the mirror of the $O_+(k)$ gauge theory with $N$ odd
and $q=-1$ has
\begin{equation}
\left( \begin{array}{c} (N-1)/2 \\ (k-1)/2 \end{array} \right)
\end{equation}
vacua, the same number as the mirror of the corresponding
$SO(N-k+1)$ gauge theory with $\tilde{q}=-1$.  In this case,
all of the vacua arise in untwisted sectors.

Finally, for $k$ odd, consider the $O_+(k)$ gauge theory with $N$ even
and $q=+1$.  We have seen that the mirror has
\begin{equation}
2 \left( \begin{array}{c} N/2 \\ (k-1)/2 \end{array} \right)
\end{equation}
vacua, the same number as the mirror of the corresponding $SO(N-k+1)$ gauge
theory with $\tilde{q}=-1$.  

To summarize, we have verified that our mirror construction correctly
reproduces the
\begin{displaymath}
O_+(k) \: \leftrightarrow \: SO(N-k+1)
\end{displaymath}
duality between regular theories described in \cite{Hori:2011pd},
at least at the level of vacua.

\subsubsection{Central charges}
\label{sect:o-so:c}

In the case that all of the twisted masses vanish, these dual
gauge theories are believed to flow to a nontrivial SCFT in the IR.
The central charge of that IR SCFT is computed in 
\cite{Hori:2011pd}[equ'n (4.28)] to be
\begin{equation}
\frac{c}{3} \: = \: 
Nk \: - \: \frac{1}{2} k (k-1).
\end{equation}
In this section, we will describe how the same result can be
derived from the mirror theories, as another consistency check on the
mirrors.

Our computation is based on the fact that for a Landau-Ginzburg
model with a quasi-homogeneous potential, one can determine the
central charge.  Specifically, if the superpotential is
quasi-homogeneous in the form
\begin{equation}
W \left( \lambda^{q_i} \Phi_i \right) \: = \: \lambda^2 W\left( \Phi \right),
\end{equation}
then it is believed that there is a nontrivial SCFT in the IR with
central charge \cite{Vafa:1988uu}
\begin{equation}
\frac{c}{3} \: = \: \sum_i \left( 1 - q_i \right).
\end{equation}
As a consistency check, if a Landau-Ginzburg model has a massive
field, then the superpotential is of the form
$W = x^2$, which is quasi-homogeneous with $q = 1$, but since it is
massive, it should not survive to the IR, and indeed from the formula
above, its contribution to the central charge is $1-1 = 0$.

First, consider the mirror to the $O_+(k)$ theory,
described in section~\ref{sect:o+k:mirror}.  
The superpotential~(\ref{eq:o+k:mirror:sup-final}), in the
special case that all twisted masses $\tilde{m}_i$, vanish, is 
quasi-homogeneous under the following symmetry:
\begin{eqnarray}
Y & \mapsto & Y - \ln \lambda^2,
\\
X & \mapsto & \lambda^2 X.
\end{eqnarray}
This determines multiplicative charges $q$ and central charge
contributions as follows:
\begin{center}
\begin{tabular}{c|cc}
Field & $q$ & $c/3 = 1-q$ \\ \hline
$Y$ & $0$ & $1$ \\
$X$ & $2$ & $-1$
\end{tabular}
\end{center}

To compute the central charge, we need to count the number of each
type of field:
\begin{itemize}
\item $Y_{i,2a}$:  $(N-1) M$ fields,
\item $Y_{i,2a-1}$:  $N M$ fields,
\item $Y_{i,2M+1}$:  $N$ fields, in the case that $k$ is odd,
\item $X_{\mu \nu}$:  $(1/2) k (k-1) - M$ fields.
\end{itemize}
Adding up these contributions, we get the following results for
the central charge.  First, for $k$ even, so that $M=k/2$, we get
\begin{eqnarray}
\frac{c}{3} & = & (N-1) M + NM - \frac{1}{2} k(k-1) + M,
\\
& = & 2 N M - \frac{1}{2} k (k-1),
\\
& = & N k - \frac{1}{2} k (k-1),
\end{eqnarray}
matching \cite{Hori:2011pd}[equ'n (4.28)].
For $k$ odd, so that $M=(k-1)/2$, we get
\begin{eqnarray}
\frac{c}{3} & = & (N-1) M + N M + N - \frac{1}{2} k(k-1) + m,
\\
& = & 2 N M + N - \frac{1}{2} k(k-1),
\\
& = &  N k - \frac{1}{2} k (k-1),
\end{eqnarray}
again matching \cite{Hori:2011pd}[equ'n (4.28)].

Now, let us turn to the mirror to the dual $SO(N-k+1)$ gauge
theory, described in section~\ref{sect:sonk1:mirror}, again in the
special case that all twisted masses vanish.
Here, the fundamental fields, their charges, and their contributions
to the central charge are as follows:
\begin{center}
\begin{tabular}{c|cccc}
Field & R-symmetry & $q$ & $c/3=1-q$ & Number of fields \\ \hline
$W_+^{i,a}$ & $W \mapsto \lambda^2 W$ & $2$ & $-1$ & $N M$ \\
$\ln W_-^{i,a}$ & invariant & $0$ & $1$ & $(N-1) M$ \\
$W^{i,2M+1}$ & $W \mapsto \lambda W$ & $1$ & $0$ & $N$ if $k$ is odd,
$0$ else \\
$X_{\mu \nu}$ & $X \mapsto \lambda^2 X$ & $2$ & $-1$ & 
$(1/2) (N-k+1)(N-k) - M$ \\
$T_{ij}$ & $T \mapsto T + \ln \lambda^2$ & $0$ & $1$ &
$(1/2) N (N+1)$
\end{tabular}
\end{center}
Summing the contributions, we get for the total central charge,
\begin{equation}
\frac{c}{3} \: = \: N k  -  \frac{1}{2} k (k-1),
\end{equation}
which matches the result for the dual theory, as well as
\cite{Hori:2011pd}[equ'n (4.28)], as expected.

\subsection{Prototype:  $O_+(1) \leftrightarrow SO(N)$}
\label{sect:proto:o+1}

In this section, we will discuss the mirror to an $O_+(1)$ gauge theory with
$N$ vectors $x_1, \cdots, x_N$, 
which is dual to an $SO(N)$ gauge theory with $N$ vectors
$\tilde{x}^1, \cdots, \tilde{x}^N$, $(1/2)N(N+1)$ singlets $x_{ij} = + x_{ji}$,
and a superpotential
\begin{equation}
W \: = \: \sum_{i j} s_{ij} \tilde{x}^i \cdot \tilde{x}^j.
\end{equation}
We will compute the vacua in the mirrors to both theories.

\subsubsection{Mirror to $O_+(1)$}

The original gauge theory is a ${\mathbb Z}_2$ orbifold of
$N$ chiral superfields.  The mirror is one or two copies of 
the mirror to $N$ free fields, from the proposal in
section~\ref{sect:orb-odd}.

The mirror to $N$ free fields with twisted masses is simply
a set of $Y_i$ with superpotential \cite{Hori:2000kt,Hori:2001ax}
\begin{equation}
W \: = \: \sum_i \exp\left( - Y_i \right) \: - \:
\sum_i \tilde{m}_i Y_i.
\end{equation}

Now, the critical locus of the superpotential above is
trivially easy to compute, and is given by
\begin{equation}
\exp\left( - Y_i \right) \: = \: - \tilde{m}_i,
\end{equation}
hence the vacuum is unique.

From the proposal in section~\ref{sect:orb-odd},
if $N$ is even, the mirror is two copies of the theory above,
and hence has two vacua.  If $N$ is odd, the mirror is one copy
of the theory above, and hence has one vacuum.
For an $O(1)$ theory, as there does not exist a nontrivial discrete
theta angle, we necessarily have $q=+1$, for which regularity
(in the sense of \cite{Hori:2011pd}) requires $N$ even,
for which we have argued two vacua.
This matches the result for the number of vacua for this case
in \cite{Hori:2011pd}[table~(4.20)] and table~\ref{table:vac}.

\subsubsection{Mirror to $SO(N)$}

This mirror is a variation on that computed in
section~\ref{sect:sonk1:mirror}.
For brevity, we refer the reader to that section, and only
discuss the analysis of the vacua.

First, the Coulomb branch relation takes the form
\begin{equation}
\prod_{i=1}^N \left( \sigma_a - \tilde{m}_i \right)
\: = \:
(-)^N \tilde{q}
\prod_{i=1}^N \left( - \sigma_a - \tilde{m}_i \right).
\end{equation}
The excluded locus is given by
\begin{eqnarray}
\sigma_a & \neq & \pm \tilde{m}_i,
\\
\sigma_a & \neq & \pm \sigma_b 
\: \: \: \mbox{ for }a \neq b,
\end{eqnarray}
and for $N$ odd,
\begin{equation}
\sigma_a \: \neq \: 0.
\end{equation}

First, consider the case that $(-)^N \tilde{q} = +1$.

For $N$ odd, the Coulomb branch relation is a degree $N$ polynomial,
with roots
\begin{equation}
\sigma \: = \: 0, \pm \tilde{\sigma}_1, \cdots, 
\pm \tilde{\sigma}_{(N-1)/2}
\end{equation}
(for generic $\tilde{m}_i$).
The $\sigma=0$ root lies on the excluded locus, so taking into account the
Weyl orbifold, we find
\begin{equation}
\left( \begin{array}{c} (N-1)/2 \\ (N-1)/2 \end{array} \right)
\: = \: 1
\end{equation}
vacuum.  This theory is regular in the sense of \cite{Hori:2011pd}.

For $N$ even, the Coulomb branch relation is a degree $N-1$ polynomial,
with roots
\begin{equation}
\sigma \: = \: \pm \tilde{\sigma}_1, \cdots,
\pm \tilde{\sigma}_{N/2}.
\end{equation}
Taking into account the Weyl orbifold, we find
\begin{equation}
2 \left( \begin{array}{c} N/2 \\ N/2 \end{array} \right)
\: = \: 2
\end{equation}
vacua.  (The factor of 2 is due to the fact that the Weyl orbifold
only exchanges signs of pairs of $\sigma$'s, not individual $\sigma$'s.)
This theory is not regular in the sense of \cite{Hori:2011pd}.

Next, consider the case that $(-)^N \tilde{q} = -1$.

For $N$ odd, the Coulomb branch relation is a degree $N-1$ polynomial,
with roots
\begin{equation}
\sigma \: = \: \pm \tilde{\sigma}_1, \cdots,
\pm \tilde{\sigma}_{(N-1)/2}.
\end{equation}
Taking into account the Weyl orbifold, we find
\begin{equation}
\left( \begin{array}{c} (N-1)/2 \\ (N-1)/2 \end{array} \right)
\: = \: 1
\end{equation}
vacuum.  This theory is not regular in the sense of \cite{Hori:2011pd}.

For $N$ even, the Coulomb branch relation is a degree $N$ polynomial,
with roots
\begin{equation}
\sigma \: = \: \pm \tilde{\sigma}_1, \cdots,
\pm \tilde{\sigma}_{N/2}.
\end{equation}
Taking into account the Weyl orbifold, we find
\begin{equation}
2 \left( \begin{array}{c} N/2 \\ N/2 \end{array} \right)
\: = \: 2
\end{equation}
vacua.  This theory is regular in the sense of 
\cite{Hori:2011pd}.

Finally, we compare to the results for the mirrors to the
$O_+(1)$ theories.  We argued in the previous subsection
that the mirror to $O_+(1)$ with $N$ even has 2 vacua,
which matches the number of vacua in the mirror above to the
regular $SO(N)$ theory with $N$ even.  Similarly, we argued in the
previous subsection that the mirror to $O_+(1)$ with $N$ odd has
only 1 vacuum, which matches the number of vacua in the mirror
above to the regular $SO(N)$ theory with $N$ odd.  Thus, we see
that for regular theories, the proposed mirror is consistent with
the duality of \cite{Hori:2011pd}.

\subsection{Prototype: $O_+(3) \leftrightarrow SO(2)$}
\label{sect:proto:o+3}

In this section we will discuss the mirror to the $O_+(3)$ gauge theory with
$4$ vectors $x_1, \cdots, x_4$, 
which is dual to an $SO(2)$ gauge theory with $4$ vectors
$\tilde{x}^1, \cdots, \tilde{x}^4$,
$10$ singlets $s_{ij} = + s_{ji}$, 
and a superpotential
\begin{equation}
W \: = \: \sum_{i j} s_{ij} \tilde{x}^i \cdot \tilde{x}^j.
\end{equation}
We will compute and compare the vacua in the mirrors to both of
the theories.

\subsubsection{Mirror to $O_+(3)$}

In this section we will compute the mirror to the $O_+(3)$ gauge
theory with 4 vectors of twisted masses $\tilde{m}_i$.

Briefly, following the proposal in section~\ref{sect:orb-odd},
this is two copies of the mirror to the corresponding $SO(3)$ theory,
so we analyze that first.  The $SO(3)$ mirror is
an orbifold of a Landau-Ginzburg model with fields
\begin{itemize}
\item $Y_{i,\alpha}$, $i \in \{1, \cdots, 4 \}$, $\alpha \in \{1, 2, 3\}$,
\item $X_{1 3}$, $X_{2 3}$,
\item $\sigma$,
\end{itemize}
with superpotential
\begin{eqnarray}
W & = &
\sigma \left( \sum_{i=1}^4 \left( Y_{i,2} - Y_{i,1} \right)
\: - \: \ln \left( \frac{ X_{2 3} }{ X_{1 3} } \right)
\: - \: t \right)
\nonumber \\
& & 
\: + \:
\sum_{i=1}^4 \sum_{\alpha = 1}^3 \exp\left( - Y_{i,\alpha} \right)
\: + \: X_{1 3} \: + \: X_{2 3}
\: - \: \sum_{i, \alpha} \tilde{m}_i Y_{i, \alpha}.
\end{eqnarray}

Integrating out $\sigma$, we get the constraint
\begin{equation}
 \sum_{i=1}^4 \left( Y_{i,2} - Y_{i,1} \right)
\: - \: \ln \left( \frac{ X_{2 3} }{ X_{1 3} } \right)
\: = \: t.
\end{equation}
As before, we make the change of variables
\begin{eqnarray}
Y^+_{i} & \equiv & \frac{1}{2} \left( Y_{i,2} + Y_{i,1} \right),
\\
Y^-_i & \equiv & \frac{1}{2} \left( Y_{i,2} - Y_{i,1} \right),
\end{eqnarray}
so that the constraint becomes
\begin{equation}
\sum_{i=1}^4 Y^-_i \: - \: \ln \left( \frac{ X_{2 3} }{ X_{1 3} } \right)^{1/2}
\: = \: \frac{t}{2}.
\end{equation}
We use this constraint to eliminate $Y^-_4$,
and define $\Upsilon \equiv \exp(- Y^-_4)$:
\begin{eqnarray}
\Upsilon & \equiv & \exp\left( - Y^-_4 \right),
\\
& = & q^{1/2} \left( \prod_{i=1}^3 \exp\left( +  Y^-_i \right)
\right) \left( \frac{X_{13}}{X_{23}} \right)^{1/2},
\end{eqnarray}
where $q = \exp(-t)$.  The superpotential can then be written
\begin{eqnarray}
W & = & \sum_{i=1}^3 \left( \exp\left( - Y_i^+ - Y_i^- \right)
\: + \: \exp\left( - Y_i^+ + Y_-^- \right) \right)
\nonumber \\
& & 
\: + \: \exp\left( - Y_4^+ \right) \left( \Upsilon + \Upsilon^{-1} \right)
\: + \: \sum_{i=1}^4 \exp\left( - Y_{i,3} \right)
\: + \: X_{13} \: + \: X_{23}
\nonumber \\
& & 
\: - \: \sum_{i=1}^4 2 \tilde{m}_i Y_i^+ 
\: - \: \sum_{i=1}^4 \tilde{m}_i Y_{i,3}.
\end{eqnarray}

The critical locus is given as follows.
\begin{eqnarray}
Y_i^+: & & 
\exp\left( - Y_i^+ - Y_i^- \right) \: + \: 
\exp\left( - Y_i^+ + Y_i^- \right) \: = \: - 2 \tilde{m}_i
\: \: \: \mbox{ for }i < 4,
\\
Y_4^+: & & 
\exp\left( - Y_4^+ \right) \left( \Upsilon + \Upsilon^{-1} \right)
\: = \: 2 \tilde{m}_4,
\\
Y_i^-: & & 
\exp\left( - Y_i^+ - Y_i^- \right) \: - \:
\exp\left( - Y_i^+ + Y_i^- \right) 
\nonumber \\
& & \hspace*{0.5in} \: = \:
\exp\left( - Y_4^+ \right) \left( \Upsilon - \Upsilon^{-1} \right)
\: \: \: \mbox{ for }i < 4,
\\
Y_{i,3}: & &
\exp\left( - Y_{i,3} \right) \: = \: - \tilde{m}_i,
\\
X_{13}: & &
2 X_{13} \: = \: \exp\left( - Y_4^+ \right) \left( - \Upsilon +
\Upsilon^{-1} \right),
\\
X_{23}: & &
2 X_{23} \: = \: \exp\left( - Y_4^+ \right) \left( \Upsilon -
\Upsilon^{-1}\right) \: = \: - 2 X_{13}.
\end{eqnarray}

On the critical locus, define
\begin{eqnarray}
\sigma & \equiv & \frac{1}{2} \left(
\exp\left( - Y_i^+ - Y_i^- \right) \: - \:
\exp\left( - Y_i^+ + Y_i^- \right) \right)
\: \: \: \mbox{ for }i < 4,
\\
& = & \frac{1}{2} \exp\left( - Y_4^+ \right) \left(
\Upsilon - \Upsilon^{-1} \right).
\end{eqnarray}
(The fact that these definitions agree is a consequence of the
third critical locus equation above.)

It is then straightforward to show that on the critical locus,
\begin{equation}  \label{eq:o+3:mirror:qc}
\prod_{i=1}^4 \left( \sigma - \tilde{m}_i \right) 
\: = \: 
(-) q
\prod_{i=1}^4 \left( - \sigma - \tilde{m}_i \right),
\end{equation}
and it is also straightforward to compute the excluded locus
\begin{equation}
\sigma \: \neq \: 0, \: \pm \tilde{m}_i.
\end{equation}

Now, we will compute the number of vacua.

Consider the case $q=+1$.  
The Coulomb branch relation~(\ref{eq:o+3:mirror:qc}) is a degree
four polynomial, with roots
\begin{equation}
\sigma \: = \: \pm \tilde{\sigma}_1, \pm \tilde{\sigma}_2,
\end{equation}
where for generic twisted masses, $\tilde{\sigma} \neq 0$.
The Weyl group exchanges each sign, so we count
\begin{equation}
\left( \begin{array}{c} 2 \\ 1 \end{array} \right) \: = \: 2
\end{equation}
vacua.  This is the regular case, in the language of \cite{Hori:2011pd}.

Next, consider the case $q=-1$.  
In this case, the Coulomb branch relation~(\ref{eq:o+3:mirror:qc}) is
a degree three polynomial, with roots
\begin{equation}
\sigma \: = \: 0, \pm \tilde{\sigma}_1.
\end{equation}
The zero root is on the excluded locus, and the Weyl group exchanges signs,
so we count
\begin{equation}
\left( \begin{array}{c} 1 \\ 1 \end{array} \right) \: = \: 1
\end{equation}
vacuum.
This case is not regular in the sense of \cite{Hori:2011pd}.

So far, we have focused on computing the $SO(3)$ mirror.
The $O_+(3)$ mirror is two copies of the $SO(3)$ mirror,
so in the regular case ($q=+1$), the mirror to the $O_+(3)$ 
theory has four vacua, which matches the corresponding entry in
\cite{Hori:2011pd}[table~(4.20)] and table~\ref{table:vac}.

\subsubsection{Mirror to $SO(2)$}

In this section we discuss the mirror to an $SO(2)$ gauge theory with 
4 doublets $\tilde{x}^1, \cdots, \tilde{x}^4$ of R-charge one
and twisted masses $\tilde{m}_i$, 10 singlets
$s_{ij} = + s_{ji}$ of R-charge zero and twisted masses $- \tilde{m}_i
- \tilde{m}_j$, and a superpotential
\begin{equation}
W \: = \: \sum_{i j} s_{ij} \tilde{x}^i \cdot \tilde{x}^j.
\end{equation}

Happily, this is a special case of the mirror constructed
in section~\ref{sect:sonk1:mirror}, so we can simply read off results.
The critical locus is given by the Coulomb branch relation
\begin{equation}
\prod_{i=1}^4 \left( \sigma - \tilde{m}_i \right) \: = \:
(-)^2 \tilde{q} \prod_{i=1}^4 \left( - \sigma - \tilde{m}_i \right),
\end{equation}
with excluded locus
\begin{eqnarray}
\sigma & \neq & \pm \tilde{m}_i.
\end{eqnarray}
(In this theory, there is only one $\sigma$.)

If $\tilde{q} = +1$, then this is a degree three polynomial, with roots
\begin{equation}
\sigma \: = \: 0, \pm \tilde{\sigma}_1.
\end{equation}
These roots corresond to three distinct vacua, all in the untwisted sector.

If $\tilde{q} = -1$, then this is a degree four polynomial, with roots
\begin{equation}
\sigma \: = \: \pm \tilde{\sigma}_1, \pm \tilde{\sigma}_2.
\end{equation}
These roots correspond to four distinct vacua, all in the untwisted sector.
This is the mirror of a regular theory, and the number of vacua
matches that given in \cite{Hori:2011pd}[table~(4.20)] and 
table~\ref{table:vac}.

Finally, comparing to the $O_+(3)$ mirror in the previous section,
we found there that that mirror has four vacua, matching the number
of vacua in the regular theory here.  Thus, our mirror proposal
is consistent with this example of the duality of
\cite{Hori:2011pd}, at least in terms of counting vacua.

\subsection{$SO(k) \leftrightarrow O_+(N-k+1)$ duality}
\label{sect:sok:o+n-k+1}

\subsubsection{Mirror to $SO(k)$ gauge theory}

In this section we will discuss the mirror to the $SO(k)$ gauge
theory with $N \geq k$ massless vectors $x_1, \cdots, x_N$
with twisted masses $\tilde{m}_i$.  In fact, this mirror was
previously constructed in \cite{Gu:2018fpm}[sections 9, 10], 
and a closely related
mirror was constructed in section~\ref{sect:o+k:mirror}, so here we will be
brief, and refer the reader to thsoe references for further details
of the mirror construction for this case.

The critical locus of the mirror superpotential is defined by the
Coulomb branch relation
\begin{equation} \label{eq:sok:mirror:qc}
\prod_{i=1}^N \left( \sigma_a - \tilde{m}_i \right) \: = \: 
(-)^k q
\prod_{i=1}^N \left( - \sigma_a - \tilde{m}_i \right),
\end{equation}
with excluded loci
\begin{eqnarray}
\sigma_a & \neq & \pm \tilde{m}_i,
\\
\sigma_a & \neq & \pm \sigma_b \: \: \: \mbox{ for }a \neq b,
\end{eqnarray}
and in the case that $k$ is odd,
\begin{equation}
\sigma_a \: \neq \: 0.
\end{equation}
We will restrict to mirrors of regular theories, in the sense of
\cite{Hori:2011pd}, which implies $q = (-)^{N-k+1}$.

First, we consider the case that $k$ is even.

If $N$ is even (for even $k$), then the Coulomb branch 
relation~(\ref{eq:sok:mirror:qc}) reduces to 
a degree $N$ polynomial with roots
\begin{equation}
\sigma \: = \: \pm \tilde{\sigma}_1, \cdots,
\pm \tilde{\sigma}_{N/2}.
\end{equation}
Since the orbifold group acts by rearrangement and by
flipping pairs of signs of $\sigma_a$, 
the number
of vacua is then
\begin{equation}
2 \left( \begin{array}{c} N/2 \\ k/2 \end{array} \right).
\end{equation}
This result matches the corresponding value for the original gauge theory in
\cite{Hori:2011pd}[table~(4.20)] and table~\ref{table:vac}.
All of these vacua lie in the untwisted sector of the orbifold,
and more generally, for an $SO(k)$ mirror, since the gauge group is
connected, the vacua will always
lie in the untwisted sector, as the fixed-point locus does not
intersect the critical locus outside of the excluded locus.

Now, consider the case that $N$ is odd (for even $k$).
The Coulomb branch relation~(\ref{eq:sok:mirror:qc}) reduces to 
a degree $N$ polynomial again, with roots
\begin{equation}
\sigma \: = \: 0, \pm \tilde{\sigma}_1, \cdots,
\pm \tilde{\sigma}_{(N-1)/2}.
\end{equation}
Since the orbifold group acts by rearrangement and
by signs on pairs of $\sigma_a$,
a single zero root is allowed amongst the $\sigma_a$
and not on the excluded locus.  

Now, sets of $\sigma_a$ in which only nonzero roots appear
all come in pairs, since the Weyl orbifold group acts by rearranging and by
flipping
signs of pairs of $\sigma_a$.  For example, if $k=4$, then one 
has vacua of the form
\begin{eqnarray*}
(+ \tilde{\sigma}_1, + \tilde{\sigma}_2) & \sim &
( - \tilde{\sigma}_1, - \tilde{\sigma}_2),
\\
(+ \tilde{\sigma}_1, - \tilde{\sigma}_2) & \sim &
(- \tilde{\sigma}_1, + \tilde{\sigma}_2).
\end{eqnarray*}
On the other hand, if the set of $\sigma_a$ includes the zero root,
then since zero is invariant under sign flips, one only gets as many
vacua as distinct roots up to sign.  For example, for $k=4$, 
there are vacua of the form
\begin{eqnarray*}
(0, + \tilde{\sigma}_1) & \sim & (0, - \tilde{\sigma}_1),
\\
(0, + \tilde{\sigma}_2) & \sim & (0, - \tilde{\sigma}_2).
\end{eqnarray*}
As a result, the total number of vacua in this case is
\begin{equation}
2 \left( \begin{array}{c} (N-1)/2 \\ k/2 \end{array} \right)
\: + \:
\left( \begin{array}{c} (N-1)/2 \\ k/2 - 1 \end{array} \right),
\end{equation}
where the two terms count, respectively, the number of vacua defined
by solely non-zero roots, and the number of vacua with one zero root.
This result matches the corresponding result for the original
gauge theory in \cite{Hori:2011pd}[table~(4.20)] and table~\ref{table:vac}.

Next, we consider the case that $k$ is odd.

If $N$ is even (and $k$ odd), then for the mirror to a regular theory,
the Coulomb branch relation~(\ref{eq:sok:mirror:qc}) reduces to
a degree $N$ polynomial, with roots
\begin{equation}
\sigma \: = \: \pm \tilde{\sigma}_1, \cdots,
\pm \tilde{\sigma}_{N/2}.
\end{equation}
The Weyl orbifold group of the mirror to $SO(k)$ for $k$ odd acts
by rearranging and on individual $\sigma_a$ by sign flips, 
so we see that there are
\begin{equation}
\left( \begin{array}{c} N/2 \\ (k-1)/2 \end{array} \right)
\end{equation}
vacua, matching the result for the original gauge theory in
\cite{Hori:2011pd}[table~(4.20)] and table~\ref{table:vac}.

If $N$ is odd (and $k$ odd), then for the mirror to a regular theory,
the Coulomb branch relation~(\ref{eq:sok:mirror:qc}) reduces to
a degree $N$ polynomial, with roots
\begin{equation}
\sigma \: = \: 0, \pm \tilde{\sigma}_1, \cdots,
\pm \tilde{\sigma}_{(N-1)/2}.
\end{equation}
In this case, since $k$ is odd, the zero root lies on the excluded locus,
and so can be ignored.  Furthermore, for the mirror to $SO$(odd),
the Weyl orbifold group acts by rearranging and by
sign flips on individual $\sigma_a$,
not pairs, so there is no two-fold degeneracy amongst vacua.
The number of vacua is then
\begin{equation}
\left( \begin{array}{c} (N-1)/2 \\ (k-1)/2 \end{array} \right),
\end{equation}
which matches the result for the original gauge theory in
\cite{Hori:2011pd}[table~(4.20)] and table~\ref{table:vac}.

In the next section, we will perform the analogous computations in the
mirror to the dual gauge theory.

\subsubsection{Mirror to $O_+(N-k+1)$ gauge theory}

In this section we describe the mirror to the
$O_+(N-k+1)$ gauge theory arising as the dual of the $SO(k)$ gauge
theory of the previous subsection.  This $O_+(N-k+1)$ gauge theory
has $N \geq k$ vectors $\tilde{x}^1, \cdots, \tilde{x}^N$
of twisted mass $- \tilde{m}_i$ and R-charge $1$,
plus $(1/2)N(N+1)$ singlets $s_{ij} = + s_{ji}$ of twisted
mass $+ \tilde{m}_i + \tilde{m}_j$ and R-charge $0$, and a superpotential
\begin{equation}
W \: = \: \sum_{ij} s_{ij} \tilde{x}^i \cdot \tilde{x}^j.
\end{equation}

The mirror to this theory has the same fields and superpotential
as the $SO(N-k+1)$ mirror discussed in section~\ref{sect:sonk1:mirror},
and we refer the reader there for those details, omitting them here
for brevity.
The $O_+(N-k+1)$ mirror differs from the $SO(N-k+1)$ mirror in having
a different orbifold structure:
\begin{itemize}
\item If $N-k+1$ is even, then the orbifold is a ${\mathbb Z}_2$
extension of the Weyl group orbifold of the $SO(N-k+1)$ mirror,
implementing in effect the Weyl group of $SO(N-k+2)$, as discussed in
section~\ref{sect:orb-even}.
\item If $N-k+1$ is odd, then the theory is either one or two copies
of the corresponding $SO(n-k+1)$ mirror, as discussed in
section~\ref{sect:orb-odd}.
\end{itemize}

As discussed in section~\ref{sect:sonk1:mirror},
the critical locus of the superpotential is defined by the
Coulomb branch relation
\begin{equation}  \label{eq:o+nk1:mirror:qc}
\prod_{i=1}^N \left( \sigma_a - \tilde{m}_i \right)
\: = \:
(-)^{N-k+1} \tilde{q} \prod_{i=1}^N \left( - \sigma_a - \tilde{m}_i
\right),
\end{equation}
and the excluded loci are given by
\begin{eqnarray}
\sigma_a & \neq & \pm \tilde{m}_i,
\\
\sigma_a & \neq & \pm \sigma_b \: \: \: \mbox{ for }a \neq b,
\end{eqnarray}
and if $N-k+1$ is odd,
\begin{equation}
\sigma_a \: \neq \: 0.
\end{equation}
We shall assume in this section that the original gauge
theory is regular in the sense of \cite{Hori:2011pd},
which requires $\tilde{q} = (-)^{k}$.

First, we consider the case that $k$ is even.

Suppose that $N$ is even (for $k$ even), so that $N-k+1$ is odd.
The Coulomb branch relation~(\ref{eq:o+nk1:mirror:qc}) reduces to
a degree $N$ polynomial, with roots
\begin{equation}
\sigma \: = \: \pm \tilde{\sigma}_1, \cdots,
\pm \tilde{\sigma}_{N/2}.
\end{equation}
The Weyl orbifold group of $SO(N-k+1)$ acts by rearrangement and
on individual
$\sigma_a$ by sign flips, so we find that the number of vacua 
of the $SO(N-k+1)$ mirror is
\begin{equation}
\left( \begin{array}{c} N/2 \\ (N-k)/2 \end{array} \right)
\: = \:
\left( \begin{array}{c} N/2 \\ k/2 \end{array} \right).
\end{equation}
Now, the mirror of $O_+(N-k+1)$ is two copies of the $SO(N-k+1)$
mirror, from section~\ref{sect:orb-odd}, so the total number of
vacua in the $O_+(N-k+1)$ mirror is
\begin{equation}
2 \left( \begin{array}{c} N/2 \\ k/2 \end{array} \right),
\end{equation}
which precisely matches the result of the previous section for
the mirror of the dual $SO(k)$ gauge theory.

Now, suppose that $N$ is odd (for $k$ even), so that $N-k+1$ is even.
The Coulomb branch relation~(\ref{eq:o+nk1:mirror:qc}) reduces to
a degree $N$ polynomial, with roots
\begin{equation}
\sigma \: = \: 0, \pm \tilde{\sigma}_1, \cdots,
\pm \tilde{\sigma}_{(N-1)/2}.
\end{equation}
The zero root is not excluded, since $N-k+1$ is even.
The orbifold group of the mirror to $O_+(N-k+1)$ flips signs of
individual $\sigma_a$, as discussed in section~\ref{sect:orb-even},
so the zero root is on the fixed-point locus,
and there may exist twisted sectors.

Following section~\ref{sect:orb-even}, in the mirror to $O_+$(even),
the extra ${\mathbb Z}_2$ in the orbifold group is $\tau (-)^F$,
so from section~\ref{sect:proto:o+2}, since there are zero chiral
multiplets with twisted mass in the mirror, there are invariant
ground states in both the twisted and untwisted sectors, for both
RR and NS-NS.  (The fields in the mirror have complex masses,
mirror to fields in the original gauge theory with twisted masses.)

There are
\begin{equation}
\left( \begin{array}{c} (N-1)/2 \\ k/2-1 \end{array} \right)
\end{equation}
vacua corresponding to nonzero roots, all in the untwisted sector,
and 
\begin{equation}
2 \left( \begin{array}{c} (N-1)/2 \\ k/2 \end{array} \right)
\end{equation}
vacua in the twisted and untwisted sectors, corresponding to vacua
utilizing one copy of the zero root, for a total of
\begin{equation}
2 \left( \begin{array}{c} (N-1)/2 \\ k/2 \end{array} \right)
\: + \:
\left( \begin{array}{c} (N-1)/2 \\ k/2-1 \end{array} \right)
\end{equation}
twisted and untwisted sector vacua.  This total matches the number
of vacua in the mirror to the dual $SO(k)$ gauge theory.
The reader should note that this matching mixes twisted and untwisted
sectors:  in the mirror to the corresponding $SO(k)$ gauge
theory, all of the vacua were in untwisted sectors, whereas here
they arise in both.

Next, we consider the case that $k$ is odd.

Suppose that $N$ is even (for $k$ odd), so that $N-k+1$ is even.
The Coulomb branch relation~(\ref{eq:o+nk1:mirror:qc}) reduces to
a degree $N$ polynomial, with roots
\begin{equation}
\sigma \: = \: \pm \tilde{\sigma}_1, \cdots,
\pm \tilde{\sigma}_{N/2}.
\end{equation}
The orbifold group of the $O_+(N-k+1)$ mirror acts on the $\sigma_a$
by individual sign flips and rearrangements, so we see that there are
\begin{equation}
\left( \begin{array}{c} N/2 \\ (k-1)/2 \end{array} \right)
\end{equation}
vacua, matching the number of vacua in the mirror of the dual
$SO(k)$ theory described in the last section.

Finally, suppose that $N$ is odd (for $k$ odd), so that $N-k+1$ is odd.
The Coulomb branch relation~(\ref{eq:o+nk1:mirror:qc}) reduces to
a degree $N$ polynomial, with roots
\begin{equation}
\sigma \: = \: 0, \pm \tilde{\sigma}_1, \cdots,
\pm \tilde{\sigma}_{(N-1)/2}.
\end{equation}
Here, the zero root is excluded, and the orbifold group acts by
rearrangements and by
sign flips on the individual $\sigma_a$.
The mirror to the $SO(N-k+1)$ theory then has
\begin{equation}
\left( \begin{array}{c} (N-1)/2 \\ (k-1)/2 \end{array} \right)
\end{equation}
vacua.  From section~\ref{sect:orb-odd}, the mirror of $O_+$(odd) with
an odd number of vectors is one copy of the mirror to the
corresponding $SO$ gauge theory, so we see that the mirror to
the $O_+(N-k+1)$ gauge theory has
\begin{equation}
\left( \begin{array}{c} (N-1)/2 \\ (k-1)/2 \end{array} \right)
\end{equation}
vacua, which matches the result for the number of vacua for
the mirror of the dual $SO(k)$ gauge theory.

Thus, we have found that the mirrors to the dual $SO(k)$ and
$O_+(N-k+1)$ gauge theories have the same number of vacua
(in regular cases).

\section{$O_--O_-$ duality}
\label{sect:o-o-}

In this section, we will study mirrors to both sides of the
duality \cite{Hori:2011pd}[section 4.6]
\begin{equation}
O_-(k) \: \leftrightarrow \: O_-(N-k+1)
\end{equation}
for $N \geq k$.  As in the $SO-O$ dualities,
\begin{itemize}
\item the theory on the left has
$N$ massless vectors $x_1, \cdots, x_N$,
with twisted masses $\tilde{m}_i$,
and 
\item
the theory on the right has $N$ vectors
$\tilde{x}^1, \cdots, \tilde{x}^N$, of twisted masses $\tilde{m}_i$,
along with
$(1/2)N(N+1)$ singlets $s_{ij} = + s_{ji}$,
$1 \leq i, j \leq N$, of twisted mass $-\tilde{m}_i - \tilde{m}_j$,
and a superpotential
\begin{equation}
W \: = \: \sum_{i,j} s_{ij} \tilde{x}^i \cdot \tilde{x}^j.
\end{equation}
\end{itemize}
The mesons in the two theories are related by
\begin{equation}
s_{ij} \: = \: x_i \cdot x_j.
\end{equation}
The dualities are only claimed to exist when all the theories
in question are regular, in the sense of \cite{Hori:2011pd}
(and as reviewed in section~\ref{sect:regular-review}), which constrains the
discrete theta angle.

\subsection{Prototype: $O_-(2) \leftrightarrow O_-(N-1)$}

In this section, we will consider mirrors to the $O_--O_-$ duality in the
special case $k=2$, relating 
\begin{itemize}
\item an $O_-(2)$ gauge theory with $N$ chiral multiplets in the
doublet representation, of twisted mass $\tilde{m}_i$, and
\item an $O_-(N-1)$ gauge theory with $N$ chiral multiplets 
$\tilde{x}^1, \cdots, \tilde{x}^N$ in
the vector representation, of twisted mass $\tilde{m}_i$,
plus $(1/2) N (N+1)$ singlets $s_{ij} = + s_{ji}$,
$1 \leq i, j \leq N$, of twisted mass $- \tilde{m}_i - \tilde{m}_j$, and a 
superpotential
\begin{equation}
W \: = \: \sum_{i j} s_{ij} \tilde{x}^i \cdot \tilde{x}^j.
\end{equation}
\end{itemize}

\subsubsection{Mirror of $O_-(2)$ gauge theory}

Here we consider the mirror to an $O_-(2)$ gauge theory with $N$ chirals in
the doublet representation.  This will be closely related to the $O_+(2)$ mirror
discussed in section~\ref{sect:proto:o+2}.  We will assume $N$ is odd.

Briefly, the mirror to the theory with $q=+1$ is an orbifold of the
Landau-Ginzburg model defined by the superpotential
\begin{equation}
W \: = \: \sigma \left( - \sum_{i=1}^N Y_1^i + \sum_{i=1}^3 Y_2^i
 + t \right)
\: - \: \sum_{i=1}^N \tilde{m}_i \left(  Y_1^i + Y_2^i \right)
\: + \: \sum_{i=1}^N \exp\left( - Y_1^i \right) 
\: + \: \sum_{i=1}^N \exp\left( - Y_2^i \right).
\end{equation}
The orbifold in the mirror is a $\tau$ orbifold, that acts on the fields as
\begin{equation}
\sigma \: \mapsto \: - \sigma, \: \: \:
Y_1^i \: \leftrightarrow Y_2^i.
\end{equation}

Proceeding as in section~\ref{sect:proto:o+2}, we define
\begin{eqnarray}
Y_+^i & = & \frac{1}{2} \left(Y_1^i \: + \: Y_2^i \right), \\
Y_-^i & = & \frac{1}{2} \left( - Y_1^i \: + \: Y_2^i \right),
\end{eqnarray}
so that the constraint arising from integrating out $\sigma$ is
\begin{equation}
\sum_i Y_-^i \: = \: 0.
\end{equation}
We use the constraint to eliminate $Y_-^N$:
\begin{equation}
Y_-^N \: = \: - \sum_{i=1}^{N-1} Y_-^i ,
\end{equation}
and then the superpotential can be rewritten as
\begin{eqnarray}
W 
& = & - 2 \sum_{i=1}^N \tilde{m}_i Y_+^i \: + \:
\sum_{i=1}^{N-1} \exp\left( - Y_+^i + Y_-^i \right) \: + \:
\sum_{i=1}^{N-1} \exp\left( - Y_+^i - Y_-^i \right) 
\nonumber \\
& & 
\: + \:  \exp\left( - Y_+^N - \sum_{i=1}^{N-1} Y_-^i  \right) \: + \:
 \exp\left( - Y_+^N + \sum_{i=1}^{N-1} Y_-^i  \right).
\end{eqnarray}
The (untwisted) critical locus is defined by
\begin{eqnarray*}
\lefteqn{
\exp\left( - Y_+^i + Y_-^i \right) \: + \:
 \exp\left( - Y_+^N + \sum_{i=1}^{N-1} Y_-^i  \right)
} \nonumber \\
& = &
\exp\left( - Y_+^i - Y_-^i \right) \: + \:
 \exp\left( - Y_+^N - \sum_{i=1}^{N-1} Y_-^i  \right)
\: \: \: \mbox{ for } i \in \{1, \cdots, N-1\},
\end{eqnarray*}
\begin{eqnarray*}
\exp\left( - Y_+^i + Y_-^i \right) \: + \:
\exp\left( - Y_+^i - Y_-^i \right) & = &
-2 \tilde{m}_i
\: \: \: \mbox{ for } i\in \{1, \cdots, N-1\} ,
\\
 \exp\left( - Y_+^N - \sum_{i=1}^{N-1} Y_-^i  \right) \: + \:
 \exp\left( - Y_+^N + \sum_{i=1}^{N-1} Y_-^i  \right)
& = & -2 \tilde{m}_N.
\end{eqnarray*}
Along the critical locus, define
\begin{eqnarray}
\sigma & = & \frac{1}{2} \left( \exp\left( - Y_+^i + Y_-^i \right) \: - \:
\exp\left( - Y_+^i - Y_-^i \right) \right)
\: \: \: \mbox{ for }i=1, 2, \\
& = & \frac{1}{2} \left(
\exp\left( - Y_+^N - \sum_{i=1}^{N-1} Y_-^i  \right) \: - \:
\exp\left( - Y_+^N + \sum_{i=1}^{N-1} Y_-^i  \right) \right),
\end{eqnarray}
and it is straightforward to verify that this satisfies
\begin{equation}  \label{eq:o-2:mirror:qc}
\prod_{i=1}^N \left( \sigma - \tilde{m}_i \right) \: = \:
\prod_{i=1}^N \left( -\sigma - \tilde{m}_i \right).
\end{equation}

In this section we focus on the special case $N$ is odd,
so that the theory is regular for $q=+1$.
When $N$ is odd, equation~(\ref{eq:o-2:mirror:qc}) is a degree
$N$ polynomial, that is symmetric under $\sigma \mapsto -\sigma$.
It has roots
\begin{equation}
0, \pm \tilde{\sigma}_1, \pm \tilde{\sigma}_2, \cdots, 
\pm \tilde{\sigma}_{(N-1)/2}.
\end{equation}
The orbifold relates $\sigma \sim -\sigma$, so there are only $(N+1)/2$ distinct
roots of the Coulomb branch relation, in the untwisted sector.
The root at $\sigma=0$ intersects the fixed point locus of the orbifold,
and so there can be twisted sector ground states.

Here, we have $N$ fields with complex masses ($Y^i_-$)
that are acted upon by the orbifold (including $Y^N_-$, which we
previously integrated out).  (We do not include the $\sigma$ field
in this counting, as in the $e^2 \rightarrow \infty$ limit, it is merely
an auxiliary field.)  From section~\ref{sect:proto:o+2}
and \cite{Hori:2011pd}, in a $\tau$ orbifold, we know that
since $N$ is odd, only the RR twisted sector and NS-NS untwisted sector
will have an invariant ground state.
As a result, the total number of vacua in this theory, for $N$ odd,
both untwisted and twisted, is
\begin{equation}
(N-1)/2 \: + \: 1 \: = \: (N+1)/2.
\end{equation}
This matches the result for a regular $O_-$(even) theory with $N$ odd
given in \cite{Hori:2011pd}[table~(4.20)] and table~\ref{table:vac}.

\subsubsection{Mirror of $O_-(N-1)$ gauge theory}

In this section, we discuss the mirror to the dual
$O_-(N-1)$ gauge theory with $N$ chiral multiplets of R-charge $1$
and twisted mass $\tilde{m}_i$ and $(1/2)N(N+1)$ singlets
$s_{ij} = + s_{ji}$ of twisted mass $-\tilde{m}_i - \tilde{m}_j$.

As in the $O_-(2)$ discussion, we assume that $N$ is odd.
For $N$ odd, the mirror theory is an orbifold of a Landau-Ginzburg model
with fields
\begin{itemize}
\item $W^{i \alpha}$, $i \in \{1, \cdots, N\}$, $\alpha \in \{1, \cdots, N-1\}$,
\item $T_{ij} = + T_{ji}$,
\item $X_{\mu \nu} = X_{\nu \mu}^{-1}$, $\mu, \nu \in \{1, \cdots, N-1\}$,
\item $\sigma_a$, $a \in \{1, \cdots, M = (N-1)/2 \}$,
\end{itemize}
with superpotential
\begin{eqnarray}
W & = & \sum_{a=1}^M \sigma_a \Biggl( - \sum_{i=1}^N \ln \left( W^{i,2a}
\right)^2 \: + \: \sum_{i=1}^N \ln \left( W^{i,2a-1} \right)^2
\: + \: \sum_{\nu > 2a} \ln \left( \frac{ X_{2a-1,\nu} }{ X_{2a, \nu} }
\right)^2
\nonumber \\
& & \hspace*{2in}
 \: + \: \sum_{\mu < 2a-1} \ln \left(
\frac{ X_{\mu, 2a-1} }{ X_{\mu, 2a} } \right) \: - \:
\tilde{t} \Biggr)
\nonumber \\
& & \: + \: \sum_{i \alpha} \left( W^{i \alpha} \right)^2 \: + \:
\sum_{\mu < \nu} X_{\mu \nu} \: + \: 
\sum_{i \leq j} \exp\left( - T_{ij} \right)
\nonumber \\
& & \: + \: \sum_{i \alpha} \tilde{m}_i \ln \left( W^{i, \alpha} \right)^2
\: + \: \sum_{i \leq j} \left( \tilde{m}_i + \tilde{m}_j \right) T_{ij}.
\end{eqnarray}
The first part of the orbifold group is the extension of the
Weyl group discussed in section~\ref{sect:orb-even}.
Here, to describe the $O_-$ mirror rather than the $O_+$ theory,
we take the extra ${\mathbb Z}_2$ to be $\tau$, in the conventions of
that section.  In addition, for each $W^{i, \alpha}$,
there is an additional ${\mathbb Z}_2$ orbifold that maps
$W^{i, \alpha} \mapsto - W^{i, \alpha}$.

The untwisted sector analysis then closely follows the same
pattern as in section~\ref{sect:sonk1:mirror} for the mirror to an $SO(N-k+1)$
gauge theory, for $N-k+1$ even.  We only summarize the pertinent results here.
First, the Coulomb branch relation is
\begin{equation}
\prod_{i=1}^N \left( \sigma_a - \tilde{m}_i \right) \: = \:
(-)^{N-1} \tilde{q} \prod_{i=1}^N \left( - \sigma_a - \tilde{m}_i
\right),
\end{equation}
where the $\sigma_a$ describe the critical locus, subject to the
following excluded loci:
\begin{eqnarray}
\sigma_a & \neq & \pm \tilde{m}_i, 
\\
\sigma_a & \neq & \pm \sigma_b \: \: \: \mbox{ for }a \neq b.
\end{eqnarray}

Repeating the analysis of section~\ref{sect:sonk1:mirror},
if $N$ is odd and $(-)^{N-1} \tilde{q} = +1$, or more simply
$\tilde{q} = +1$, then 
the Coulomb branch equation has $N$ roots of the form
\begin{equation}
\sigma \: = \: 0, \pm \tilde{\sigma}_1, \cdots, \pm \tilde{\sigma}_{(N-1)/2},
\end{equation}
Precisely one zero root is allowed by the excluded locus conditions,
which may or may not have twisted sector contributions.
For a $\tau$ orbifold acting on $n$ fields with nonzero complex masses,
as discussed in section~\ref{sect:proto:o+2},
if $n$ is even, there will be invariant ground states in both
untwisted and twisted sectors, in both RR and NS-NS,
whereas if $n$ is odd, only the RR twisted sector and NS-NS untwisted
sector will have an invariant ground state.

In the present case, if we take the extra ${\mathbb Z}_2$ extending
Weyl to act on the $a$th $\sigma$, then it also acts as
\begin{eqnarray}
W^{i,2a-1} & \leftrightarrow & W^{i,2a},
\\
X_{\mu,2a-1} & \leftrightarrow & X_{\mu, 2a} 
\: \: \: \mbox{ for } \mu < 2a-1,
\\
X_{2a-1,\nu} & \leftrightarrow & X_{2a,\nu}
\: \: \: \mbox{ for }\nu > 2a,
\end{eqnarray}
As a result, there are $N$ $W$ fields of eigenvalue $-1$
under the ${\mathbb Z}_2$ (possibly including the previously
integrated-out $W^{i,2M}$),
and $2(M-1)$ $X$ fields of eigenvalue $-1$.  Since $N$ is odd,
$N+2(M-1)$ is always odd.

As a consequence, if $N$ is odd, then we only get contributions
from RR twisted and NS-NS untwisted sectors, hence there are
\begin{equation}
\left( \begin{array}{c} (N+1)/2 \\ (N-1)/2 \end{array} \right)
\: = \:
(N+1)/2  
\end{equation}
vacua.
This particular case ($N$ odd, $k=N-1$ even, $\tilde{q}=+1$)
is regular in the sense of
\cite{Hori:2011pd}, and the result above for the number of vacua in
the mirror correctly matches an entry
in \cite{Hori:2011pd}[table~(4.20)] and table~\ref{table:vac}.
As expected, this matches the result for the number of vacua
in the $O_-(2)$ mirror for $N$ odd, as discussed in the last section,
and so again we have confirmed that the mirror proposal is consistent
with the dualities in \cite{Hori:2011pd}.

\subsection{$O_-(k) \leftrightarrow O_-(N-k+1)$ duality}

In this section, we will compute the mirrors to either side
of the duality between 
\begin{itemize}
\item an $O_-(k)$ gauge theory with $N$ massless vectors
$x_1, \cdots, x_N$, with twisted masses $\tilde{m}_i$, and
\item an $O_-(N-k+1)$ gauge theory with $N$ vectors
$\tilde{x}^1, \cdots, \tilde{x}^N$ with twisted masses
$\tilde{m}_i$, singlets $s_{ij} = + s_{ji}$ of twisted masses
$- \tilde{m}_i - \tilde{m}_j$, and the superpotential
\begin{equation}
W \: = \: \sum_{i, j} s_{ij} \tilde{x}^i \cdot \tilde{x}^j.
\end{equation}
\end{itemize}
As before, since the original gauge theories are only equivalent
in the IR, we only expect the mirrors to be equivalent in the IR,
so we compare vacua for generic twisted masses, and verify that the
mirrors have the same number of vacua as one another and as the
original gauge theories.

The analysis in this section will be very similar to analyses done
previously.  The main novelty is that here we will discuss
mirrors of $O_-$ gauge theories, rather than $SO$ or $O_+$ gauge
theories.

\subsubsection{Mirror to $O_-(k)$ gauge theory}

In this section we will give the mirror to an $O_-(k)$ gauge theory
with $N \geq k$ vectors.

Now, the mirror to the $O_-(k)$ theory is closely related to the mirror
of an $SO(k)$ theory with the same matter.  From 
sections~\ref{sect:orb-even}, \ref{sect:orb-odd}:
\begin{itemize}
\item If $k$ is even, the $O_-(k)$ mirror is nearly the same as the $O_+(k)$
mirror, in which the Weyl group has been extended by ${\mathbb Z}_2$,
specifically a $\tau$ orbifold.
\item If $k$ is odd, the $O_-(k)$ mirror will be one or two copies of
the $SO(k)$ mirror:  one copy if $N$ is even, two copies if $N$ is odd.
\end{itemize}

In any event, the starting point is the mirror to an $SO(k)$ gauge
theory with $N$ vectors.  This is discussed in 
\cite{Gu:2018fpm}[sections 9, 10] and also reviewed in
section~\ref{sect:o+k:mirror}, so we will focus on vacua,
referring readers to those references for further details.
For the mirror to $SO(k)$, the critical loci of the mirror superpotential
are defined by the Coulomb branch relation
\begin{equation}  \label{eq:o-k:mirror:qc}
\prod_{i=1}^N \left( \sigma_a - \tilde{m}_i \right) \: = \:
(-)^k q \prod_{i=1}^N \left( - \sigma_a - \tilde{m}_i \right),
\end{equation}
where $a \in \{1, \cdots, M\}$ for $k = 2M$ if even or $2M+1$ if odd,
and with excluded loci
\begin{eqnarray}
\sigma_a & \neq & \pm \tilde{m}_i,
\\
\sigma_a & \neq & \pm \sigma_b \: \: \: \mbox{ for }a \neq b,
\end{eqnarray}
and if $k$ is odd, then in addition,
\begin{equation}
\sigma_a \: \neq \: 0.
\end{equation}

Next, we shall count vacua, in the case $k$ is even.
For brevity, we restrict to regular cases in the sense of \cite{Hori:2011pd},
meaning that $q = (-)^{N+1}$ for $k$ even.

If $N$ is even (and $k$ is even), then in the regular case,
the Coulomb branch relation~(\ref{eq:o-k:mirror:qc}) reduces
to a degree $N$ polynomial, with roots
\begin{equation}
\sigma \: = \: \pm \tilde{\sigma}_1, \cdots,
\pm \tilde{\sigma}_{N/2}.
\end{equation}
The Weyl orbifold group, for $O_-(k)$, is extended to that of
$SO(k+1)$, and so acts by flipping the sigms of each $\sigma$
individually, and by rearrangement of the $\sigma_a$.  
As a result, taking into account the excluded locus
requirement that different $\sigma_a$ must be distinct, and the
$S_k$ quotient, we see that there are
\begin{equation}
\left( \begin{array}{c} N/2 \\ k/2 \end{array} \right)
\end{equation}
vacua, all in the untwisted sector, which matches the counting
in \cite{Hori:2011pd}[table~(4.20)] and table~\ref{table:vac}.

If $N$ is odd (and $k$ is even), then in the regular case,
the Coulomb branch relation~(\ref{eq:o-k:mirror:qc}) reduces to
a degree $N$ polynomial, with roots
\begin{equation}
\sigma \: = \: 0, \pm \tilde{\sigma}_1, \cdots,
\pm \tilde{\sigma}_{(N-1)/2}.
\end{equation}
The Weyl orbifold group, for $O_-(k)$, is extended to that of $SO(k+1)$,
and so acts by flipping the signs of each $\sigma$ individually, as well
as by rearrangement of the $\sigma_a$.
However, since $k$ is even, the zero root is not excluded, and is on
the fixed-point locus of the orbifold.

Specifically, the zero root lies in the fixed-point locus of the
extra ${\mathbb Z}_2$ orbifold, which from section~\ref{sect:orb-even}
is a $\tau$ orbifold.  As discussed
in section~\ref{sect:proto:o+2}, the number of ground states that
survive a $\tau$ orbifold projection depends upon whether the number of
chiral multiplets with nonzero complex mass is even or odd.
In this case, if we think of the extra ${\mathbb Z}_2$ as acting
on the `last' component\footnote{
The Weyl group makes this choice equivalent to any other.
}, then for generic $\tilde{m}_i$, there are $k-2$ massive $X$ fields
which are acted upon nontrivially (the antiinvariant parts of 
$X_{\mu,k-1} \leftrightarrow X_{\mu,k}$),
which for $k$ even is always even, and $N$ massive $Y$ fields which
are acted upon nontrivially, which is odd.
(We include the $Y$ field we integrated out, as it is an example of
a massive field, but omit the $\sigma$ fields, which in the limit
$e^2 \rightarrow \infty$ are auxiliary fields.)

As a result, the number of fields with a complex mass in this
(mirror) theory is odd, so from the analysis of
section~\ref{sect:proto:o+2}, only the RR twisted sector and NS-NS untwisted
sector will have an invariant ground state.
Combining twisted and untwisted sector ground states, corresponding
to different roots, we find a total of
\begin{equation}
\left( \begin{array}{c} (N-1)/2 \\ k/2 \end{array} \right)
\: + \:
 \left( \begin{array}{c} (N-1)/2 \\ k/2 - 1 \end{array} \right)
\: = \:
\left( \begin{array}{c} (N+1)/2 \\ k/2 \end{array} \right)
\end{equation}
vacua, which matches the appropriate entry in
\cite{Hori:2011pd}[table~(4.20)] and table~\ref{table:vac}.

Next, we turn to the case that $k$ is odd.
For brevity, we restrict to regular cases in the sense of
\cite{Hori:2011pd}, which means that $q = (-)^N$.
The mirror to $O_-(k)$ for $k$ odd will be 
\begin{itemize}
\item one copy of the $SO(k)$ mirror if $N$ is even,
\item two copies of the $SO(k)$ mirror if $N$ is odd,
\end{itemize}
so we first analyze the $SO(k)$ mirror.

If $N$ is even (and $k$ odd), the Coulomb
branch relation~(\ref{eq:o-k:mirror:qc}) reduces to a degree $N$
polynomial, with roots
\begin{equation}
\sigma \: = \: \pm \tilde{\sigma}_1, \cdots,
\pm \tilde{\sigma}_{N/2}.
\end{equation}
Since $k$ is odd, the Weyl orbifold group acts by rearrangement and
individual
sign flips, so the number of vacua of the $SO(k)$ mirror is
\begin{equation}
\left( \begin{array}{c} N/2 \\ (k-1)/2 \end{array} \right).
\end{equation}
Since the $O_-(k)$ mirror is one copy of the $SO(k)$ mirror,
we see that the $O_-(k)$ mirror also has
\begin{equation}
\left( \begin{array}{c} N/2 \\ (k-1)/2 \end{array} \right)
\end{equation}
vacua, which matches the appropriate entry in
\cite{Hori:2011pd}[table~(4.20)] and table~\ref{table:vac}.

If $N$ is odd (and $k$ odd), the Coulomb branch
relation~(\ref{eq:o-k:mirror:qc}) reduces to a degree $N$ polynomial,
with roots
\begin{equation}
\sigma \: = \: 0, \pm \tilde{\sigma}_1, \cdots,
\pm \tilde{\sigma}_{(N-1)/2}.
\end{equation}
Since $k$ is odd, the Weyl orbifold group acts by rearrangement and
by individual
sign flips, and furthermore the zero root lies on the excluded
locus.  As a result, the number of vacua of the $SO(k)$ mirror is
\begin{equation}
\left( \begin{array}{c} (N-1)/2 \\ (k-1)/2 \end{array} \right).
\end{equation}
The $O_-(k)$ mirror is two copies of the $SO(k)$ mirror since $N$
is odd, so we see that the $O_-(k)$ mirror has
\begin{equation}
2 \left( \begin{array}{c} (N-1)/2 \\ (k-1)/2 \end{array} \right)
\end{equation}
vacua, which matches \cite{Hori:2011pd}[table~(4.20)] and table~\ref{table:vac}.

Next, we will analyze the mirror of the dual, and compare
vacua.

\subsubsection{Mirror to $O_-(N-k+1)$ gauge theory}

In this section, we will compute the mirror to the $O_-(N-k+1)$
gauge theory with $N \geq k$ vectors $\tilde{x}^1, \cdots,
\tilde{x}^N$ of twisted masses
$\tilde{m}_i$ and $(1/2) N (N+1)$ singlets $s_{ij} = + s_{ji}$ of
twisted mass $- \tilde{m}_i - \tilde{m}_j$, with superpotential
\begin{equation}
W \: = \: \sum_{ij} s_{ij} \tilde{x}^i \cdot \tilde{x}^j.
\end{equation}

Now, the mirror to an $O_-$(even) theory is merely the
mirror to $SO$(even) with an enlarged Weyl orbifold,
and the mirror to an $O_-$(odd) theory is merely copies of the
mirror to $SO$(odd).  As a result, we can re-use the results of
section~\ref{sect:sonk1:mirror} on mirrors to closely related
$SO(N-k+1)$ gauge theories.

In section~\ref{sect:sonk1:mirror}, briefly, it was argued that the
critical locus of the mirror superpotential is given by
the Coulomb branch relation
\begin{equation}  \label{eq:o-nk1:mirror:qc}
\prod_{i=1}^N \left( \sigma_a - \tilde{m}_i \right) \: = \:
(-)^{N-k+1} \tilde{q} \prod_{i=1}^N \left( - \sigma_a - \tilde{m}_i \right),
\end{equation}
with excluded locus
\begin{eqnarray}
\sigma_a & \neq & \pm \tilde{m}_i,
\\
\sigma_a & \neq & \pm \sigma_b \: \: \: \mbox{ for }a \neq b,
\end{eqnarray}
and if $N-k+1$ is odd, then in addition,
\begin{equation}
\sigma_a \: \neq \: 0.
\end{equation}
We shall restrict to cases in which the original gauge theory is regular,
meaning $\tilde{q} = (-)^k$.

Now, we shall count vacua, beginning with the case that $k$ is even.

If $N$ is even (and $k$ is even), then $N-k+1$ is odd, and in the regular
case, the Coulomb branch relation~(\ref{eq:o-nk1:mirror:qc}),
reduces to a degree $N$ polynomial with roots
\begin{equation}
\sigma \: = \: \pm \tilde{\sigma}_1, \cdots,
\pm \tilde{\sigma}_{N/2}.
\end{equation}
The Weyl group orbifold of $SO(N-k+1)$ acts by flipping the signs of
each $\sigma$ individually, so as a result, in the $SO(N-k+1)$ mirror,
there are
\begin{equation}
\left( \begin{array}{c} N/2 \\ (N-k)/2 \end{array} \right)
\: = \: 
\left( \begin{array}{c} N/2 \\ k/2 \end{array} \right)
\end{equation}
vacua, all in the untwisted sector.  Since $N-k+1$ is odd, and $N$ is even,
the mirror of the $O_-(N-k+1)$ guage theory is one copy of the
$SO(N-k+1)$ gauge theory, from section~\ref{sect:orb-odd},
and so we see that the $O_-(N-k+1)$ gauge theory has
\begin{equation}
\left( \begin{array}{c} N/2 \\ k/2 \end{array} \right)
\end{equation}
vauca.  This precisely matches the result for $N$ even and $k$ even for
the $O_-(k)$ mirror, as expected.

If $N$ is odd (and $k$ is even), then $N-k+1$ is even,
and in the regular case, the Coulomb branch relation~(\ref{eq:o-nk1:mirror:qc})
reduces to a degree $N$ polynomial with roots
\begin{equation}
\sigma \: = \: 0, \pm \tilde{\sigma}_1, \cdots,
\pm \tilde{\sigma}_{(N-1)/2}.
\end{equation}
The orbifold group for the $O_-(N-k+1)$ mirror extends the 
Weyl group orbifold of $SO(N-k+1)$ to act by, for example,
flipping signs of all $\sigma$, not just pairs, in addition to
rearrangment of the $\sigma_a$.
In addition, the zero root is not excluded (since $N-k+1$ is even),
and so represents an intersection of the critical locus with the
fixed-point locus, hence implies possible twisted sector contributions.

For generic $\tilde{m}_i$, there should be $N-k-1$ massive
$X$ fields which are acted upon nontrivially by the extra ${\mathbb Z}_2$
(the antiinvariant part of
$X_{\mu,N-k} \leftrightarrow X_{\mu,N-k+1}$), which is always even, and $N$ 
massive $Y$ fields which are acted upon nontrivially
(the antiinvariant part of $Y_{i,N-k} \leftrightarrow Y_{i,N-k+1}$),
which is odd.
(As before, we include the $Y$ field we integrated out, as it is a massive
field, but omit the $\sigma$ fields, which are auxiliary in the limit.)

As a result, there is an odd number of fields with nonzero complex
mass which are acted upon by the $\tau$ orbifold in the mirror to the
$O_-(N-k+1)$ theory, so from the analysis of section~\ref{sect:proto:o+2}
there will only be an invariant ground state in the RR twisted sector and
NS-NS twisted sector.
Combining twisted and untwisted sector ground
states, corresponding to different roots, we find a total of
\begin{equation}
\left( \begin{array}{c} (N-1)/2 \\ k/2 \end{array} \right)
\: + \:
 \left( \begin{array}{c} (N-1)/2 \\ k/2-1 \end{array} \right)
\: = \:
\left( \begin{array}{c} (N+1)/2 \\ k/2 \end{array} \right)
\end{equation}
vacua, which matches the result for $N$ odd and $k$ even for
the $O_-(k)$ mirror, as expected.

Next, we turn to the case that $k$ is odd.

If $N$ is even (and $k$ odd), then $N-k+1$ is even,
and in the regular case, the Coulomb branch relation~(\ref{eq:o-nk1:mirror:qc})
reduces to a degree $N$ polynomial, with roots
\begin{equation}
\sigma \: = \: \pm \tilde{\sigma}_1, \cdots,
\pm \tilde{\sigma}_{N/2}.
\end{equation}
The Weyl group orbifold of $O_-(N-k+1)$ acts by flipping signs of
individual $\sigma$ (a ${\mathbb Z}_2$ extension of the Weyl group of
$SO(N-k+1)$), so in the $O_-(N-k+1)$ mirror, there are
\begin{equation}
\left( \begin{array}{c} N/2 \\ (N-k+1)/2 \end{array} \right)
\: = \:
\left( \begin{array}{c} N/2 \\ (k-1)/2 \end{array} \right)
\end{equation}
vacua.  This precisely matches the result for $N$ even and $k$ odd
for the $O_-(k)$ mirror, as expected.

Finally, if $N$ is odd (and $k$ is odd), then $N-k+1$ is odd,
and in the regular case, the Coulomb branch relation~(\ref{eq:o-nk1:mirror:qc})
reduce to a degree $N$ polynomial, with roots
\begin{equation}
\sigma \: = \: 0, \pm \tilde{\sigma}_1, \cdots,
\pm \tilde{\sigma}_{(N-1)/2}.
\end{equation}
Since $N-k+1$ is odd, the zero root is excluded, and for $SO(N-k+1)$, the
Weyl group acts by signs on individual $\sigma$s, so the number of vacua
in the $SO(N-k+1)$ mirror is
\begin{equation}
\left( \begin{array}{c} (N-1)/2 \\ (k-1)/2 \end{array} \right).
\end{equation}
Since $N-k+1$ is odd and $N$ is odd, the mirror of the $O_-(N-k+1)$
gauge theory is two copies of the $SO(N-k+1)$ mirror,
from section~\ref{sect:orb-odd}, so we see that the number of vacua
in the mirror to the $O_-(N-k+1)$ gauge theory is
\begin{equation}
2 \left( \begin{array}{c} (N-1)/2 \\ (k-1)/2 \end{array} \right).
\end{equation}
This precisely matches the number of vacua of the mirror of the
dual $O_-(k)$ gauge theory, as expected.

Thus, we see that the mirror of the dual $O_-(N-k+1)$ theory has the
same number of vacua as the mirror to the $O_-(k)$ theory, confirming
our mirror construction.

\section{Other properties of $SO$ theories}
\label{sect:other}

\subsection{Supersymmetry breaking: $N \leq k-2$}

It was argued in \cite{Hori:2011pd}[section 4.4] that for
$SO(k)$ gauge theories with $N \leq k-2$ vectors,
supersymmetry is broken.  We can see that explicitly in the mirror.
Briefly, applying the $SO(k)$ mirrors described in
\cite{Gu:2018fpm}[sections 9, 10] and reviewed in 
section~\ref{sect:o+k:mirror}, the perturbative vacua, in the untwisted
sector, are defined by solutions to the Coulomb branch relation
\begin{equation}
\prod_{i=1}^N \left( \sigma_a - \tilde{m}_i \right) \: = \:
(-)^q \prod_{i=1}^N \left( - \sigma_a - \tilde{m}_i \right),
\end{equation}
subject to excluded loci
\begin{eqnarray}
\sigma_a & \neq & \pm \tilde{m}_i, 
\\
\sigma_a & \neq & \pm \sigma_b \: \: \: \mbox{ for }a \neq b,
\end{eqnarray}
and for $k$ odd,
\begin{equation}
\sigma_a \: \neq \: 0.
\end{equation}

Put briefly, for $N \leq k-2$, no solutions can be found that are not
excluded, hence there are no supersymmetric vacua.

Depending upon the value of $q$ and whether $N$ is even or odd,
the Coulomb branch relation above will have $N$ roots, at most
one of which will vanish, for which a vacuum is defined as a set
of either $k/2$ ($k$ even) or $(k-1)/2$ ($k$ odd) distinct
values of $\sigma$ coinciding with roots.  
Below is a table of possibilities for mirrors to $SO(k)$,
culled from the results in section~\ref{sect:o+k:mirror}:
\begin{center}
\begin{tabular}{c||c|c||c|c}
& \multicolumn{2}{c}{$q=+1$} & \multicolumn{2}{c}{$q=-1$} \\ \hline
& $N$ even & $N$ odd & $N$ even & $N$ odd \\ \hline
$k$ even & 0, $(N-2)/2$ & 0, $(N-1)/2$ & $N/2$
& $(N-1)/2$ \\
$k$ odd & $N/2$ & $(N-1)/2$ & 0, $(N-2)/2$ & 0, $(N-1)/2$
\end{tabular}
\end{center}
Entries including a `0' have a zero root; the other number gives the
number of nonzero entries, up to signs.

For $k$ even, to have a supersymmetric vacuum,
we need at least $k/2$ distinct (after signs) roots,
to be consistent with the excluded locus conditions.
However, since $N \leq k-2$, $N/2 \leq (k/2)-1$, it is straightforward
to see that not enough distinct $\sigma$ can be found,
hence no supersymmetric vacua exist in the $SO(k)$ mirror.

For $k$ odd, to have a supersymmetric vacuum,
we need at least $(k-1)/2$ distinct (after signs) roots,
to be consistent with the excluded locus conditions.
However, since $N \leq k-2$, we have $N/2 \leq (k/2)-1$, and examining
the table above we again see that not enough distinct roots exist,
hence no supersymmetric vacua exist in the $SO(k)$ mirror.

Thus, for $N \leq k-2$, the mirror to an $SO(k)$ gauge theory with
$N$ vectors does not admit a supersymmetric vacuum,
and so supersymmetry is broken.

\subsection{$N=k-1$ and free field theories}

Consider an $SO(k)$ gauge theory with $N$ chirals in the vector
representation.  For $N=k-1$, according to
\cite{Hori:2011pd}, these theories flow in the
IR to a free theory of $(1/2) k (k-1)$ mesons on the Higgs branch.
In this section, we will perform some consistency checks of this claim
in the mirror:  we will find the free field mirrors in the structure
of $SO(2)$ and $SO(3)$ theories, and will check central charge computations
for more general $k$.  We will leave detailed verifications for general
$SO(k)$ to future work.

\subsubsection{Prototype:  $SO(2)$, $N=1$}

As a simple model of later analyses, we consider the case that $k=2$,
namely the mirror to an $SO(2)$ gauge theory with $N=1$ doublet.
The mirror is described by three fields, $\sigma$, $Y_1$, $Y_2$,
with superpotential
\begin{equation}
W \: = \: \sigma\left( Y_2 - Y_1 - t \right) \: + \: \exp\left( - Y_1 \right)
\: + \: \exp\left( - Y_2 \right).
\end{equation}
(We assume all twisted masses vanish.)
As we are interested in generalizations to other $SO(k)$ groups,
we take $t \in \{0, \pi i \}$.

Integrating out $\sigma$ gives the constraint
\begin{equation}
Y_2 \: = \: Y_1 + t,
\end{equation}
hence the resulting superpotential
\begin{equation}
W \: = \: \exp\left( - Y_1 \right) \: + \: e^{-t} \exp\left( - Y_1 \right)
\: = \: \left( 1 + e^{-t} \right) \exp\left( - Y_1 \right).
\end{equation}
In the case that the original gauge theory is regular, $t = 0$,
and the superpotential $W \propto \exp\left( - Y_1 \right)$,
which is the same as the mirror to a single free chiral multiplet
\cite{Hori:2000kt,Hori:2001ax}.

Thus, in this case, we see that the mirror to a regular $SO(2)$ gauge
theory with $N=1$ doublet is the same as the mirror to a free field
theory of one chiral multiplet, consistent with the expectation that the
gauge theory flows to a theory of $(1/2) k (k-1) = 1$ chirals.

\subsubsection{Prototype:  $SO(3)$, $N=2$}

As a slightly more complicated example, we next turn to the case
that $k=3$, the mirror to an $SO(3)$ gauge theory with $N=2$
vectors.  This theory is predicted to have $(1/2)k(k-1) = 3$
free fields in the IR.  The mirror is described by the fields
$Y^i_a$ ($a \in \{1, 2, 3\}$, $i \in \{1, 2\}$), $X_{13}$, $X_{23}$,
and $\sigma$, and has the superpotential
\begin{eqnarray}
W & = & \sigma\left( \sum_{i=1}^2 \left( Y_2^i - Y_1^i \right)
\: - \: \ln \left( \frac{X_{23}}{X_{13}} \right) \right)
\: + \: X_{13} \: + \: X_{23} \: + \:
\sum_{i=1}^2 \sum_{a=1}^3 \exp\left( - Y^i_a \right).
\end{eqnarray}
We assume the theory is regular, so we have set $t=0$ for simplicity.

Now, we will first integrate out the $X$ fields.  Doing so will generate
a measure factor we should compute.  It is straightforward to show that
\begin{equation}
\frac{\partial^2 W}{\partial X_{13}^2} \: = \: 
- \frac{\sigma}{X_{13}^2}, \: \: \:
\frac{\partial^2 W}{\partial X_{23}^2} \: = \: + \frac{\sigma}{X_{23}^2},
\: \: \:
\frac{\partial^2 W}{\partial X_{13} \partial X_{23} } \: = \: 0,
\end{equation}
so
\begin{equation}
\det \partial^2 W \: = \: - \frac{ \sigma^2 }{ X_{13}^2 X_{23}^2 },
\end{equation}
which after taking into account the operator mirror map relation
\begin{equation}
X_{23} \: = \: + \sigma \: = \: - X_{13},
\end{equation}
implies that
\begin{equation}
\det \partial^2 W \: = \: - \frac{1}{\sigma^2},
\end{equation}
and so we get the measure factor
\begin{equation}
\left( \det \partial^2 W \right)^{-1} \: = \: - \sigma^2.
\end{equation}
After integrating out the two $X$ fields, the superpotential becomes
\begin{equation}
W \: = \: \sigma\left( \sum_{i=1}^2 \left( Y_2^i - Y_1^i \right)
\: - \: \pi i \right)  \: + \:
\sum_{i=1}^2 \sum_{a=1}^3 \exp\left( - Y^i_a \right).
\end{equation}

Next, we will integrate out two $Y$ fields, so as to generate a measure
factor that will cancel out the one obtained from integrating out $X$ fields.
Specifically, we will integrate out $Y^2_1$ and $Y^2_2$.
We find, for the superpotential above,
\begin{equation}
\frac{\partial W}{\partial Y^2_1} \: = \: - \sigma - \exp\left( - Y^2_1 \right),
\: \: \:
\frac{\partial W}{\partial Y^2_2} \: = \: + \sigma - \exp\left( - Y^2_2 \right),
\end{equation}
so that along the critical locus we can identify
\begin{equation}
\exp\left( - Y^2_1 \right) \: = \: - \sigma,
\: \: \:
\exp\left( -Y^2_2 \right) \: = \: + \sigma.
\end{equation}
To get the measure factor, note that
\begin{equation}
\frac{ \partial^2 W}{\partial Y^2_1 \partial Y^2_1} \: = \:
+ \exp\left( - Y^2_1 \right),
\: \: \:
\frac{\partial^2 W}{\partial Y^2_2 \partial Y^2_2 } \: = \:
+ \exp\left( - Y^2_2 \right),
\: \: \:
\frac{\partial^2 W}{\partial Y^2_1 \partial Y^2_2 } \: = \: 0,
\end{equation}
from which we compute
\begin{equation}
\det \partial^2 W \: = \: \exp\left( - Y^2_1 \right) 
\exp\left( - Y^2_2 \right) \: = \: - \sigma^2,
\end{equation}
which precisely cancels out the measure factor resulting from integrating
out the $Y$ fields.

Now, solving for $Y^2_{1,2}$ along the critical locus, we find
\begin{equation}
Y^2_1 \: = \: \ln \left( - \sigma \right), 
\: \: \:
Y^2_2 \: = \: \ln \left( + \sigma \right),
\end{equation}
so after integrating out these two $Y$ fields, the superpotential becomes
\begin{eqnarray}
W & = & \sigma\left( Y^1_2 - Y^1_1 + \ln \left( \frac{- \sigma}{\sigma} \right)
- \pi i \right) \: + \: \sigma \: - \sigma \: + \: 
\exp\left( - Y^1_1 \right) \: + \:
\exp\left( - Y^1_2 \right)
\nonumber \\
& &
 \: + \:
\exp\left( - Y^1_3 \right) \: + \:
\exp\left( - Y^2_3 \right),
\\
& & \nonumber \\
& = & \sigma\left( Y^1_2 - Y^1_1 \right)
\: + \: 
\exp\left( - Y^1_1 \right) \: + \:
\exp\left( - Y^1_2 \right)
\nonumber \\
& &
 \: + \:
\exp\left( - Y^1_3 \right) \: + \:
\exp\left( - Y^2_3 \right).
\end{eqnarray}
From the last line of the equation above, we see that $Y^1_3$ and
$Y^2_3$ already are visible as the mirrors to two free fields,
so it remains to make explicit the third.

Next, we integrate out $\sigma$, which gives the constraint
\begin{equation}
 Y_2^1 - Y_1^1 
\: = \: 0.
\end{equation}
Define
\begin{eqnarray}
Y_+^i & = & \frac{1}{2} \left( Y_1^i + Y_2^i \right),
\\
Y_-^i & = & \frac{1}{2} \left( - Y_1^i + Y_2^i \right),
\end{eqnarray}
then the constraint is simply
\begin{equation}
Y_-^1  \: = \: 0 .
\end{equation}

Eliminating $Y^1_-$, 
the superpotential becomes 
\begin{eqnarray}
W & = & 
2 \exp\left( - Y^1_+ \right)
\: + \: 
\exp\left( - Y_3^1 \right) \: + \: 
\exp\left( - Y_3^2 \right).
\end{eqnarray}
Modulo a field redefinition to rescale the first term, this is precisely
the mirror superpotential for three free fields, as expected for an
$SO(3)$ theory with $N=2$.

\subsubsection{Central charges}

In this section, we will briefly perform a consistency
check by computing the central charge predicted by the mirror theory.
Our analysis will be very similar to the central charge computation
in section~\ref{sect:o-so:c}, and so we will be very brief.

If one sets all twisted masses to zero,
the mirror to an $SO(k)$ theory, as discussed in
\cite{Gu:2018fpm}[sections 9, 10], has a superpotential which is
quasi-homogeneous with respect to the symmetry
\begin{equation}
x_{\mu \nu} \: \mapsto \: \lambda^2 X_{\mu \nu},
\: \: \:
Y_{i \alpha} \: \mapsto \: Y_{i \alpha} - \ln \lambda^2.
\end{equation}
As a result, these fields contribute to the central charge of a 
possible nontrivial IR limit as follows:
\begin{center}
\begin{tabular}{c|cc}
Field & $q$ & $c/3 = 1-q$ \\ \hline
$Y$ & $0$ & $1$ \\
$X$ & $2$ & $-1$.
\end{tabular}
\end{center}

Suppose first that $k$ is even.  In that case, there are
\begin{displaymath}
\frac{1}{2} k (k-1) \: - \: \frac{1}{2} k \: = \: 
\frac{1}{2} k (k-2)
\end{displaymath}
$X$ fields, and 
\begin{displaymath}
(N-1) \frac{k}{2} \: + \: N \frac{k}{2} \: = \:
\frac{1}{2} k (k-2) \: + \: \frac{1}{2} k (k-1)
\end{displaymath}
$Y$ fields, so the total predicted central charge is
\begin{equation}
\frac{c}{3} \: = \: \frac{1}{2} k (k-1),
\end{equation}
which matches the predicted number of free mesons in the IR.

Next, suppose that $k$ is odd.  In this case,
there are
\begin{displaymath}
\frac{1}{2} k (k-1) \: - \: \frac{1}{2} (k-1) \: = \: \frac{1}{2} (k-1)^2
\end{displaymath}
$X$ fields, and
\begin{displaymath}
(N-1) \frac{ (k-1)}{2} \: + \: N \frac{ (k-1)}{2} \: + \: N
\: = \:
\frac{1}{2} (k-1)(k-2) \: + \: \frac{1}{2} (k-1)^2 \: + \: k-1
\end{displaymath}
$Y$ fields, so the total predicted central charge is
\begin{equation}
\frac{c}{3} \: = \: \frac{1}{2} (k-1)(k-2) \: + \: (k-1) \: = \:
\frac{1}{2} k (k-1),
\end{equation}
again matching the predicted number of free mesons in the IR.

\section{Symplectic-symplectic duality}
\label{sect:sp-sp}

In \cite{Hori:2011pd}[section 5.6], it was proposed for that
$N$ odd, $N \geq 2k+3$, the following two-dimensional (2,2) supersymmetric
gauge theories are dual in the IR:
\begin{itemize}
\item $Sp(2k)$ gauge theory with $N$ fundamentals $x_1, \cdots, x_N$, 
with twisted masses $\tilde{m}_i$,
\item $Sp(N-2k-1)$ gauge theory with $N$ fundamentals
$\tilde{x}^1, \cdots \tilde{x}^N$, with twisted masses $\tilde{m}_i$, 
and $(1/2)N(N-1)$ singlets $a_{ij} = - a_{ji}$,
$1 \leq i, j \leq N$, with twisted masses $- \tilde{m}_i - \tilde{m}_j$
and superpotential
\begin{equation}
W \: = \: \sum_{i,j} a_{ij} [ \tilde{x}^i \tilde{x}^j ].
\end{equation}
\end{itemize}
(In the notation used here, 
$Sp(2) = SU(2)$.)
The mesons in the two theories are related as
\begin{equation}
[x_i x_j] \: = \: a_{ij}.
\end{equation}
In the expressions above,
the bracket notation indicates $Sp$-invariants:
\begin{equation}
[\tilde{x}^i \tilde{x}^j] \: = \: J^{ab} \tilde{x}^i_a \tilde{x}^j_b,
\end{equation}
where $J^{ab}$ is the antisymmetric $Sp$ symplectic form.

In this section we will compute the number of vacua of the mirrors
to each side of the duality above, and check that they match.
Unlike previous cases, here on both sides of the duality the
gauge group is connected, so this will not be a test of mirrors of
gauge theories with non-connected gauge groups; nevertheless,
as this is another duality in \cite{Hori:2011pd}, this paper does seem
the appropriate place to test it.

\subsection{Mirror to $Sp(2k)$ gauge theory}

The mirror to the first theory is 
discussed in \cite{Gu:2018fpm}[section 11].
For brevity, we refer the reader to that reference for details of
the mirror and its analysis.

It was shown in \cite{Gu:2018fpm}[section 11]
that
the critical loci are defined by the equation
\begin{equation}   \label{eq:spk:qc}
\prod_{i=1}^N \left( \sigma_a \: - \: \tilde{m}_i \right) \: = \:
\prod_{i=1}^N \left( - \sigma_a \: - \: \tilde{m}_i \right),
\end{equation}
and the excluded loci are defined by \cite{Gu:2018fpm}[section 11]
\begin{eqnarray}
\sigma_a & \neq & 0, \pm \tilde{m}_i,
\\
\sigma_a & \neq & \pm \sigma_b \: \: \: \mbox{ for }a \neq b.
\end{eqnarray}
The Weyl group orbifold $W$ has the same form as for $SO(2k+1)$:
it is an extension
\begin{equation}
1 \: \longrightarrow \: \left( {\mathbb Z}_2 \right)^k \: \longrightarrow \:
W \: \longrightarrow \: S_k \: \longrightarrow \: 1,
\end{equation}
where the symmetric group $S_k$ acts by exchanging $\sigma_a$,
and each ${\mathbb Z}_2$ acts by flipping signs of $\sigma_a$'s.

The Coulomb branch relation~(\ref{eq:spk:qc}) can be rewritten as
\begin{equation}
\sigma_a^N \: + \: \left( \sum_{i < j} \tilde{m}_i \tilde{m}_j \right)
\sigma_a^{N-2} \: + \: \left( \sum_{i_1 < i_2 < i_3 < i_4}
\tilde{m}_{i_1} \tilde{m}_{i_2} \tilde{m}_{i_3} \tilde{m}_{i_4} \right)
\sigma_a^{N-4} \: + \: \cdots \: = \: 0,
\end{equation}
which is symmetric under $\sigma \mapsto - \sigma$.

In this theory, $N$ is always odd, so
this has roots
\begin{equation}
\sigma \: = \: 0, \pm \tilde{\sigma}_1, \cdots, \pm \tilde{\sigma}_{(N-1)/2},
\end{equation}
where for generic $\tilde{m}_i$, none of the $\tilde{\sigma}_i$ vanish.
The zero root is not an allowed solution, as it sits on the
excluded locus.
The Weyl group orbifold acts on each $\sigma_a$ by sign flips, and
also exchanges different $\sigma_a$, so as a result, since there
are no twisted sector contributions, we see there are
\begin{equation}
\left( \begin{array}{c} (N-1)/2 \\ k \end{array} \right)
\end{equation}
vacua.

\subsection{Mirror to $Sp(N-2k-1)$ gauge theory}

The mirror to the second theory is a Weyl-group orbifold of a Landau-Ginzburg
model with fields
\begin{itemize}
\item $W^i_{\mu} = \exp( - Y^i_{\mu}/2)$, 
$ 1 \leq i \leq N$, $1 \leq \mu \leq N-2k-1$,
mirror to the fundamentals $\tilde{x}^i$, which we take to have R-charge $1$,
\item $A_{ij} = - A_{ji}$, mirror to the singlets $a_{ij}$, which we
take to have R-charge $0$,
\item $X_{\mu \nu}$, 
$\mu \leq \nu$,
$1 \leq \mu, \nu \leq N-2k-1$,
excluding $X_{2a-1,2a}$ (which would be mirror to the Cartan subalgebra),
\item $\sigma_a$, $1 \leq a \leq (1/2)(N-2k-1)$,
\end{itemize}
with superpotential
\begin{eqnarray}
W & = & \sum_{a} \sigma_a \Biggl( - 2 \sum_{i, \mu} 
\rho^a_{i \mu} 
\ln W^i_{ \mu}
\: - \: \sum_{\mu} \alpha^a_{\mu \mu} 
\ln X_{\mu \mu}
\nonumber \\
& & \hspace*{0.5in} \: - \: \sum_{b<c} \bigl(
\alpha^a_{2b,2c} \ln X_{2b,2c} + \alpha^a_{2b-1,2c-1} \ln X_{2b-1,2c-1}
\nonumber \\
& & \hspace*{1in}
+ \alpha^a_{2b-1,2c} \ln X_{2b-1,2c} +
 \alpha^a_{2b,2c-1} \ln X_{2b,2c-1}
\bigr) \Biggr)
\nonumber \\
& &
\: + \:
\sum_{i \mu} \left( W^i_{\mu} \right)^2 \: + \: 
2 \sum_{i \mu} \tilde{m}_i \ln W^i_{\mu}
\nonumber \\
& &
\: + \: \sum_{\mu} X_{\mu \mu} \: + \:
\sum_{a < b} \left( X_{2a,2b} + X_{2a-1,2b-1} +
X_{2a-1,2b} + X_{2a,2b-1} \right)
\nonumber \\
& &
\: + \: 
\sum_{i < j} \exp( - A_{ij} )
\: + \: \sum_{i < j} \left( \tilde{m}_i + \tilde{m}_j \right) A_{ij},
\end{eqnarray}
where
\begin{eqnarray}
\rho_{i \mu}^a & = & \delta_{\mu,2a} - \delta_{\mu,2a-1},
\\
\alpha^a_{\mu\nu} & = & \delta_{\mu,2a} - \delta_{\mu,2a-1}
+ \delta_{\nu,2a} - \delta_{\nu,2a-1}.
\end{eqnarray}
In addition to the Weyl orbifold, 
there is an orbifold by $({\mathbb Z}_2)^{N(N-k-1)}$,
where the generator of each ${\mathbb Z}_2$ acts as
\begin{equation}
W^i_{\mu} \: \mapsto \: -
W^i_{\mu}.
\end{equation}
The extra superpotential term of the dual theory
defines R-charges for the fields, but does not otherwise appear in the mirror.

Integrating out the $\sigma_a$ gives the constraints
\begin{equation}
\ln \left( \prod_{i=1}^N \frac{ W^i_{2a-1} }{W^i_{2a}} \right)^2 \: + \:
\ln \left( \prod_{\mu \leq \nu} \frac{ X_{2a-1,\nu} }{ X_{2a,\nu} } 
\frac{ X_{\mu,2a-1} }{ X_{\mu,2a} } \right) \: = \: 0.
\end{equation}
We define
\begin{eqnarray}
W_+^{i,a} & \equiv & W^{i,2a-1} W^{i,2a},
\\
W_-^{i,a} & \equiv & \frac{ W^{i,2a-1} }{ W^{i,2a} },
\end{eqnarray}
in terms of which the constraint becomes
\begin{equation}
\ln\left( \prod_{i=1}^N W_-^{i,a} \right)^2 
\: + \:
\ln \left( \prod_{\mu \leq \nu} \frac{ X_{2a-1,\nu} }{ X_{2a,\nu} } 
\frac{ X_{\mu,2a-1} }{ X_{\mu,2a} } \right) \: = \: 0.
\end{equation}
We use this constraint to eliminate $W_-^{N,a}$:
\begin{equation}
\ln \left( W_-^{N,a} \right)^2 \: = \: - \sum_{i=1}^{N-1} \ln 
\left( W_-^{i,a} \right)^2 \: - \:
\sum_{\mu \leq \nu} \ln \left(  \frac{ X_{2a-1,\nu} }{ X_{2a,\nu} } 
\frac{ X_{\mu,2a-1} }{ X_{\mu,2a} } \right).
\end{equation}
To make our notation consistent with earlier sections, define
\begin{eqnarray}
\Upsilon_a & \equiv & W_-^{N,a},
\\
& = & \left( \prod_{i=1}^{N-1} W_-^{i,a} \right)^{-1}
\left[ \prod_{\mu \leq \nu} \frac{ X_{2a,\nu} }{ X_{2a-1,\nu} }
\frac{ X_{\mu,2a} }{ X_{\mu,2a-1} } \right]^{1/2}.
\end{eqnarray}

The superpotential now becomes
\begin{eqnarray}
W & = &
\sum_{i=1}^{N-1} \sum_a W_+^{i,a} \left( W_-^{i,a} \: + \: 
\frac{1}{W_-^{i,a}} \right)
\: + \: \sum_a W_+^{N,a} \left( \Upsilon_a + \Upsilon_a^{-1} \right)
\nonumber \\
& & 
\: + \: \sum_{i=1}^N \sum_a \tilde{m}_i \ln \left( W_+^{i,a} \right)^2
\nonumber \\
& &
\: + \: \sum_{\mu} X_{\mu \mu} \: + \:
\sum_{a < b} \left( X_{2a,2b} + X_{2a-1,2b-1} +
X_{2a-1,2b} + X_{2a,2b-1} \right)
\nonumber \\
& &
\: + \: 
\sum_{i < j} \exp( - A_{ij} )
\: + \: \sum_{i < j} \left( \tilde{m}_i + \tilde{m}_j \right) A_{ij}.
\end{eqnarray}

The critical locus is then given by
\begin{eqnarray}
W_-^{ia}: & &
W_+^{i,a} \left( W_-^{i,a} \: + \: \frac{1}{W_-^{i,a}} \right) \: = \:
- 2 \tilde{m}_i
\: \: \: \mbox{ for }i < N,
\\
W_+^{N,a}: & &
W_+^{N,a} \left( \Upsilon_a + \Upsilon_a^{-1} \right) \: = \: - 2
\tilde{m}_N,
\\
W_-^{i,a}: & &
W_+^{i,a} \left( W_-^{i,a} \: - \: \frac{1}{W_-^{i,a}} \right)
\: = \: 
W_+^{N,a} \left( \Upsilon_a - \Upsilon_a^{-1} \right),
\: \: \: \mbox{ for }i < N,
\\
A_{ij}: & &
\exp\left( - A_{ij} \right) \: = \: \tilde{m}_i + \tilde{m}_j,
\\
X_{\mu \nu}: & &
X_{2a,2a} \: = \: W_+^{N,a} \left( - \Upsilon_a + \Upsilon_a^{-1} \right),
\\
& &
X_{2a-1,2a-1} \: = \: W_+^{N,a} \left( \Upsilon_a - \Upsilon_a^{-1} \right)
\: = \: - X_{2a,2a},
\\
& &
2 X_{2a,2b} \: = \: W_+^{N,a} \left( \Upsilon_a - \Upsilon_a^{-1} \right)
\: + \: W_+^{N,b} \left( \Upsilon_b - \Upsilon_b^{-1} \right)
\: \: \: \mbox{ for }a<b,
\\
& & 
2 X_{2a,2b-1} \: = \: W_+^{N,a} \left( \Upsilon_a - \Upsilon_a^{-1} \right)
\: + \: W_+^{N,b} \left( - \Upsilon_b + \Upsilon_b^{-1} \right),
\\
& & 
2 X_{2a-1,2b} \: = \: W_+^{N,a} \left( - \Upsilon_a + \Upsilon_a^{-1} \right)
\: + \: W_+^{N,b} \left( \Upsilon_b - \Upsilon_b^{-1} \right)
\\
& & \hspace*{1in}
\: = \: - 2 X_{2a, 2b-1},
\\
& & 
2 X_{2a-1,2b-1} \: = \: W_+^{N,a} \left( - \Upsilon_a + \Upsilon_a^{-1}
\right)
\nonumber \\
& & \hspace*{1.5in}
 \: + \: W_+^{N,b} \left( - \Upsilon_b + \Upsilon_b^{-1} \right)
\: \: \: \mbox{ for }a < b,
\\
& & \hspace*{1in}
\: = \: - 2 X_{2a,2b}.
\end{eqnarray}

On the critical locus, define
\begin{eqnarray}
\sigma_a & \equiv & \frac{1}{2} W_+^{i,a} \left( W_-^{i,a} \: - \: 
\frac{1}{W_-^{i,a}} \right) 
\: \: \: \mbox{ for }i < N,
\\
& = & \frac{1}{2} W_+^{N,a} \left( \Upsilon_a - \Upsilon_a^{-1} \right).
\end{eqnarray}
These expressions match due to the third critical locus equation above.

It is straightforward to verify the Coulomb branch
relation
\begin{equation}
\prod_{i=1}^N \left( \sigma_a - \tilde{m}_i \right) \: = \:
\prod_{i=1}^N \left( - \sigma_a - \tilde{m}_i \right).
\end{equation}

The excluded locus is given by
\begin{eqnarray}
\sigma_a & \neq & \pm \tilde{m}_i,
\\
\sigma_a & \neq & \pm \sigma_b 
\: \: \: \mbox{ for }a \leq b,
\end{eqnarray}
hence, for example, $\sigma_a \neq 0$.

Now, by assumption, $N$ is odd,
so the Coulomb branch relation above reduces to a degree $N$ polynomial
with roots
\begin{equation}
\sigma \: = \: 0, \pm \tilde{\sigma}_1, \cdots,
\pm \tilde{\sigma}_{(N-1)/2}.
\end{equation}
The zero root is excluded, and as the Weyl orbifold group exchanges
vacua and flips signs of individual $\sigma_a$, we find that the
total number of vacua is
\begin{equation}
\left( \begin{array}{c} (N-1)/2 \\ k \end{array} \right),
\end{equation}
all in the untwisted sector, which matches our previous result
for the number of vacua in the mirror to the corresponding
$Sp(2k)$ gauge theory.

Thus, we see that at the level of vacua, the nonabelian mirror proposal
is consistent with the symplectic-symplectic duality of
\cite{Hori:2011pd}.

\section{Conclusions}

In this paper we have described an extension of the nonabelian
mirrors proposal of \cite{Gu:2018fpm}, from mirrors of two-dimensional
(2,2) supersymmetric gauge theories
with connected gauge groups to include the gauge groups $O_{\pm}$.
We have checked the proposal by comparing mirrors to each side of the
gauge theories dualities of \cite{Hori:2011pd}.  Since the original
gauge duality relates gauge theories in the IR, the proposal here
relates the mirrors in the IR, and so we compared the number of vacua.
For the groups $O_{\pm}$, unlike connected gauge groups \cite{Gu:2018fpm},
the orbifold group acting on the Landau-Ginzburg mirror often has
fixed points intersecting the critical loci of the mirror superpotential,
and so one must take into account twisted sectors.  The mirror dualities
often relate untwisted sector vacua to twisted sector contributions,
making for a rather intricate test of the nonabelian mirror proposal.
In cases with nontrivial IR limits, we have also checked central charges,
and in addition we have also checked mirrors to some other unrelated properties
discussed in \cite{Hori:2011pd}.

\section{Acknowledgements}

We would like to thank J.~Knapp and Y.~Tachikawa for useful discussions.
E.S. was partially supported by NSF grant PHY-1720321.

\appendix

\section{Higgs branches and chiral rings}
\label{app:chiral}

Most of this paper and previous work \cite{Gu:2018fpm,Chen:2018wep}
has focused on understanding Coulomb branches of gauge theories in
mirror constructions.  Briefly, we suggest in this section that the
mirror to a Higgs branch might be encoded in asymptotic limits of
the fields, and based on that idea, we will outline a proposal for
mirrors to the chiral rings discussed in \cite{Hori:2011pd}.

Gauge-invariant operators in a (2,2) GLSM can roughly be characterized
into two types:
\begin{itemize}
\item Products of $\sigma$s, along a Coulomb branch.
These correspond to equivariant cohomology of the ambient space,
as discussed in detail in e.g. \cite{Jia:2014ffa}[appendix A].
\item Gauge-invariant combinations of matter fields, along a Higgs branch.
\end{itemize}

For a GLSM describing a compact space such as a hypersurface,
the gauge-invariant matter field combinations typically corresopnd to
algebraic complex structure deformations.  For example,
for the quintic hypersurface in ${\mathbb P}^4$,
the gauge-invariant field combinations have the form
\begin{displaymath}
p f_5(\phi),
\end{displaymath}
and products thereof, where $f_5(\phi)$ is a degree five polynomial
in the charge-one matter fields.  If the target is a Calabi-Yau threefold,
these can be identified with elements of $H^{2,1}$, modulo the usual
subtleties of overcounting and non-algebraic deformations.

For a GLSM with vanishing superpotential,
the same gauge-invariant field combinations merely correspond to
functions on the target space.  For example, for the GLSM with gauge
group $U(1)$ and fields
\begin{center}
\begin{tabular}{c|ccc}
 & $x_0$ & $x_1$ & $p$ \\ \hline
$U(1)$ & $1$ & $1$ & $-2$
\end{tabular}
\end{center}
the gauge-invariant field combinations are
\begin{displaymath}
p f_2(x_0, x_1),
\end{displaymath}
which simply correspond to the invariant function ring on
${\mathbb C}^2 / {\mathbb Z}_2$:
if we identify
\begin{equation}
x \: \equiv \: p x_0^2, \: \: \:
y \: \equiv \: p x_1^2, \: \: \:
z \: \equiv \: p x_0 x_1,
\end{equation}
then the ring of gauge-invariant polynomials is 
\begin{equation}
{\mathbb C}[x,y,z] / (xy = z^2).
\end{equation}
As the GLSM itself is describing a resolution of ${\mathbb C}^2/{\mathbb Z}_2$,
this is entirely appropriate.

In any event, due to the fact that for GLSMs describing compact spaces,
the rings of gauge invariant operators corresond to algebraic complex
structure deformations, this ring of GLSM operators is sometimes referred
to as the GLSM ``(c,c)'' ring, generalizing the notion of chiral rings
in (2,2) SCFTs \cite{Lerche:1989uy}.

In the remainder of this section, we will briefly outline a suggestion
for how to see such chiral rings for $(S)O(k)$ theories, from
\cite{Hori:2011pd}[section 4.7], in the mirror presented here, in the
special case $k=2$.
As discussed there, the chiral ring in the untwisted sector of an $O(k)$
theory is 
\begin{equation}
{\mathbb C}[ (x_i x_j) ] / J_1,
\end{equation}
and the chiral ring of an $SO(k)$ theory is
\begin{equation}
{\mathbb C}[ (x_i x_j), [x_{i_1} \cdots x_{i_k}] ] / (J_1, J_2, J_3),
\end{equation}
where $J_1$ denotes relations of the form
\begin{equation}
\det \left[ \begin{array}{ccc}
(x_{i_0} x_{j_0} ) & \cdots & (x_{i_0} x_{j_k} ) \\
\vdots & & \vdots \\
(x_{i_k} x_{j_0}) & \cdots & (x_{i_k} x_{j_k} ) \end{array} \right],
\end{equation}
$J_2$ denotes relations of the form
\begin{equation}
[ x_{i_1} \cdots x_{i_k} ] [ x_{j_1} \cdots x_{j_k} ] \: = \:
\det \left[ \begin{array}{ccc}
(x_{i_1} x_{j_1}) & \cdots & (x_{i_1} x_{j_k}) \\ 
\vdots & & \vdots \\
(x_{i_k} x_{j_1} ) & \cdots & (x_{i_k} x_{j_k} ) \end{array} \right],
\end{equation}
and $J_3$ denotes relations of the form
\begin{equation}
\sum_{p=0}^k (-)^p [ x_{i_0} \cdots \widehat{ x_{i_p} } \cdots
x_{i_k} ] ( x_{i_p} x_j ) \: = \: 0,
\end{equation}
where
\begin{eqnarray}
\left( x_i x_j \right) & = & \sum_{ab} x_i^a x_j^b \delta_{ab},
\\
\left[ x_{i_1} \cdots x_{i_k} \right] & = &
\det\left[ \begin{array}{cccc}
x_{i_1}^1 & x_{i_2}^1 & \cdots & x_{i_k}^1 \\
x_{i_1}^2 & x_{i_2}^2 & \cdots & x_{i_k}^k \\
\vdots & & & \vdots \\
x_{i_1}^k & x_{i_2}^k & \cdots & x_{i_k}^k
\end{array} \right].
\end{eqnarray}

Formally, we will consider a region where all the $Y^i \rightarrow \infty$.
As a result, defining
\begin{eqnarray}
Y_{-}^i & = & \frac{1}{2} \left( Y_{2}^i - Y_{1}^i \right),
\\
Y_{+}^i & = & \frac{1}{2} \left( Y_{2}^i + Y_{1}^i \right),
\end{eqnarray}
we see that each $Y_{+}^i \rightarrow \infty$, but we will take limits
in such a way that each $Y_{-}^i$ remains finite.

First, consider an $O(2)$ theory.  We make the ansatz that the mirror
of $(x_i x_j)$ is
\begin{equation}
(x_i x_j) \: \leftrightarrow \:
\exp\left( - Y_-^i \right) + \exp\left( - Y_-^j \right) + 
\exp\left( + Y_-^i \right) + \exp\left( + Y_-^j \right).
\end{equation}
It is straightforward to show that these quantities have relations in
the ideal $J_1$, which here is generated by relations of the form
\begin{equation}
\det \left[ \begin{array}{ccc}
( x_{i_0} x_{j_0} ) & ( x_{i_0} x_{j_1} ) & ( x_{i_0} x_{j_2} ) \\
( x_{i_1} x_{j_0} ) & ( x_{i_1} x_{j_1} ) & ( x_{i_1} x_{j_2} ) \\
( x_{i_2} x_{j_0} ) & ( x_{i_2} x_{j_1} ) & ( x_{i_2} x_{j_2} )
\end{array} \right].
\end{equation}

It is straightforward to check that the mirror operators have
vanishing determinants of $3 \times 3$ matrices of the form above,
meaning for example
\begin{displaymath}
\det \left[ \begin{array}{ccc}
x_1 + x_1^{-1} + y_1 + y_1^{-1} &
x_1 + x_1^{-1} + y_2 + y_2^{-1} &
x_1 + x_1^{-1} + y_3 + y_3^{-1} \\
x_2 + x_2^{-1} + y_1 + y_1^{-1} &
x_2 + x_2^{-1} + y_2 + y_2^{-1} &
x_2 + x_2^{-1} + y_3 + y_3^{-1} \\
x_3 + x_3^{-1} + y_1 + y_1^{-1} &
x_3 + x_3^{-1} + y_2 + y_2^{-1} &
x_3 + x_3^{-1} + y_3 + y_3^{-1} 
\end{array} \right] \: = \: 0,
\end{displaymath}
for $x_i = \exp(- Y_-^i )$, $y_j = \exp(- Y_-^j)$,
but for example corresponding $2 \times 2$ matrices do not have vanishing
determinant:
\begin{displaymath}
\det \left[ \begin{array}{cc}
x_1 + x_1^{-1} + y_1 + y_1^{-1} &
x_1 + x_1^{-1} + y_2 + y_2^{-1} \\
x_2 + x_2^{-1} + y_1 + y_1^{-1} &
x_2 + x_2^{-1} + y_2 + y_2^{-1} 
\end{array} \right] \: \neq \: 0,
\end{displaymath}
exactly as expected for the relations to be encoded in the ideal $J_1$.

Next, consider an $SO(2)$ theory.  We claim the mirror map acts on
chiral ring generators as
\begin{eqnarray}
\left(x_i x_j \right) & \leftrightarrow & \exp\left( - Y_-^i \right) +
\exp\left( - Y_-^j \right),
\\
\left[ x_i x_j \right]
 & \leftrightarrow & i \left( \exp\left( - Y_-^i \right)
- \exp\left( - Y_-^j \right) \right).
\end{eqnarray}

It is straightforward to check that the mirror operators obey the relations
in the ideals $J_1$, $J_2$, $J_3$.  For example, for
$x_i = \exp(- Y_-^i )$, $y_j = \exp(- Y_-^j)$,
it is straightforward to show that
\begin{displaymath}
\det \left[ \begin{array}{ccc}
x_1 + y_1 & x_1 + y_2 & x_1 + y_3 \\
x_2 + y_1 & x_2 + y_2 & x_2 + y_3 \\
x_3 + y_1 & x_3 + y_2 & x_3 + y_3 
\end{array} \right] \: = \: 0,
\end{displaymath}
but they do not obey relations encoded in determinants of smaller matrices,
for example
\begin{displaymath}
\det \left[ \begin{array}{cc}
x_1 + y_1 & x_1 + y_2 \\
x_2 + y_1 & x_2 + y_2
\end{array} \right] \: \neq \: 0.
\end{displaymath}
Hence, the mirror operators are in the ideal $J_1$, but do not obey a smaller
subset of those relations.
Similarly, it is straightforward to show that
\begin{displaymath}
- (x_1 - x_2) (y_1 - y_2) \: = \: \det \left[ \begin{array}{cc}
x_1 + y_1 & x_1 + y_2 \\
x_2 + y_1 & x_2 + y_2 \end{array} \right],
\end{displaymath}
which generates the ideal $J_2$, and also 
\begin{displaymath}
i (x_1 - x_2) (x_0 + y) \: - \: i (x_0 - x_2)(x_1 + y) \: + \:
i (x_0 - x_1) (x_2 + y) \: = \: 0,
\end{displaymath}
which generates the ideal $J_3$.

We leave proposals for mirrors to chiral ring structures in $(S)O(k)$
theories for $k>2$ to future work.

\end{document}